\pdfoutput=1

\documentclass{article}

\usepackage{microtype}
\usepackage{graphicx}
\usepackage{subcaption}
\usepackage{booktabs} 
\usepackage{minitoc}
\usepackage{hyperref}
\usepackage{url}
\usepackage{graphicx} 
\usepackage{algorithm}
\usepackage{algorithmic}
\usepackage[table]{xcolor}
\usepackage{graphicx}
\usepackage{caption}
\usepackage{subcaption}
\usepackage{placeins}
\usepackage{newfloat}
\usepackage{listings}
 \usepackage{multirow}
\usepackage{adjustbox}
\usepackage{xcolor}
\usepackage{hyperref}

\usepackage[accepted]{icml2026}

\usepackage{amsmath}
\usepackage{amssymb}
\usepackage{mathtools}
\usepackage{amsthm}

\usepackage[capitalize,noabbrev]{cleveref}

\theoremstyle{plain}
\newtheorem{theorem}{Theorem}[section]

\newtheorem{lemma}[theorem]{Lemma}

\theoremstyle{definition}

\theoremstyle{remark}

\usepackage[textsize=tiny]{todonotes}

\icmltitlerunning{Breaking the Capacity Bottleneck in Model-Heterogeneous Federated Learning
via Gradual Model Restoration}

\begin{document}

\twocolumn[
  \icmltitle{Breaking the Capacity Bottleneck in Model-Heterogeneous Federated Learning via Gradual Model Restoration}




  \begin{icmlauthorlist}
    \icmlauthor{Chengjie Ma}{yyy}
    \icmlauthor{Seungeun Oh}{yyy,comp}
    \icmlauthor{Jihong Park}{comp}
    \icmlauthor{Seong-Lyun Kim}{yyy}
  
  \end{icmlauthorlist}

  \icmlaffiliation{yyy}{School of Electrical and Electronic Engineering, Yonsei University, Seoul, South Korea}
  \icmlaffiliation{comp}{Information Systems Technology and Design Pillar, Singapore University of Technology and Design, Singapore, Singapore}

  \icmlcorrespondingauthor{Seong-Lyun Kim}{slkim@yonsei.ac.kr}

  \icmlkeywords{Model-heterogeneous, Federated Learning, Model Capacity, Convergence Analysis}

  \vskip 0.3in
]



\printAffiliationsAndNotice{}  

\begin{abstract}
Federated learning (FL) enables distributed model training, yet in heterogeneous deployments,
Bandwidth-Constrained Clients (BCCs) often contribute inefficiently due to limited uplink bandwidth.
In model-heterogeneous FL with fixed small sub-models, BCCs with sub-models may improve quickly in early rounds but become under-parameterized later,
resulting in slow convergence and poor generalization.
To address this challenge, we propose \textbf{FedGMR}, a federated learning framework centered around
\emph{Gradual Model Restoration} (GMR), where GMR progressively increases each client’s sub-model density during training, allowing BCCs to remain effective contributors throughout optimization.
To make GMR practical under real-world heterogeneity, FedGMR is realized as an
end-to-end workflow with asynchronous coordination and stable, mask-aware aggregation.
We further establish convergence guarantees, showing that the aggregation error scales with the
\emph{average} sub-model density across clients and rounds, and that GMR provably narrows the gap toward full-model FL.
Extensive experiments on FEMNIST, CIFAR-10, ImageNet-100, and StackOverflow demonstrate that FedGMR improves both convergence speed and final accuracy,
especially under severe heterogeneity and non-IID data distributions.
Code is available at \url{https://github.com/machengjie321/FedGMR}.
\end{abstract}

\section{Introduction}
Federated learning (FL) enables multiple clients to collaboratively train a global model without sharing local data \cite{pei2024review,hu2024overview}. In practical deployments, clients frequently exhibit substantial variability in computation and communication capabilities \cite{zhang2022scalable}. This motivates model-heterogeneous FL (MHFL) \cite{diao2021heterofl, wu2024fiarse,liu2025fedgraft,kim2023depthfl} , where bandwidth-limited clients (BCCs) are assigned compact sub-models to reduce their training and transmission cost.

However, an important limitation of MHFL has received much less attention:
\textbf{small sub-models may work well early on, yet their contribution often fades as training progresses.}
Early in training, compact sub-models can capture a coarse decision boundary and yield fast gains.
As the global model moves toward a higher-capacity solution, these sub-models become under-parameterized, making sub-models updates weak, noisy, or uninformative.
This is not merely an optimization artifact, but a consequence of an intrinsic \emph{capacity gap}:
sub-model training restricts learning to a smaller hypothesis class (e.g., width-reduced or mask-restricted parameterizations), whose best achievable risk typically plateaus earlier as the target boundary becomes more complex \cite{yu2019universally}.
Moreover, updating only a subset of parameters projects local gradients onto a constrained subspace, further reducing gradient fidelity in later rounds \cite{sung2021training}.

This capacity-induced plateau leads to \emph{client marginalization} reminiscent of asynchronous FL (AFL) \cite{xie2019asynchronous, lan2024asynchronous, chen2025advances, zhou2024towards}, but for a different reason:
in AFL, BCCs are marginalized because they contribute fewer (often stale) updates,
whereas in MHFL, BCCs may participate regularly yet become marginalized because their sub-models lack sufficient capacity to keep improving.
As a result, aggregation is increasingly dominated by full-capacity clients, biasing the global model toward full-model solutions rather than the true global optimum, especially under non-IID data.
Restoring capacity is the most direct way to overcome this limitation,
This motivates a natural question:
\emph{Can we design an MHFL framework in which small sub-models are used when they are
beneficial—i.e., fast in early rounds—but are gradually replaced by larger ones once their limited
capacity prevents further progress?}  
Equivalently, the optimal sub-model density for BCCs is \emph{dynamic} and should gradually increase as training progresses.
\begin{figure*}[htbp]                
    \centering
    \includegraphics[width=0.90\textwidth, height=0.18\textheight]{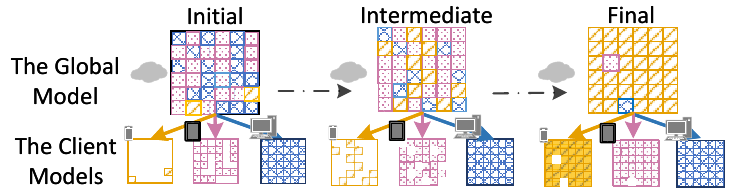} 
\caption{\textbf{GMR} overview across training stages. The server maintains a full global model (top row), while clients train heterogeneous masked sub-models (bottom row) with different densities. GMR progressively restores pruned coordinates, increasing client densities from the initial to the final stage, so lightweight sub-models accelerate early training and larger models sustain late-stage convergence. Arrows indicate extracting masked sub-models from the global model and aggregating client updates back to the server.}

\label{fig1}
\end{figure*}

To overcome this limitation, we propose \textbf{FedGMR}, a model-heterogeneous
federated learning framework built around a core mechanism:
\emph{Gradual Model Restoration} (GMR).
Unlike prior MHFL methods with static sub-model structures,
GMR allows client model capacity to \emph{evolve with training progress},
restoring parameters only when additional expressivity becomes necessary.
This enables BCCs to benefit from fast early-stage training while avoiding
capacity saturation in later rounds.
Importantly, dynamic restoration reintroduces latency imbalance and
heterogeneous updates, but this tradeoff is worthwhile to avoid late-round
marginalization of BCCs.
FedGMR therefore realizes GMR through a dedicated training workflow that
supports asynchronous coordination and stable aggregation under evolving
sub-model structures.
The resulting framework provides a principled way to reconcile efficiency,
capacity, and robustness in heterogeneous federated learning.

\paragraph{Contributions.}
This work makes four contributions:
\begin{itemize}
\item We reveal an overlooked limitation of model-heterogeneous FL: static
sparse sub-models may accelerate early training but inevitably saturate in
later rounds, leading to capacity-induced marginalization of BCCs.
\item We propose FedGMR, a model-heterogeneous FL framework driven by Gradual
Model Restoration (GMR), together with a supporting asynchronous workflow that
enables dynamic density scheduling under practical heterogeneity.
\item We provide convergence guarantees showing that the global optimization
bias scales with the average sub-model density across clients and rounds, and
that GMR monotonically reduces this bias toward full-model FL. We further
analyze aggregation during restoration and show that mask-aware aggregation
better preserves gradient stability.
\item Extensive experiments on FEMNIST, CIFAR-10, ImageNet-100, and StackOverflow benchmarks demonstrate faster
and more stable convergence, clear gains brought by GMR, and improved
robustness under strong structural and data heterogeneity. We further apply
GMR to multiple MHFL methods and observe consistent gains across different
pruning and sub-model construction strategies, suggesting that model capacity
restoration is broadly effective.
\end{itemize}

\noindent\textbf{Conflict of Interest Disclosure.}
The authors declare that they have no financial conflicts of interest related to this work.

\section{Related Works}
\paragraph{Heterogeneous-client federated learning.}
Client heterogeneity is pervasive in practical FL, arising from diverse computation and communication capabilities.
System- and protocol-level solutions mitigate stragglers and stale updates via asynchronous or semi-asynchronous coordination,
often combined with staleness-aware aggregation and drift correction
\cite{xie2019asynchronous, lan2024asynchronous, chen2025advances, zhou2024towards}.
Beyond protocol-level mitigation, another major direction is \emph{MHFL},
which assigns models of different sizes to match client resource budgets, thereby reducing local training and uplink costs.
Representative strategies include width/depth scaling
\cite{diao2021heterofl, ilhan2023scalefl, kim2023depthfl,horvath2021fjord,wu2024fiarse},
distillation-based aggregation across heterogeneous architectures
\cite{Lin_Kong_Stich_Jaggi_2020},
and pruning-based sub-model construction
\cite{jiang2022fedmp, vahidian2021personalized}.
While these methods significantly improve resource-efficiency and participation of low-end clients,
most of them still employ \emph{static or budget-limited} sub-models, which can saturate as training proceeds.

\paragraph{Model capacity and sub-model saturation.}
Despite these advances, most MHFL methods rely on \emph{static} sub-model sizes throughout training.
As optimization progresses and the global decision boundary becomes more complex,
compact sub-models may become under-parameterized and their updates can saturate, weakening their contribution in later rounds.
This issue is closely related to uneven parameter training under sub-model participation (e.g., FedRolex)
\cite{alam2022fedrolex},
and is also supported theoretically: \cite{zhou2023every} establishes convergence bounds for MHFL and shows that
compact sub-models can inherently converge to sub-optimal solutions compared to full-model training.
Related evidence from sub-network learning further reports an accuracy--width trade-off \cite{yu2019universally},
and shows that training with fixed sparse masks/sparse updates effectively restricts optimization to a constrained parameter subspace
\cite{sung2021training}.

\paragraph{Summary.}
Protocol-level heterogeneity methods mitigate stragglers and staleness, whereas MHFL reduces per-round cost by assigning compact sub-models.
However, static MHFL may suffer from late-stage saturation due to limited sub-model capacity.
Our work addresses this unresolved efficiency--capacity tension by gradually restoring sub-model density during training.

\section{Preliminary: Model-Heterogeneous Federated Learning}

We consider a FL system with a central server and $C$ heterogeneous clients, with client set $\mathcal{C}=\{1,\dots,C\}$. Each client $i\in\mathcal{C}$ holds a local dataset $D_i$ and possesses a different communication budget. To keep training speeds comparable across devices, MHFL assigns \emph{smaller sub-models} to BCCs, while bandwidth-abundant clients train larger ones.

\paragraph{Sub-model representation.}
Let $w\in\mathbb{R}^N$ denote the global model with $N$ parameters. At round $k$, client $i$ trains a masked sub-model defined by $m_{i,k}\in\{0,1\}^N$, where $m_{i,k}^{(n)}=1$ indicates that parameter $n$ is retained. For each coordinate $n$, its coverage set is $C_k^{(n)}=\{\,i\in\mathcal{C}\mid m_{i,k}^{(n)}=1\,\}$ with $1\le|C_k^{(n)}|<C$. We denote the sub-model density by $\rho_{i,k}\in(0,1]$, so that  $\rho_{i,k}=\frac{\|m_{i,k}\|_0}{N}$ gives the number of trainable parameters (i.e., a capacity budget). We keep this notation throughout the paper, including the convergence analysis. From a hypothesis-class viewpoint, masking constrains optimization to a smaller parameterization, which may plateau earlier when higher capacity is needed \cite{shalev2014understanding}.
\paragraph{Local objective and updates.}
At round $k$, client $i$ trains a masked sub-model specified by $m_{i,k}$.
It starts from $w_{i,k,0}=W_k\odot m_{i,k}$ and minimizes the masked objective
$F_i(w;m_{i,k})=\mathbb{E}_{\xi\sim\mathcal{D}_i}[\ell(w\odot m_{i,k};\xi)]$.
The client runs $\mathcal{T}$ steps of SGD with learning rate $\gamma>0$.
At step $\tau$, it applies $w_{i,k,\tau}=w_{i,k,\tau-1}-\gamma\,\nabla F_i(w_{i,k,\tau-1},\xi_{i,\tau-1})\odot m_{i,k}$, so pruned coordinates receive zero updates.
This is equivalent to projecting the (full-model) stochastic gradient
onto the feasible subspace induced by $m_{i,k}$ \cite{sung2021training}.

\paragraph{MaskFedAvg.}
  Clients train sub-models with different masks, so their local updates contain zeros on pruned coordinates. Naively averaging such masked updates, as in FedAvg~\cite{mcmahan2017communication}, would treat these
  zeros as contributions and dilute parameters updated only by bandwidth-abundant clients toward zero. To avoid this issue, we use the term \textit{MaskFedAvg} to denote an aggregation rule that averages each
  coordinate only over the clients that retain it. Although a similar rule appears in \citet{vahidian2021personalized}, that work did not introduce it in the MHFL setting. Concretely, the server computes $w^{(n)}
  = \big( \sum_{i \in C_k^{(n)}} w_i^{(n)} \big) / |C_k^{(n)}|$ for $n=1,\dots,N$, ensuring that each coordinate is updated under structural heterogeneity.

\section{FedGMR}
We now present the details of GMR, then introduce the full process of the FedGMR framework. 

\subsection{Gradual model Restore}
Under MHFL, assigning low-density sub-models with density $\rho_i$ allows BCCs to communicate and update more frequently.
However, as optimization progresses, overly sparse sub-models become capacity-limited and their updates gradually saturate.
Increasing $\rho_i$ improves model capacity but also increases per-update latency, reducing the effective update rate in wall-clock time.
Therefore, the most beneficial density is \emph{stage-dependent}: early training favors smaller $\rho_i$ for fast progress, while later training requires larger $\rho_i$ to avoid capacity-induced stagnation.
This motivates \emph{Gradual Model Restoration (GMR)}, which progressively increases sub-model density as training deepens.

\paragraph{Empirical validation.}
To examine this hypothesis in FL, we conduct multiple FL runs where each run fixes the sub-model density throughout training and keeps all other settings identical. All runs share the same client sampling, optimizer, and aggregation schedule; only the density differs. We then compute the \emph{accuracy--time slope} to quantify the learning speed at different stages, where time is measured by the number of server aggregation steps, and stages correspond to when accuracy improves rapidly in the early rounds versus when progress slows and approaches a plateau in later rounds.

\begin{figure*}[t]
    \centering
    \includegraphics[width=0.98\textwidth,keepaspectratio,trim={0 0 0 0},clip]{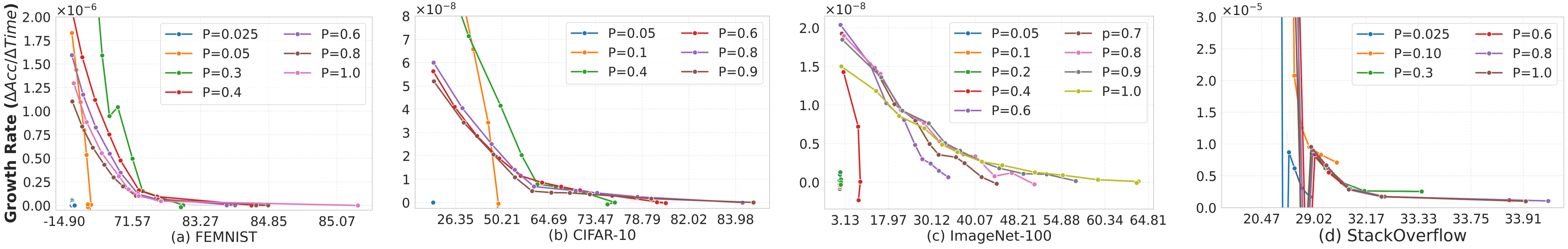}
    \caption{Accuracy growth rates under different model densities. Smaller models improve faster in early stages, while larger models become more advantageous later.}
    \label{fig:optimal_density_curve}
\end{figure*}

As shown in Fig.~\ref{fig:optimal_density_curve}, low-density sub-models achieve the steepest early-stage improvement (e.g., Conv2D: $\leq 0.3$--$0.4$, VGG-11: $\leq 0.3$--$0.4$, ResNet: $\leq 0.6$, Transformer: $\leq 0.3$), whereas high densities ($0.9$--$1.0$) dominate in later stages as optimization requires higher capacity. Moreover, extremely small densities (e.g., $\rho=0.025$) consistently plateau quickly, whereas near-full densities ultimately converge to substantially higher final accuracy. These results provide direct evidence that the optimal density shifts over training, motivating a gradual restoration strategy that increases model capacity as training proceeds.

\paragraph{When to restore sub-models?}
As discussed earlier, low-density sub-models cannot accelerate FL training
indefinitely. To maintain learning effectiveness, their capacity can be
gradually restored. However, \emph{when} to initiate restoration is critical.
Restoring too early causes sub-models to rapidly approach the full model,
effectively degenerating MHFL into asynchronous FL. Conversely, restoring too
late makes the process resemble fixed-density MHFL, where limited sub-model
capacity prevents further convergence.

Ideally, one would quantify the trade-off among model capacity, convergence speed, and training latency. In practice, this relationship is highly task- and architecture-dependent, making explicit modeling difficult. Instead, we consider an idealized but insightful scenario. Under the hypothesis-class view that masked sub-models optimize over a constrained parameterization of the full model \cite{shalev2014understanding}, a capacity-limited sub-model may reach a sub-optimal stationary point where its training loss decreases very slowly (i.e., its updates plateau), even though additional capacity would enable further improvement.
To make this notion concrete, we define the \emph{average convergence speed} over an interval $[t_1,t_2]$ as $\mathcal{A}(t_1,t_2):=-\big(F(w_{t_2})-F(w_{t_1})\big)/(t_2-t_1)$. Once a sub-model enters a plateau, restoring its capacity yields a strictly higher subsequent speed, i.e., $\mathcal{A}(t_2,t_3)>\mathcal{A}(t_1,t_2)$ for some $t_1<t_2<t_3$, indicating renewed descent after stagnation. The formal proof is provided in Appendix~\ref{sec:GMR_effectness}.

Motivated by this, GMR triggers density restoration when the training progress enters a \emph{plateau phase}.
Specifically, the server runs a unified \textbf{Early Stopping Mechanism (ESM)} at a fixed check interval,
monitoring a validation metric over a sliding window of recent checkpoints.
If the best validation performance does not improve for $k_{\text{rest}}$ consecutive checks,
the corresponding sub-model is deemed saturated and restoration is activated.
Furthermore, inspired by Once-for-All training~\cite{cai2020once}, FedGMR adopts a \emph{multi-level} density schedule.
Given a density ladder $\mathcal{R}=\{r_1<\cdots<r_M\}$, a sub-model at density $\rho=r_m$ is restored to $\rho\leftarrow r_{m+1}$
once the ESM condition is satisfied (e.g., $0.1\rightarrow0.2$ for $\{0.05,0.1,0.2,0.5,1.0\}$).
This stepwise schedule enables gradual capacity expansion, stabilizing optimization during restoration.

\subsection{Workflow of FedGMR}

\paragraph{Server-side asynchronous coordination.}
FedGMR follows a server-driven training workflow.
After introducing density restoration, a key challenge re-emerges:
once some clients restore to higher-density sub-models, their per-round
latency increases, reintroducing severe training-speed heterogeneity.
To prevent restored clients from becoming stragglers and blocking
global progress, FedGMR adopts a \emph{semi-asynchronous} coordination
scheme at the server.
Notably, asynchrony is most critical \emph{after} restoration begins.
At early stages, sub-models are small and similar in cost, and updates
arrive nearly synchronously even under an asynchronous protocol. The detailed process is specified in Algorithm~\ref{alg:agmr}.

\paragraph{Buffer-based mask-aware aggregation.}
Under asynchronous arrivals and heterogeneous sub-models, naive aggregation
can easily become unstable.
FedGMR therefore employs a buffer-based, mask-aware aggregation rule,
denoted as \textbf{BuffMaskFedAvg} (Algorithm~\ref{algorithm:MaskFedAvg}).
The server maintains a buffer of the most recently received client models
$\{\mathbf{w}_{i,k'_i}\}$, which mitigates \emph{asynchronous forgetting}
by preventing frequent updates from fast clients from immediately overwriting
information contributed by slower ones.
To further handle out-of-order arrivals, staleness-aware weighting is applied
at the neuron level.
For neuron $n$, the staleness weight is defined as
\(s^{(n)}_{i,k}=\bigl(1+(k-k'_i)\bigr)^{-\alpha},\qquad
\beta^{(n)}_{i,k}=
\frac{s^{(n)}_{i,k}}{\sum_{j\in C_k^{(n)}}s^{(n)}_{j,k}},\)
where $C_k^{(n)}$ denotes the set of clients whose sub-models retain neuron
$n$.
The global aggregation is then performed neuron-wise as
\begin{equation}
W_k^{(n)}=
\begin{cases}
\displaystyle
\frac{\sum_{i\in C_k^{(n)}}\beta^{(n)}_{i,k}\, w^{(n)}_{i,k}}
{\sum_{i\in C_k^{(n)}}\beta^{(n)}_{i,k}},
&\text{if }\sum_{i\in C_k^{(n)}}\beta^{(n)}_{i,k}>0,\\[5pt]
W_{k-1}^{(n)}, &\text{otherwise}.
\end{cases}
\end{equation}
This preserves neuron-wise consistency by normalizing only over
clients that cover the corresponding coordinates.
\paragraph{Density restoration via GMR.}
Within this asynchronous aggregation framework, GMR acts as a
\emph{stage-dependent density controller} at the server.
As training progresses, the server monitors whether a client’s current
sub-model has entered a saturation phase.
Once saturation is detected, GMR increases the client’s model density to the
next available level, restoring additional parameters while keeping existing
ones unchanged.
Through this mechanism, sub-model capacity is gradually expanded only when
needed, allowing clients to escape expressivity bottlenecks without destabilizing
the aggregation process.

\paragraph{Client-side sub-model execution.}
At each aggregation round, after the server updates the global model
$W_{k+1}$ and the client-density vector $\mathbf{P}_{k+1}$,
it prepares client-specific sub-models according to the current densities.
Concretely, the server extracts sub-models
$\{w_{i,k+1}\}_{i=1}^C$ induced by $\mathbf{P}_{k+1}$ (Algorithm~\ref{algorithm:Model_Pruning}) and dispatches them to clients.
To reduce server--to--client communication overhead during density restoration,
FedGMR optionally applies \emph{incremental model splitting} (IMS; Appendix~\ref{sec:IMS}) to reduce server-to-client communication. Since sub-models are nested across densities, IMS splits a sub-model into a base part and incremental blocks, the server transmits only the increment rather than resending the full sub-model.
Clients reconstruct their local sub-models by merging these increments with
previously received parameters.
IMS does not alter the optimization logic of GMR and is therefore orthogonal
to the core mechanism.
Each client then trains only its assigned sub-model on local data,
performing standard local updates independently.
Upon completion, the updated sub-model is asynchronously uploaded to the server
and stored in the buffer for subsequent aggregation.
\begin{algorithm}[t]
\caption{FedGMR}
\label{alg:agmr}
\footnotesize
\begin{algorithmic}
\STATE \textbf{Initialize:} global model $W_0$; client densities $\mathbf{P}_0=\{\rho_{i,0}\}_{i=1}^C$; buffer $\mathcal{B}\leftarrow\emptyset$; aggregation times $T_1=0$, and $T_k=T_{k-1}+\Delta T$ for $k>1$
\WHILE{$T < T_{\max}$}
  \STATE
  \STATE \textbf{--- Client-side local updates ---}
  \FORALL{client $i$}
    \IF{client $i$ is free}
      \STATE $w_{i,k+1} \leftarrow \mathrm{LocalTrain}(w_{i,k}, \mathcal{D}_i)$
      \STATE \textit{Upload}($w_{i,k+1}$)
    \ENDIF
  \ENDFOR
  \STATE
  \STATE \textbf{--- Server-side semi-asynchronous aggregation ---}
  \IF{a client update $w_{i,k}$ arrives}
    \STATE $\mathcal{B} \leftarrow \mathcal{B} \cup \{ w_{i,k}\}$
  \ENDIF
  \IF{$T = T_k$}
    \STATE $W_{k+1} \leftarrow \textit{BuffMaskFedAvg}(\mathcal{B}, W_k)$
    \FORALL{client $i$}
      \IF{\textit{ESM}$(w_i, D_{\text{val}})$}
        \STATE $\rho_{i,k+1} \leftarrow \min\!\Bigl\{1,\; \min\{\rho \in \mathbf{P}_k \mid \rho > \rho_{i,k}\}\Bigr\}$
      \ELSE
        \STATE $\rho_{i,k+1} \leftarrow \rho_{i,k}$
      \ENDIF
    \ENDFOR
    \STATE $\mathbf{P}_{k+1} \leftarrow \{\rho_{i,k+1}\}_{i=1}^C$
    \STATE $\{w_{i,k+1}\}_{i=1}^C \leftarrow \textit{Extract}(W_{k+1}, \mathbf{P}_{k+1})$
  \ENDIF
\ENDWHILE
\end{algorithmic}
\end{algorithm}

\section{Convergence Analysis}
We analyze how sub-model sparsity shapes the convergence behavior of
MHFL, aiming to characterize the optimization gap relative to
full-model FL and clarify why increasing density (as in FedGMR) narrows this
gap.
Sparsity impacts aggregation in two fundamental ways:  
(1) each client supplies only partial gradients, shrinking the effective
optimization subspace; and  
(2) heterogeneous masks induce uneven coverage across clients, creating
aggregation discrepancy.  
These effects accumulate over rounds, producing a density-dependent gap that
monotonically decreases as model density increases—providing the intuition
behind GMR.

To isolate the effect of sparsity, we analyze the synchronous 
setting and reduce BuffMaskFedAvg to its core \emph{mask-aware aggregation}
(MA) operator.  
This abstraction allows us to focus on how masked gradients are aggregated and
how coverage across density groups determines the resulting convergence gap.
We now state the assumptions and present the formal results.
Following standard FL practice, each pruned sub-model is treated as the full
model composed with a binary masking operator; for client $i$ at round $k$, the
density $\rho_{i,k}$ extends classical smoothness, variance, and heterogeneity
assumptions to masked structures. For a structural group $g$, we write
$\rho_{g,k}$ for its common density at round $k$.

\textbf{Assumption 5.1.} (\textit{Smoothness}). All client cost functions \( F_i \) are \( L \)-smooth. That is, for any \( w, \phi \in \mathbb{R}^N \) and any client \( i \), there exists a constant \( L > 0 \) such that:
\begin{align}
     \| \nabla F_i(w) - \nabla F_i(\phi) \| &\leq L \| w - \phi \|, \\
   \| \nabla F_i(w \odot m_1) - \nabla F_i(\phi \odot m_2) \| 
    &\leq L \| w \odot m_1 - \phi \odot m_2 \|.\notag
    \label{eq:lipschitz_pruned}
\end{align}
\textbf{Assumption 5.2.} (\textit{Bounded Gradient}).  
Define the stochastic full gradient
\(g_{i,k,\tau} := \nabla F_i(W_{i,k,\tau}, \xi_{i,k,\tau}).\)
Then for any client $i$, round $k$, and local step $\tau$, it holds that
\begin{align}
   \mathbb{E}\| g_{i,k,\tau} \|^2 &\leq G^2, \\
   \mathbb{E}\| g_{i,k,\tau} \odot m_{i,k} \|^2 &\leq f_1^2(\rho_{i,k}) G^2. \notag
\end{align}
\textbf{Assumption 5.3.} (\textit{Gradient Noise}).  
Define the gradient noise
\(n_{i,k,\tau} := \nabla F_i(W_{i,k,\tau}, \xi_{i,k,\tau}) - \nabla F_i(W_{i,k,\tau}).\)
Then for any $i,k,\tau$, it holds that
\begin{align}
   \mathbb{E}\| n_{i,k,\tau} \|^2 &\leq \sigma^2,\\
   \mathbb{E}\| n_{i,k,\tau} \odot m_{i,k} \|^2 &\leq f^2_2(\rho_{i,k}) \sigma^2.\notag
\end{align}
\textbf{Assumption 5.4.} (\textit{Bounded Non-IID Bias}).  
Define the Non-IID bias
\(b_i(W) := \nabla F_i(W) - \nabla F(W).\)
Then it holds that
\begin{align}
   \mathbb{E}\| b_i(W) \|^2 &\leq \zeta^2, \\
   \mathbb{E}\| b_i(W) \odot m_{i,k} \|^2 &\leq f_3^2(\rho_{i,k}) \zeta^2. \notag
\end{align}
where $f_i(\cdot)\!\in\![0,1]$ for $i\!\in\!\{1,2,3\}$ are non-decreasing
sparsity–scaling functions that capture how density affects the drift,
variance, and non-IID bias.

The notation used in the proofs is summarized in Table~\ref{tab:conv_notation}
at the start of Appendix~\ref{sec:convergence_analysis}. We adopt the group-wise regrouping notation formalized in
Appendix~\ref{sec: two_stage}. In particular, $S_{g,k}:=\{n:m_{g,k}^{(n)}=1\}$
denotes the retained-coordinate set of density group $g$, and
$c_{g,k}^*:=\min_{n\in S_{g,k}} |C_k^{(n)}|$ denotes the minimum coordinate
coverage within that group.  
This decomposition enables tighter and more structured convergence bounds by
separating intra-group averaging from inter-group mask-aware aggregation.
Appendix~\ref{sec:convergence_analysis} reorganizes the full proof around this
decomposition.

\paragraph{Proof sketch.}
The proof has four steps. First, we apply smoothness to one communication round
under MA and obtain a master descent inequality consisting of a descent term
plus three error terms: gradient drift caused by local masked training,
stochastic variance, and the coverage mismatch introduced by heterogeneous
masks. Second, we regroup clients by structural group and rewrite each
coordinate-wise average as a weighted combination of group averages. This is
the key step that introduces the coverage factors
\(\frac{|C_g|}{c_{g,k}^*}\) and \(\frac{|C_g|}{(c_{g,k}^*)^2}\). Third,
Lemmas~\ref{lem:2}--\ref{lem:5} upper-bound the drift, variance, and Non-IID
bias terms using the density-scaling functions \(f_1,f_2,f_3\). Finally,
substituting these bounds yields the one-round result in Theorem~\ref{thm:ma_one_round}, and
telescoping over rounds gives Theorem~\ref{thm:ma_telescope}. Under GMR, densities increase over
time, coordinate coverage improves, and the resulting density-dependent
penalties shrink accordingly.
\paragraph{Theorem 5.5.}
\label{thm:ma_one_round}
Under Assumptions~5.1--5.3, suppose all client masks satisfy
$|C_{k}^{(n)}| \ge 1$.
Then the expected one-round descent bound $\mathbb{E}[F(W_{k+1})]-\mathbb{E}[F(W_k)]$ satisfies
\begin{align}
&\mathbb{E}\big[F(W_{k+1})\big]-\mathbb{E}\big[F(W_k)\big]  \notag\\
\le &
- \frac{G_0}{2} \, \mathbb{E}\|\nabla F(W_k)\|^{2} \notag \\
&+ \frac{3G_0J_0}{2} \, \mathbb{E}\sum_{n=1}^{N}\Bigl\|\tfrac{1}{|C_k^{(n)}|}\sum_{j\in C_k^{(n)}}\nabla F_j^{(n)}(W_k)\Bigr\|^{2} \notag \\
&+ \frac{G_0 H_0}{6} \sum_{g=1}^{\mathcal{G}} \frac{|C_g|}{c_{g,k}^{*}} f_1^2(\rho_{g,k}) G^2 \\
&+ \frac{3 G_0 I_0}{2} \sum_{g=1}^{\mathcal{G}} \frac{|C_g|}{(c_{g,k}^{*})^{2}} f_2^2(\rho_{g,k}) \sigma^2.
\label{eq:main_descent_bound}
\end{align}
where
\(G_0=\mathcal{T}\gamma\),
\(H_0=L^{2}\mathcal{T}^{2}\gamma^{2}(3L\mathcal{T}\gamma+1)\),
\(I_0=L\gamma\),
and \(J_0=L\mathcal{T}\gamma\).
Full proof is provided in Appendix~\ref{sec:convergence_threome_1}.

\textbf{Interpretation.}
Theorem~\ref{thm:ma_one_round} isolates the effect of mask-aware aggregation (MA) on a single
optimization step.
The second term,
\(\mathbb{E}\sum_{n=1}^{N}\bigl\|\tfrac{1}{|C_k^{(n)}|}\sum_{j\in C_k^{(n)}}\nabla F_j^{(n)}(W_k)\bigr\|^{2}\),
arises because MA reconstructs the full-model gradient by normalizing each
coordinate only over the clients that retain it.
The last two terms scale with
\(\sum_{g=1}^{\mathcal{G}} \frac{|C_g|}{c_{g,k}^{*}} f_1^2(\rho_{g,k})\)
and
\(\sum_{g=1}^{\mathcal{G}} \frac{|C_g|}{(c_{g,k}^{*})^{2}} f_2^2(\rho_{g,k})\),
which are inverse-coverage-weighted penalty factors. Since
\(c_{g,k}^{*}\le |C_g|\), the factor \(|C_g|/c_{g,k}^{*}\) is always at least
one, while \(|C_g|/(c_{g,k}^{*})^{2}\) is not necessarily larger than one.
More importantly, both factors increase monotonically as \(c_{g,k}^{*}\)
decreases. Therefore, when coverage becomes worse, the corresponding drift and
variance error terms become larger, which leads to a looser convergence
neighborhood; when coverage improves, these penalties shrink accordingly.
Compared with naive weight or gradient averaging
(Appendix~\ref{different_aggregation_method_effect}),
MA preserves consistent gradient directions and stable weight evolution.
This stability is crucial for model restoration: newly restored coordinates
integrate smoothly into optimization, avoiding oscillation and enabling
progress beyond the capacity-limited plateau (empirically confirmed in
Section~\ref{sec:aggre_method}).

\paragraph{Theorem 5.6.}
\label{thm:ma_telescope}
Under Assumptions~5.1--5.4, suppose all client masks satisfy
$|C_{k}^{(n)}| \ge 1$,
and choose the local learning rate \(\gamma < \frac{1}{6L\mathcal{T}}\)
for IID data and \(\gamma < \frac{1}{12L\mathcal{T}}\) for Non-IID data.
Then MHFL converges to a small neighborhood of a stationary point of standard FL as follows:
\begin{align}
& \frac{1}{K} \sum_{k=1}^K \mathbb{E}\|\nabla F(W_k)\|^{2}
\label{eq:unified_bound} \\
\le {} & \frac{4\mathbb{E}[F(W_1)]}{G_0 K}
 + \underbrace{\frac{3LH_0}{2K}
\sum_{k=1}^K \sum_{g=1}^{\mathcal{G}} \frac{|C_g|}{c_{g,k}^{*}} f_1^2(\rho_{g,k}) G^2}_{\text{Model Drift Term}} \notag \\
&+ \underbrace{\frac{6I_0}{K}
\sum_{k=1}^K \sum_{g=1}^{\mathcal{G}} \frac{|C_g|}{(c_{g,k}^{*})^{2}} f_2^2(\rho_{g,k}) \sigma^2}_{\text{Variance Term}} \notag \\
&+ \underbrace{\frac{J_0}{K}
\sum_{k=1}^K \sum_{g=1}^{\mathcal{G}} \frac{|C_g|}{c_{g,k}^{*}} f_3^2(\rho_{g,k}) \zeta^2}_{\text{Bias Term (Non-IID only)}}.\notag
\end{align}

\textbf{Interpretation.}
Theorem~\ref{thm:ma_telescope} aggregates the per-round effects of Theorem~\ref{thm:ma_one_round} and yields a unified
convergence bound: MA converges to a density-dependent neighborhood of a
stationary point of full-model FL.
The radius of this neighborhood is governed by the averaged functionals
$
\frac{1}{K}\sum_{k=1}^{K}\sum_{g=1}^{\mathcal{G}}
\frac{|C_g|}{(c_{g,k}^{*})^{\alpha_i}} f_i^2(\rho_{g,k}),
$
with $i\in\{1,2\}$ for the IID case and additionally $i=3$ for the Non-IID
case, where $\alpha_1=\alpha_3=1$ and $\alpha_2=2$.
As client densities increase, coverage improves, these aggregated penalties shrink,
and the convergence neighborhood tightens.
The complete proof is provided in Appendix~\ref{sec:convergence_threome_2}.

From a subspace perspective, as shown in
Appendix~\ref{different_aggregation_method_effect},the aggregated update at round $k$ lies in the
feasible span
\(
\mathcal{S}_{\mathrm{feas}}^{(k)}
=\mathrm{span}\{\nabla F_i(W_k)\odot m_{i,k}\}.
\)
Model restoration ensures $m_{i,k}\le m_{i,k+1}$, leading to a monotonically
expanding subspace
\(
\mathcal{S}_{\mathrm{feas}}^{(k)}
\subseteq 
\mathcal{S}_{\mathrm{feas}}^{(k+1)},
\)
and increased coverage $|C_k^{(n)}|\!\uparrow$.  
Consequently, the optimization search domain enlarges over training and the
coverage-induced noise diminishes, allowing the solution to approach the
full-model optimum.

\section{Experiment} 

\textbf{Datasets and Tasks.}
We evaluate FedGMR on four benchmark tasks that span small, medium, and large neural
architectures, enabling a comprehensive assessment across different model
capacities:  
(1) a Conv-2 network on FEMNIST \cite{caldas2018leaf},  
(2) a VGG-11 network \cite{simonyan2014very} on CIFAR-10 \cite{krizhevsky2009learning}, and  
(3) a ResNet-18 network \cite{he2016deep} on ImageNet-100 \cite{jiang2022fedmp}. 
(4) a Transformer language model on StackOverflow \cite{bonawitz2019towards}
For the Non-IID setting, FEMNIST follows the LEAF partitioning protocol
\cite{caldas2018leaf}; for StackOverflow, we use a user-wise split;
CIFAR-10 and ImageNet-100 use Dirichlet label
partitioning with concentration parameter $\alpha=0.6$.  
We simulate a federated environment with one server and ten clients whose
communication bandwidths vary, thereby inducing High, Medium, and Low degrees
of system heterogeneity depending on the number of BCCs.  
Details are provided in
Appendix~\ref{experiemnt_settings}.

\begin{table*}[t]
\centering
\renewcommand{\arraystretch}{0.3}
\caption{Test accuracy (\%) of FedGMR and baselines across datasets and heterogeneity levels, together with MRI (\%).}
\resizebox{\textwidth}{!}{
\begin{tabular}{llccccccccll}
\toprule
\multirow{2}{*}{Hetero.} & \multirow{2}{*}{Method}
& \multicolumn{2}{c}{FEMNIST (70k)} 
& \multicolumn{2}{c}{CIFAR-10 (220k)}
& \multicolumn{2}{c}{ImageNet-100 (250k)} & \multicolumn{2}{c}{StackOverflow(70k)}\\
\cmidrule(lr){3-4}\cmidrule(lr){5-6}\cmidrule(lr){7-8}\cmidrule(lr){9-10}
& & IID & Non-IID & IID & Non-IID & IID & Non-IID  & IID&Non-IID\\
\midrule

\multirow{8}{*}{High}
 & \textbf{FedGMR}   & \textbf{82.67}{\tiny$\pm$0.21} & \textbf{81.86}{\tiny$\pm$0.19}
                    & \textbf{84.52}{\tiny$\pm$0.14} & \textbf{81.68}{\tiny$\pm$0.28}
                    & \textbf{60.84}{\tiny$\pm$0.41} & \textbf{58.01}{\tiny$\pm$0.50}
                    & \textbf{30.00}{\tiny$\pm$0.01} & \textbf{30.07}{\tiny$\pm$0.02} \\

 & FedAvg           & 75.62{\tiny$\pm$0.69} & 74.64{\tiny$\pm$0.60}
                    & 70.75{\tiny$\pm$0.74} & 68.46{\tiny$\pm$0.66}
                    & 48.65{\tiny$\pm$1.10} & 47.46{\tiny$\pm$1.40}
                    & 24.38{\tiny$\pm$0.60} & 24.38{\tiny$\pm$0.61} \\

 & FedAsync          & 81.34{\tiny$\pm$0.24} & 81.03{\tiny$\pm$0.27}
                    & 82.86{\tiny$\pm$0.22} & 79.28{\tiny$\pm$0.37}
                    & 58.19{\tiny$\pm$0.58} & 55.36{\tiny$\pm$0.98}
                    & 25.86{\tiny$\pm$0.01} & 25.83{\tiny$\pm$0.01} \\

 & HeteroFL         & 82.22{\tiny$\pm$0.21} & 79.80{\tiny$\pm$0.26}
                    & 81.06{\tiny$\pm$0.21} & 75.64{\tiny$\pm$0.25}
                    & 41.90{\tiny$\pm$0.57} & 28.01{\tiny$\pm$0.43}
                    & 29.42{\tiny$\pm$0.02} & 29.17{\tiny$\pm$0.02} \\

 & FedRolex         & 82.19{\tiny$\pm$0.23} & 77.83{\tiny$\pm$0.27}
                    & 80.85{\tiny$\pm$0.21} & 67.82{\tiny$\pm$1.30}
                    & 32.17{\tiny$\pm$0.85} & 15.26{\tiny$\pm$0.66}
                    & 25.39{\tiny$\pm$0.01} & 25.15{\tiny$\pm$0.02} \\

 & Fjord            & 82.59{\tiny$\pm$0.19} & 81.85{\tiny$\pm$0.25}
                    & 81.79{\tiny$\pm$0.24} & 81.34{\tiny$\pm$0.26}
                    & 41.18{\tiny$\pm$0.64} & 32.93{\tiny$\pm$0.66}
                    & 29.18{\tiny$\pm$0.02} & 27.73{\tiny$\pm$0.10} \\

 & Fiarse           & 81.22{\tiny$\pm$0.18} & 78.77{\tiny$\pm$0.43}
                    & 74.04{\tiny$\pm$0.75} & 69.19{\tiny$\pm$0.52}
                    & 54.00{\tiny$\pm$0.68} & 48.27{\tiny$\pm$0.73}
                    & 28.77{\tiny$\pm$0.03} & 28.57{\tiny$\pm$0.05} \\

\rowcolor{gray!15}
 & \textbf{MRI (\%)}     & \textbf{2.32} & \textbf{3.73} & \textbf{7.96} & \textbf{11.54} & \textbf{37.39} & \textbf{85.10} & \textbf{11.05} & \textbf{12.68} \\
\midrule\midrule

\multirow{8}{*}{Medium}
 & \textbf{FedGMR}   & \textbf{82.84}{\tiny$\pm$0.33} & \textbf{82.35}{\tiny$\pm$0.24}
                    & \textbf{85.31}{\tiny$\pm$0.22} & \textbf{82.92}{\tiny$\pm$0.32}
                    & \textbf{62.17}{\tiny$\pm$0.45} & \textbf{60.27}{\tiny$\pm$0.63}
                    & \textbf{30.21}{\tiny$\pm$0.01} & \textbf{30.22}{\tiny$\pm$0.01} \\

 & FedAvg           & 75.92{\tiny$\pm$0.59} & 74.95{\tiny$\pm$0.57}
                    & 71.26{\tiny$\pm$0.78} & 69.04{\tiny$\pm$0.67}
                    & 49.02{\tiny$\pm$1.20} & 48.83{\tiny$\pm$1.10}
                    & 24.38{\tiny$\pm$0.60} & 24.38{\tiny$\pm$0.61} \\

 & FedAsync          & 81.99{\tiny$\pm$0.25} & 81.84{\tiny$\pm$0.19}
                    & 83.89{\tiny$\pm$0.21} & 80.77{\tiny$\pm$0.37}
                    & 60.47{\tiny$\pm$0.37} & 57.74{\tiny$\pm$0.52}
                    & 26.09{\tiny$\pm$0.01} & 26.06{\tiny$\pm$0.02} \\

 & HeteroFL         & 76.95{\tiny$\pm$0.36} & 79.62{\tiny$\pm$0.13}
                    & 81.01{\tiny$\pm$0.19} & 76.63{\tiny$\pm$0.35}
                    & 41.92{\tiny$\pm$0.66} & 32.65{\tiny$\pm$0.82}
                    & 29.55{\tiny$\pm$0.02} & 29.32{\tiny$\pm$0.02} \\

 & FedRolex         & 81.75{\tiny$\pm$0.24} & 79.51{\tiny$\pm$0.15}
                    & 80.56{\tiny$\pm$0.28} & 36.35{\tiny$\pm$0.74}
                    & 30.99{\tiny$\pm$1.20} & 14.63{\tiny$\pm$0.51}
                    & 25.51{\tiny$\pm$0.02} & 25.28{\tiny$\pm$0.03} \\

 & Fjord            & 82.78{\tiny$\pm$0.16} & 82.20{\tiny$\pm$0.23}
                    & 82.28{\tiny$\pm$0.19} & 81.93{\tiny$\pm$0.16}
                    & 41.88{\tiny$\pm$0.78} & 38.44{\tiny$\pm$1.10}
                    & 27.40{\tiny$\pm$0.01} & 27.64{\tiny$\pm$0.10} \\

 & Fiarse           & 81.64{\tiny$\pm$0.23} & 79.37{\tiny$\pm$0.41}
                    & 73.00{\tiny$\pm$1.60} & 72.21{\tiny$\pm$0.53}
                    & 59.38{\tiny$\pm$0.50} & 55.74{\tiny$\pm$0.64}
                    & 28.83{\tiny$\pm$0.03} & 28.71{\tiny$\pm$0.03} \\

\rowcolor{gray!15}
 & \textbf{MRI (\%)}     & \textbf{3.45} & \textbf{3.57} & \textbf{8.86} & \textbf{29.19} & \textbf{38.62} & \textbf{81.55} & \textbf{12.57} & \textbf{12.85} \\
\midrule\midrule

\multirow{8}{*}{Low}
 & \textbf{FedGMR}   & \textbf{83.85}{\tiny$\pm$0.19} & \textbf{82.89}{\tiny$\pm$0.30}
                    & \textbf{85.99}{\tiny$\pm$0.18} & \textbf{83.76}{\tiny$\pm$0.23}
                    & \textbf{63.89}{\tiny$\pm$0.50} & \textbf{62.37}{\tiny$\pm$0.42}
                    & \textbf{30.55}{\tiny$\pm$0.01} & \textbf{30.43}{\tiny$\pm$0.01} \\

 & FedAvg           & 76.01{\tiny$\pm$0.64} & 75.04{\tiny$\pm$0.51}
                    & 71.71{\tiny$\pm$0.74} & 69.75{\tiny$\pm$0.77}
                    & 49.17{\tiny$\pm$0.93} & 48.06{\tiny$\pm$1.10}
                    & 24.48{\tiny$\pm$0.59} & 24.48{\tiny$\pm$0.60} \\

 & FedAsync          & 83.33{\tiny$\pm$0.20} & 81.90{\tiny$\pm$0.45}
                    & 85.20{\tiny$\pm$0.15} & 82.97{\tiny$\pm$0.22}
                    & 62.57{\tiny$\pm$0.42} & 61.72{\tiny$\pm$0.59}
                    & 27.06{\tiny$\pm$0.01} & 27.06{\tiny$\pm$0.01} \\

 & HeteroFL         & 80.83{\tiny$\pm$0.28} & 78.58{\tiny$\pm$0.35}
                    & 80.14{\tiny$\pm$0.24} & 75.52{\tiny$\pm$0.24}
                    & 48.80{\tiny$\pm$0.56} & 41.00{\tiny$\pm$0.70}
                    & 29.44{\tiny$\pm$0.02} & 29.42{\tiny$\pm$0.02} \\

 & FedRolex         & 80.58{\tiny$\pm$0.32} & 78.70{\tiny$\pm$0.32}
                    & 79.56{\tiny$\pm$0.56} & 75.08{\tiny$\pm$0.39}
                    & 32.58{\tiny$\pm$0.82} & 15.97{\tiny$\pm$3.70}
                    & 25.69{\tiny$\pm$0.01} & 25.70{\tiny$\pm$0.01} \\

 & Fjord            & 82.26{\tiny$\pm$0.24} & 81.04{\tiny$\pm$0.29}
                    & 81.56{\tiny$\pm$0.43} & 80.95{\tiny$\pm$0.31}
                    & 42.20{\tiny$\pm$1.10} & 41.18{\tiny$\pm$0.92}
                    & 29.26{\tiny$\pm$0.02} & 28.45{\tiny$\pm$0.04} \\

 & Fiarse           & 82.17{\tiny$\pm$0.22} & 79.87{\tiny$\pm$0.25}
                    & 82.68{\tiny$\pm$0.22} & 62.91{\tiny$\pm$1.60}
                    & 60.92{\tiny$\pm$0.59} & 58.63{\tiny$\pm$0.62}
                    & 29.14{\tiny$\pm$0.04} & 29.03{\tiny$\pm$0.02} \\

\rowcolor{gray!15}
 & \textbf{MRI (\%)}     & \textbf{3.79} & \textbf{4.76} & \textbf{7.61} & \textbf{13.35}
                    & \textbf{35.89} & \textbf{71.89} & \textbf{11.60} & \textbf{11.73} \\

\bottomrule
\end{tabular}
}
\label{tab:baselines}
\end{table*}
\subsection{Baseline Comparison: Accuracy, Speed, and MRI}
\label{sec:Baselines}
\textbf{Baselines.}
We compare FedGMR with three full-model FL approaches—
(1) \textbf{FedAvg} (synchronous FedAvg \cite{mcmahan2017communication}),
(2) \textbf{FedAsync} \cite{xie2019asynchronous}—and four state-of-the-art
MHFL methods:
\textbf{HeteroFL} \cite{diao2021heterofl},
\textbf{FjORD} \cite{horvath2021fjord},
\textbf{FedRolex} \cite{alam2022fedrolex},
\textbf{FIARSE} \cite{wu2024fiarse}.

\noindent\textbf{Evaluation protocol.}
To compare training speed, all methods are evaluated at a fixed wall-clock budget $t^\star$ chosen at the point where most methods stabilizes (e.g., $t^\star=70\text{k}\,\mathrm{s}$).
The reported accuracy at $t^\star$ is a smoothed estimate—the mean$\pm$std of 20 test evaluations in a symmetric $\pm10$ window around $t^\star$ .
To aggregate gains across settings, we report the \emph{Mean Relative Improvement (MRI)} of method $M$ over baselines $\mathcal{B}$:
\(\mathrm{MRI}(M)=\frac{100}{|\mathcal{B}|}\sum_{b\in\mathcal{B}}
\frac{\mathrm{Acc}(M)-\mathrm{Acc}(b)}{\mathrm{Acc}(b)},\)
where $\mathrm{Acc}(\cdot)$ is the smoothed accuracy at $t^\star$.
Higher MRI indicates stronger overall gains, reported in percent.

\textbf{Summary of main results.}  
From Table~\ref{tab:baselines}, \textbf{FedGMR} consistently achieves the best performance across all datasets and heterogeneity regimes, and its advantage grows as tasks become harder and heterogeneity becomes stronger. This matches the key challenge in MHFL: BCC sub-models are prone to late-round plateaus due to limited capacity, and the resulting marginalization is more severe for complex tasks with harder decision boundaries. For instance, under \emph{High} heterogeneity with \emph{Non-IID} splits, FedGMR’s MRI is \textbf{+3.73\%} on FEMNIST, \textbf{+11.54\%} on CIFAR-10, \textbf{+85.10\%} on ImageNet-100, and \textbf{+12.68\%} on StackOverflow. The especially large gain on ImageNet-100 is consistent with the stronger capacity bottleneck on deeper models (Conv2D $\rightarrow$ VGG-11 $\rightarrow$ ResNet-18). By gradually restoring capacity, FedGMR keeps BCCs effective beyond the early stage, yielding the largest gains in challenging settings. The same trend also holds for Transformer-based models, suggesting that the mechanism is largely architecture-agnostic.

Compared with baselines, fixed-capacity MHFL methods (e.g., HeteroFL/FjORD/FIARSE/FedRolex) can improve early efficiency but often struggle on harder tasks. Fixed-capacity pruning restricts clients to a limited sub-model (i.e., a constrained optimization subspace), which often leads to late-round plateaus on deep models under strong heterogeneity.
 For example, under \emph{High} heterogeneity on ImageNet-100, FedGMR improves over FIARSE from \textbf{54.00\%} to \textbf{60.84\%} under \emph{IID} splits and from \textbf{48.27\%} to \textbf{58.01\%} under \emph{Non-IID} splits. Notably, FedRolex targets a similar issue via region rotation; however, because regions are not equally informative and rotation introduces update inconsistency, its convergence can degrade sharply under the hardest settings, dropping to \textbf{32.17\%} vs.\ \textbf{60.84\%} under \emph{IID} and \textbf{15.26\%} vs.\ \textbf{58.01\%} under \emph{Non-IID} on ImageNet-100 with \emph{High} heterogeneity. FedAsync can be competitive under low heterogeneity due to training the full model, yet it cannot address the limited contribution of BCCs (few/stale updates), which bounds its gains. Overall, \textbf{FedGMR} provides robust improvements across datasets and heterogeneity levels under realistic, highly skewed FL environments.

\begin{figure*}[htbp]

\begin{minipage}{\textwidth}
\captionof{table}{Test accuracy (\%) evaluating the joint effect of GMR with various aggregation methods.}
\renewcommand{\arraystretch}{0.3}
\resizebox{\textwidth}{!}{
\begin{tabular}{llcccccccc}
\toprule
\multirow{2}{*}{Hetero.} & \multirow{2}{*}{Method}
& \multicolumn{2}{c}{FEMNIST (70k)} 
& \multicolumn{2}{c}{CIFAR-10 (220k)}
& \multicolumn{2}{c}{ImageNet-100 (250k)} & \multicolumn{2}{c}{StackOverflow(70k)}\\
\cmidrule(lr){3-4}\cmidrule(lr){5-6}\cmidrule(lr){7-8}\cmidrule(lr){9-10}
& & IID & Non-IID & IID & Non-IID & IID & Non-IID & IID & Non-IID\\
\midrule

\multirow{6}{*}{High}
& GMR + MA 
& \textbf{82.67} {\tiny$\pm$0.21} & \textbf{81.86} {\tiny$\pm$0.19} 
& \textbf{84.52} {\tiny$\pm$0.14} & \textbf{81.68} {\tiny$\pm$0.28} 
& \textbf{60.84} {\tiny$\pm$0.41} & \textbf{58.01} {\tiny$\pm$0.50} 
& \textbf{30.00} {\tiny$\pm$0.01} & \textbf{30.07} {\tiny$\pm$0.02}\\
& GMR + GA 
& 81.27 {\tiny$\pm$0.18} & 80.95 {\tiny$\pm$0.22} 
& 83.73 {\tiny$\pm$0.19} & 78.83 {\tiny$\pm$0.38} 
& 58.36 {\tiny$\pm$0.83} & 57.32 {\tiny$\pm$0.45} 
& 29.56 {\tiny$\pm$0.01} & 29.71 {\tiny$\pm$0.01}\\
& GMR + FA 
& 80.55 {\tiny$\pm$0.28} & 79.79 {\tiny$\pm$0.28} 
& 10.00 {\tiny$\pm$0.00} & 10.00 {\tiny$\pm$0.00} 
& 59.68 {\tiny$\pm$0.58} & 55.22 {\tiny$\pm$0.63} 
& 25.62 {\tiny$\pm$0.02} & 28.24 {\tiny$\pm$0.03}\\
\cmidrule(lr){2-10}
& w/o GMR + MA 
& \textbf{82.18} {\tiny$\pm$0.10} & \textbf{79.91} {\tiny$\pm$0.18} 
& \textbf{81.65} {\tiny$\pm$0.26} & 71.93 {\tiny$\pm$0.45} 
& 53.49 {\tiny$\pm$0.53} & 48.28 {\tiny$\pm$0.73} 
& \textbf{29.68} {\tiny$\pm$0.02} & \textbf{29.76} {\tiny$\pm$0.02}\\
& w/o GMR + GA 
& 80.60 {\tiny$\pm$0.17} & 78.59 {\tiny$\pm$0.31} 
& 80.66 {\tiny$\pm$0.69} & \textbf{74.37} {\tiny$\pm$0.45} 
& \textbf{60.74} {\tiny$\pm$0.59} & 56.66 {\tiny$\pm$0.49} 
& 29.00 {\tiny$\pm$0.02} & 29.11 {\tiny$\pm$0.01}\\
& w/o GMR + FA 
& 79.16 {\tiny$\pm$0.25} & 78.09 {\tiny$\pm$0.25} 
& 10.00 {\tiny$\pm$0.00} & 10.00 {\tiny$\pm$0.00} 
& 60.35 {\tiny$\pm$0.61} & \textbf{58.65} {\tiny$\pm$0.70} 
& 25.95 {\tiny$\pm$0.01} & 25.94 {\tiny$\pm$0.02}\\

\midrule
\multirow{6}{*}{Medium}
& GMR + MA
& \textbf{82.94} {\tiny$\pm$0.13} & \textbf{82.35} {\tiny$\pm$0.24}
& \textbf{85.31} {\tiny$\pm$0.22} & \textbf{82.92} {\tiny$\pm$0.32}
& \textbf{62.17} {\tiny$\pm$0.45} & \textbf{60.27} {\tiny$\pm$0.63}
& \textbf{30.21} {\tiny$\pm$0.01} & \textbf{30.22} {\tiny$\pm$0.01} \\

& GMR + GA
& 82.01 {\tiny$\pm$0.15} & 81.91 {\tiny$\pm$0.24}
& 83.48 {\tiny$\pm$0.21} & 79.83 {\tiny$\pm$0.48}
& 61.07 {\tiny$\pm$0.49} & 59.49 {\tiny$\pm$0.50}
& 29.84 {\tiny$\pm$0.01} & 29.97 {\tiny$\pm$0.02} \\

& GMR + FA
& 80.93 {\tiny$\pm$0.22} & 81.57 {\tiny$\pm$0.27}
& 10.00 {\tiny$\pm$0.00} & 10.00 {\tiny$\pm$0.00}
& 58.91 {\tiny$\pm$0.52} & 56.43 {\tiny$\pm$0.70}
& 27.62 {\tiny$\pm$0.02} & 26.09 {\tiny$\pm$0.04} \\
\cmidrule(lr){2-10}

& w/o GMR + MA
& \textbf{81.85} {\tiny$\pm$0.15} & \textbf{80.04} {\tiny$\pm$0.29}
& \textbf{84.20} {\tiny$\pm$0.21} & \textbf{80.20} {\tiny$\pm$0.30}
& 56.94 {\tiny$\pm$0.50} & 52.86 {\tiny$\pm$0.49}
& \textbf{29.92} {\tiny$\pm$0.02} & \textbf{29.91} {\tiny$\pm$0.01} \\

& w/o GMR + GA
& 80.32 {\tiny$\pm$0.21} & 78.71 {\tiny$\pm$0.29}
& 82.11 {\tiny$\pm$0.21} & 78.05 {\tiny$\pm$0.28}
& \textbf{61.11} {\tiny$\pm$0.44} & 58.56 {\tiny$\pm$0.45}
& 29.46 {\tiny$\pm$0.02} & 29.51 {\tiny$\pm$0.02} \\

& w/o GMR + FA
& 79.47 {\tiny$\pm$0.29} & 78.37 {\tiny$\pm$0.33}
& 10.00 {\tiny$\pm$0.00} & 10.00 {\tiny$\pm$0.00}
& 60.74 {\tiny$\pm$0.65} & \textbf{59.10} {\tiny$\pm$0.50}
& 26.18 {\tiny$\pm$0.01} & 26.17 {\tiny$\pm$0.02} \\

\midrule

\multirow{6}{*}{Low}
& GMR + MA
& \textbf{83.85} {\tiny$\pm$0.19} & \textbf{82.89} {\tiny$\pm$0.30}
& \textbf{85.99} {\tiny$\pm$0.18} & \textbf{83.76} {\tiny$\pm$0.23}
& \textbf{63.89} {\tiny$\pm$0.50} & \textbf{62.37} {\tiny$\pm$0.42}
& \textbf{30.55} {\tiny$\pm$0.01} & \textbf{30.43} {\tiny$\pm$0.01}\\

& GMR + GA
& 83.38 {\tiny$\pm$0.21} & 82.68 {\tiny$\pm$0.33}
& 85.59 {\tiny$\pm$0.21} & 83.73 {\tiny$\pm$0.22}
& 61.68 {\tiny$\pm$0.73} & 60.92 {\tiny$\pm$0.47}
& 30.17 {\tiny$\pm$0.01} & 30.17 {\tiny$\pm$0.01}\\

& GMR + FA
& 82.21 {\tiny$\pm$0.26} & 82.63 {\tiny$\pm$0.32}
& 10.00 {\tiny$\pm$0.00} & 48.54 {\tiny$\pm$3.60}
& 60.13 {\tiny$\pm$0.50} & 57.34 {\tiny$\pm$0.70}
& 28.57 {\tiny$\pm$0.02} & 28.82 {\tiny$\pm$0.01}\\

\cmidrule(lr){2-10}
& w/o GMR + MA
& \textbf{82.13} {\tiny$\pm$0.15} & \textbf{79.91} {\tiny$\pm$0.18}
& 84.81 {\tiny$\pm$0.14} & 81.09 {\tiny$\pm$0.31}
& 57.40 {\tiny$\pm$0.68} & 55.38 {\tiny$\pm$0.80}
& \textbf{30.14} {\tiny$\pm$0.02} & \textbf{30.14} {\tiny$\pm$0.01}\\

& w/o GMR + GA
& 81.74 {\tiny$\pm$0.25} & 79.31 {\tiny$\pm$0.22}
& \textbf{85.20} {\tiny$\pm$0.20} & \textbf{83.62} {\tiny$\pm$0.21}
& \textbf{61.45} {\tiny$\pm$0.43} & \textbf{59.58} {\tiny$\pm$0.44}
& 29.91 {\tiny$\pm$0.01} & 29.83 {\tiny$\pm$0.01}\\

& w/o GMR + FA
& 79.46 {\tiny$\pm$0.23} & 78.26 {\tiny$\pm$0.32}
& 10.00 {\tiny$\pm$0.00} & 10.00 {\tiny$\pm$0.00}
& 59.96 {\tiny$\pm$0.49} & 58.69 {\tiny$\pm$0.55}
& 26.91 {\tiny$\pm$0.02} & 26.88 {\tiny$\pm$0.02}\\

\bottomrule
\end{tabular}
}
\label{tab:Aggre}
  \end{minipage}

    \centering
    \includegraphics[width=0.99\textwidth, height=0.13\textheight]{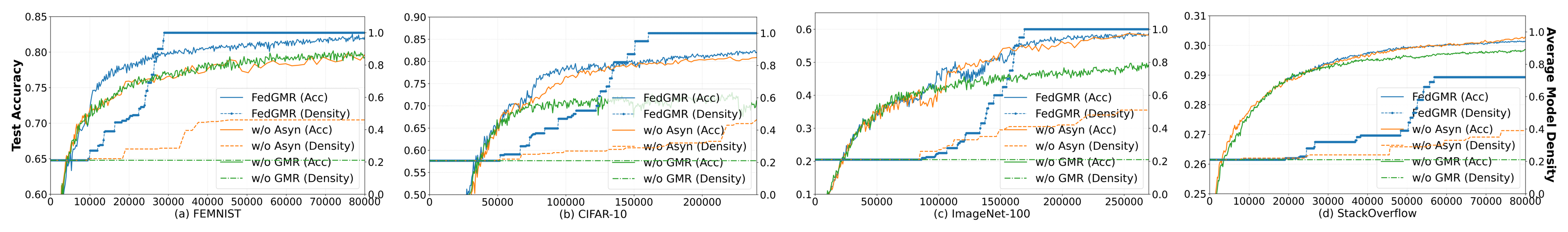}
    \caption{Ablation Study: Impact of Average Model Density on Accuracy Convergence.}
    \label{fig:Ab}

\end{figure*}

\subsection{Impact of Average Model Density on Convergence}
To isolate the effect of \emph{Gradual Model Restoration} (GMR) and the asynchronous workflow, we evaluate two variants of FedGMR: \textbf{w/o GMR} (fixed density) and \textbf{w/o Asyn} (synchronous coordination with GMR). Figure~\ref{fig:Ab} reports, under \emph{High} heterogeneity with non-IID splits, test accuracy (left y-axis) and average density (right y-axis) over server aggregation steps across four tasks.

\textbf{Effect of GMR.}
Across tasks, the performance gap between FedGMR and \textbf{w/o GMR} emerges precisely when the average density starts increasing (i.e., restoration begins), FedGMR continues improving while \textbf{w/o GMR} plateaus. This indicates that restoring capacity after saturation is critical for sustaining late-stage progress in MHFL. We also applied GMR to several other MHFL methods on FEMNIST and again observed gains in most cases, as detailed in Appendix~\ref{sec:gmr_other_methods_appendix}. This supports that the main gain comes from restoration itself, rather than a method-specific implementation detail.

\textbf{Effect of asynchrony.}
The benefit of asynchrony is task-dependent.
When density is restored to near-full capacity early (e.g., FEMNIST), asynchronous coordination significantly accelerates training by reducing idle time.
In more challenging tasks where restoration proceeds gradually, the gain from asynchrony is smaller and can even be offset by update staleness, leading \textbf{w/o Asyn} to be comparable or slightly better in some cases (e.g., ImageNet-100 and StackOverflow).
Nevertheless, we adopt an async-compatible design to remain robust under practical system heterogeneity, where client availability and latency are inherently imbalanced.
Appendices~\ref{sec:ablation} and~\ref{sec:async_extension} report additional ablations and formalize this update-frequency/staleness tradeoff.

\subsection{Comparison with different aggregation methods}
\label{sec:aggre_method} 
In our convergence analysis, we examine the effect of MaskFedAvg(MA).
For comparison, we additionally include two alternative aggregation rules—\textbf{gradient averaging}(GA) and \textbf{FedAvg}(FA)—as formally defined in Appendix~\ref{sec: two_stage}.

\paragraph{Result analysis.}
Table~\ref{tab:Aggre} leads to the following observations, highlighting how aggregation interacts with GMR.
\textbf{(1) MA is the most effective when combined with GMR.}
Across datasets and heterogeneity regimes, \textbf{GMR+MA} consistently achieves the best accuracy. Under \emph{High} heterogeneity, it reaches \textbf{82.67/81.86} on FEMNIST, \textbf{84.52/81.68} on CIFAR-10, \textbf{60.84/58.01} on ImageNet-100, and \textbf{30.00/30.07} on StackOverflow for \emph{IID/Non-IID} splits, respectively. This is consistent with Theorem~\ref{thm:ma_one_round}: as density increases, MA normalizes each coordinate by its effective participation, enabling newly restored parameters to be incorporated smoothly without distorting update magnitudes.
\textbf{(2) GMR’s gains depend on using an aggregation rule compatible with restoration.}
On CIFAR-10 and ImageNet-100, the improvements from GMR can be limited under GA/FA, and \textbf{GMR+GA} may even underperform \textbf{w/o GMR+GA}. Intuitively, restoration reactivates many previously pruned coordinates, inducing abrupt changes in the support and scale of client updates; GA/FA do not normalize such shifts adequately, which can destabilize optimization. We provide a detailed discussion in Appendix~\ref{different_aggregation_method_effect}.
\textbf{(3) Without GMR, MA can be suboptimal under fixed density.}
In the static-mask regime, MA rescales sparse updates to approximate full-model updates, but this rescaling can also amplify estimation error and variance when the mask does not change. As a result, MA may be worse than GA without restoration, since GA avoids such amplification and can accumulate smaller long-term bias. This effect is more pronounced for deeper models (e.g., ResNet-18), where the variance amplification becomes more severe (Appendix~\ref{sec:convergence_final_analysis}). Finally, FA highlights the pitfall of naive averaging under structural heterogeneity: treating missing coordinates as zeros induces weight shrinkage and leads to consistently degraded performance (e.g., near-chance $10\%$ on CIFAR-10).

\section{Conclusion}
We addressed the challenge that bandwidth-constrained clients hinder convergence and bias optimization in heterogeneous FL.  
FedGMR tackles this by gradually restoring model capacity under asynchrony and model heterogeneity, with tailored transmission and aggregation.  
Theory and experiments on image and text tasks  confirm faster convergence and higher accuracy, especially under non-IID and highly heterogeneous conditions.

\section*{Acknowledgements}
This work was supported by IITP grants funded by the Korea government (MSIT): RS-2024-00435652 (6GARROW: 6G AI-native integrated RAN-Core networks) and RS-2024-00404972 (Development of 5G-A vRAN Research Platform).

\section*{Impact Statement}

This paper presents work whose goal is to advance the field of Machine
Learning. There are many potential societal consequences of our work, none
of which we feel must be specifically highlighted here.


\FloatBarrier
\bibliography{main}

@inproceedings{yu2019universally,
  title={Universally slimmable networks and improved training techniques},
  author={Yu, Jiahui and Huang, Thomas S},
  booktitle={Proceedings of the IEEE/CVF international conference on computer vision},
  pages={1803--1811},
  year={2019}
}

@article{sung2021training,
  title={Training neural networks with fixed sparse masks},
  author={Sung, Yi-Lin and Nair, Varun and Raffel, Colin A},
  journal={Advances in Neural Information Processing Systems},
  volume={34},
  pages={24193--24205},
  year={2021}
}

@inproceedings{bonawitz2019towards,
  title={Towards Federated Learning at Scale: System Design},
  author={Keith Bonawitz and Hubert Eichner and Wolfgang Grieskamp and Dzmitry Huba and Alex Ingerman and Vladimir Ivanov and Chloe Kiddon and Jakub Kone{\v{c}}n{\'y} and Stefano Mazzocchi and H. Brendan McMahan and Timon Van Overveldt and David Petrou and Daniel Ramage and Jason Roselander},
  booktitle={Proceedings of Machine Learning and Systems (MLSys)},
  volume={1},
  pages={374--388},
  year={2019}
}

@article{horvath2021fjord,
  title={Fjord: Fair and accurate federated learning under heterogeneous targets with ordered dropout},
  author={Horvath, Samuel and Laskaridis, Stefanos and Almeida, Mario and Leontiadis, Ilias and Venieris, Stylianos and Lane, Nicholas},
  journal={Advances in Neural Information Processing Systems},
  volume={34},
  pages={12876--12889},
  year={2021}
}

@article{wu2024fiarse,
  title={FIARSE: Model-Heterogeneous Federated Learning via Importance-Aware Submodel Extraction},
  author={Wu, Feijie and Wang, Xingchen and Wang, Yaqing and Liu, Tianci and Su, Lu and Gao, Jing},
  journal={arXiv preprint arXiv:2407.19389},
  year={2024}
}

@inproceedings{diao2021heterofl,
  title={HETEROFL: COMPUTATION AND COMMUNICATION EFFICIENT FEDERATED LEARNING FOR HETEROGENEOUS CLIENTS},
  author={Diao, Enmao and Ding, Jie and Tarokh, Vahid},
  booktitle={9th International Conference on Learning Representations, ICLR 2021},
  year={2021}
}

@inproceedings{cai2020once,
  title={Once for All: Train One Network and Specialize it for Efficient Deployment},
  author={Cai, Han and Gan, Chuang and Wang, Tianzhe and Zhang, Zhekai and Han, Song},
  booktitle={International Conference on Learning Representations},
  year={2020}
}

@inproceedings{ilhan2023scalefl,
  title={Scalefl: Resource-adaptive federated learning with heterogeneous clients},
  author={Ilhan, Fatih and Su, Gong and Liu, Ling},
  booktitle={Proceedings of the IEEE/CVF Conference on Computer Vision and Pattern Recognition},
  pages={24532--24541},
  year={2023}
}

@book{shalev2014understanding,
  title={Understanding machine learning: From theory to algorithms},
  author={Shalev-Shwartz, Shai and Ben-David, Shai},
  year={2014},
  publisher={Cambridge university press}
}

@inproceedings{jiang2022fedmp,
  title={Fedmp: Federated learning through adaptive model pruning in heterogeneous edge computing},
  author={Jiang, Zhida and Xu, Yang and Xu, Hongli and Wang, Zhiyuan and Qiao, Chunming and Zhao, Yangming},
  booktitle={2022 IEEE 38th International Conference on Data Engineering (ICDE)},
  pages={767--779},
  year={2022},
  organization={IEEE}
}

@inproceedings{zhou2024towards,
  title={Towards efficient asynchronous federated learning in heterogeneous edge environments},
  author={Zhou, Yajie and Pang, Xiaoyi and Wang, Zhibo and Hu, Jiahui and Sun, Peng and Ren, Kui},
  booktitle={IEEE INFOCOM 2024-IEEE conference on computer communications},
  pages={2448--2457},
  year={2024},
  organization={IEEE}
}

@inproceedings{kim2023depthfl,
  title={DepthFL: Depthwise federated learning for heterogeneous clients},
  author={Kim, Minjae and Yu, Sangyoon and Kim, Suhyun and Moon, Soo-Mook},
  booktitle={The Eleventh International Conference on Learning Representations},
  year={2023}
}

@article{krizhevsky2009learning,
  title={Learning multiple layers of features from tiny images},
  author={Krizhevsky, Alex and Hinton, Geoffrey and others},
  year={2009},
  publisher={Toronto, ON, Canada}
}

@article{Lin_Kong_Stich_Jaggi_2020, title={Ensemble distillation for robust model fusion in federated learning}, volume={33}, journal={Advances in Neural Information Processing Systems}, author={Lin, Tao and Kong, Lingjing and Stich, Sebastian U. and Jaggi, Martin}, year={2020}, pages={2351–2363} }

@article{liu2021joint,
  title={Joint model pruning and device selection for communication-efficient federated edge learning},
  author={Liu, Shengli and Yu, Guanding and Yin, Rui and Yuan, Jiantao and Shen, Lei and Liu, Chonghe},
  journal={IEEE Transactions on Communications},
  volume={70},
  number={1},
  pages={231--244},
  year={2021},
  publisher={IEEE}
}

@inproceedings{mcmahan2017communication,
  title={Communication-efficient learning of deep networks from decentralized data},
  author={McMahan, Brendan and Moore, Eider and Ramage, Daniel and Hampson, Seth and y Arcas, Blaise Aguera},
  booktitle={Artificial intelligence and statistics},
  pages={1273--1282},
  year={2017},
  organization={PMLR}
}

@article{simonyan2014very,
  title={Very deep convolutional networks for large-scale image recognition},
  author={Simonyan, Karen and Zisserman, Andrew},
  journal={arXiv preprint arXiv:1409.1556},
  year={2014}
}

@inproceedings{vahidian2021personalized,
  title={Personalized federated learning by structured and unstructured pruning under data heterogeneity},
  author={Vahidian, Saeed and Morafah, Mahdi and Lin, Bill},
  booktitle={2021 IEEE 41st international conference on distributed computing systems workshops (ICDCSW)},
  pages={27--34},
  year={2021},
  organization={IEEE}
}

@article{xie2019asynchronous,
  title={Asynchronous federated optimization},
  author={Xie, Cong and Koyejo, Sanmi and Gupta, Indranil},
  journal={arXiv preprint arXiv:1903.03934},
  year={2019}
}

@article{chen2025advances,
  title={Advances in robust federated learning: A survey with heterogeneity considerations},
  author={Chen, Chuan and Liao, Tianchi and Deng, Xiaojun and Wu, Zihou and Huang, Sheng and Zheng, Zibin},
  journal={IEEE Transactions on Big Data},
  year={2025},
  publisher={IEEE}
}

@article{hu2024overview,
  title={An overview of implementing security and privacy in federated learning},
  author={Hu, Kai and Gong, Sheng and Zhang, Qi and Seng, Chaowen and Xia, Min and Jiang, Shanshan},
  journal={Artificial intelligence review},
  volume={57},
  number={8},
  pages={204},
  year={2024},
  publisher={Springer}
}

@article{lan2024asynchronous,
  title={Asynchronous federated reinforcement learning with policy gradient updates: Algorithm design and convergence analysis},
  author={Lan, Guangchen and Han, Dong-Jun and Hashemi, Abolfazl and Aggarwal, Vaneet and Brinton, Christopher G},
  journal={arXiv preprint arXiv:2404.08003},
  year={2024}
}

@article{liu2025fedgraft,
  title={FedGraft: Memory-Aware Heterogeneous Federated Learning via Model Grafting},
  author={Liu, Ruixuan and Hu, Ming and Xia, Zeke and Xie, Xiaofei and Xia, Jun and Zhang, Pengyu and Huang, Yihao and Chen, Mingsong},
  journal={IEEE Transactions on Mobile Computing},
  year={2025},
  publisher={IEEE}
}

@article{pei2024review,
  title={A review of federated learning methods in heterogeneous scenarios},
  author={Pei, Jiaming and Liu, Wenxuan and Li, Jinhai and Wang, Lukun and Liu, Chao},
  journal={IEEE Transactions on Consumer Electronics},
  volume={70},
  number={3},
  pages={5983--5999},
  year={2024},
  publisher={IEEE}
}

@article{zhang2022scalable,
  title={Scalable and low-latency federated learning with cooperative mobile edge networking},
  author={Zhang, Zhenxiao and Gao, Zhidong and Guo, Yuanxiong and Gong, Yanmin},
  journal={IEEE Transactions on Mobile Computing},
  volume={23},
  number={1},
  pages={812--822},
  year={2022},
  publisher={IEEE}
}

@inproceedings{molchanov2019importance,
  title={Importance estimation for neural network pruning},
  author={Molchanov, Pavlo and Mallya, Arun and Tyree, Stephen and Frosio, Iuri and Kautz, Jan},
  booktitle={Proceedings of the IEEE/CVF conference on computer vision and pattern recognition},
  pages={11264--11272},
  year={2019}
}

@article{caldas2018leaf,
  title={Leaf: A benchmark for federated settings},
  author={Caldas, Sebastian and Duddu, Sai Meher Karthik and Wu, Peter and Li, Tian and Kone{\v{c}}n{\`y}, Jakub and McMahan, H Brendan and Smith, Virginia and Talwalkar, Ameet},
  journal={arXiv preprint arXiv:1812.01097},
  year={2018}
}

@inproceedings{he2016deep,
  title={Deep residual learning for image recognition},
  author={He, Kaiming and Zhang, Xiangyu and Ren, Shaoqing and Sun, Jian},
  booktitle={Proceedings of the IEEE conference on computer vision and pattern recognition},
  pages={770--778},
  year={2016}
}

@article{zhou2023every,
  title={Every parameter matters: Ensuring the convergence of federated learning with dynamic heterogeneous models reduction},
  author={Zhou, Hanhan and Lan, Tian and Venkataramani, Guru Prasadh and Ding, Wenbo},
  journal={Advances in Neural Information Processing Systems},
  volume={36},
  pages={25991--26002},
  year={2023}
}

@article{alam2022fedrolex,
  title={Fedrolex: Model-heterogeneous federated learning with rolling sub-model extraction},
  author={Alam, Samiul and Liu, Luyang and Yan, Ming and Zhang, Mi},
  journal={Advances in neural information processing systems},
  volume={35},
  pages={29677--29690},
  year={2022}
}
\bibliographystyle{icml2026}

\newpage
\appendix
\onecolumn

\allowdisplaybreaks 
\section{Justification of GMR Improving Training Speed}
\label{sec:GMR_effectness}

To support our claim that Gradual Model Restoration (GMR) can accelerate training at certain stages, we present a simplified theoretical analysis.
\begin{lemma}
\label{lem:1}\quad Restoring Model Capacity Improves Training Speed Post-Plateau

\end{lemma}

Let $w_t^{(p)}$ denote the weight of a sub-model with density $p < 1$ at time $t$. Let $F(w)$ be the training loss and define the average convergence speed over interval $[t_1, t_2]$ as:
\begin{equation}
\mathcal{A}(t_1, t_2) := -\frac{F(w_{t_2}) - F(w_{t_1})}{t_2 - t_1}.
\end{equation}
Under the assumption that sub-models have limited representation power, during the interval $[T_{k+n}, T_{k+n+m}]$, the restored model achieves a strictly higher training speed than the prior stagnated phase $[T_k, T_{k+n}]$:
\begin{equation}
\mathcal{A}(T_{k+n}, T_{k+n+m}) > \mathcal{A}(T_k, T_{k+n}).
\end{equation}
\textbf{Assumptions A.2.:}
\begin{itemize}
\item (Smoothness) $F(\mathbf{w})$ is $L$-smooth, ensuring predictable behavior of gradient descent and enabling analysis via gradient norm.

\item (Capacity Hierarchy) Since smaller sub-models are nested within larger ones, sub-models limit representation power: for densities $p_1 < p_2 < p_3$, the corresponding optima satisfies:
\begin{equation}
F(\mathbf{w}^*_{p_3}) < F(\mathbf{w}^*_{p_2}) < F(\mathbf{w}^*_{p_1}).
\end{equation}
\end{itemize}

Then we introduce the time interval that restores the model will improve the training speed. At time $t = T_k$, the sub-model $\mathbf{w}_{T_k,p_1}$ reaches its local minimum:
\begin{equation}
F(\mathbf{w}_{T_k,p_1}) = F(\mathbf{w}^*_{p_1}).
\end{equation}
After a time interval $\Delta t$, the training gain remains zero:
\begin{equation}
\mathcal{A}(T_k, T_{k+n}) \leq 0.
\end{equation}
If we restore the model to a larger density $p_2 > p_1$ at $T_{k+n}$, the model continues improving and eventually reaches $F(\mathbf{w}^*_{p_2}) < F(\mathbf{w}^*_{p_1})$ at $T_{k+n+m}$. Thus:
\begin{equation}
\mathcal{A}(T_{k+n}, T_{k+n+m}) = -\frac{F(\mathbf{w}^*_{p_2}) - F(\mathbf{w}_{T_{k+n},p_1})}{T_{k+n+m} - T_{k+n}} > 0.
\end{equation}

\textbf{Conclusion:}
During the interval $[T_{k+n}, T_{k+n+m}]$, the restored model exhibits strictly higher training speed than the previous stagnated phase $[T_k, T_{k+n}]$:
\begin{equation}
\mathcal{A}(T_{k+n}, T_{k+n+m}) > \mathcal{A}(T_k, T_{k+n}).
\end{equation}

\textbf{Implication:}
This lemma supports the core mechanism of GMR. When a sub-model has plateaued, restoring its capacity enables further progress. Since sub-models are nested within larger models, the restoration allows smooth transitions without retraining from scratch. Although this is an idealized case, it demonstrates that restoring model capacity can improve training speed after stagnation. In other words, the optimal model size in FL is dynamic and should gradually increase as training progresses. GMR thereby enhances training efficiency and convergence in heterogeneous FL settings such as FedGMR.
\section{Pseudocode for BuffMaskFedAvg}
\label{sec:BuffMaskFedAvg_Pseu}

We adopt BuffMaskFedAvg to aggregate heterogeneous client models in asynchronous FL. Masks are used to track the pruning status of each neuron's presence across clients, enabling MA guided by a staleness-aware weighting mechanism. The pseudocode is provided in Algorithm~\ref{algorithm:MaskFedAvg}.

\begin{algorithm}[h]
\caption{BuffMaskFedAvg Aggregation}
\label{algorithm:MaskFedAvg}
\begin{algorithmic}
\STATE \textbf{Given:} previous model $\mathbf{W}_{k-1}$ and buffer $\mathcal{B}$
\FORALL{client update $\mathbf{w}_{i,k} \in \mathcal{B}$}
  \STATE compute staleness weight $\beta_{i,k}$
  \STATE derive mask $\mathbf{m}_{i,k}$ from nonzero coordinates of $\mathbf{w}_{i,k}$
  \STATE $\mathbf{W}_{\mathrm{cum}} \leftarrow \mathbf{W}_{\mathrm{cum}} + \beta_{i,k}\mathbf{w}_{i,k}$
  \STATE $\mathbf{M}_{\mathrm{cum}} \leftarrow \mathbf{M}_{\mathrm{cum}} + \beta_{i,k}\mathbf{m}_{i,k}$
\ENDFOR
\FOR{$n = 1$ \textbf{to} $N$}
  \IF{$\mathbf{M}_{\mathrm{cum}}^{(n)} \neq 0$}
    \STATE $\mathbf{W}_k^{(n)} \gets \mathbf{W}_{\mathrm{cum}}^{(n)} / \mathbf{M}_{\mathrm{cum}}^{(n)}$
  \ELSE
    \STATE $\mathbf{W}_k^{(n)} \gets \mathbf{W}_{k-1}^{(n)}$
  \ENDIF
\ENDFOR
\STATE \textbf{return} $\mathbf{W}_k$
\end{algorithmic}
\end{algorithm}

\section{Details for Extract the sub-models}
\label{sec:Extract}
First, we introduce neuron importance scores for unstructured pruning.  
Given the client density vector $\mathbf{P}_k=\{\rho_{i,k}\}_{i=1}^C$,  
the corresponding binary masks $\mathbf{M}_k=\{m_{i,k}\}_{i=1}^C$ are generated accordingly.  
Following \citet{molchanov2019importance}, pruning is guided by importance scores,  
which quantify a neuron's contribution as the squared difference in prediction error  
between the full and pruned models.  
\begin{equation}
  \mathbf{E}_k = \left( F(\mathbf{W}_k) - F(\mathbf{W}_k \odot \mathbf{m}) \right)^2 
\end{equation}
where \( \mathbf{E}_k \) represents the pruning error, \(\odot\) denotes element-wise multiplication, and \(\mathbf{m}\) is a binary mask indicating whether a neuron in \(\mathbf{W}_k\) is pruned (0) or retained (1).

To simplify computation, a first-order Taylor expansion is used \cite{molchanov2019importance, liu2021joint}, approximating neuron importance as:
\begin{equation}
    {\mathbf{I}_k} \approx \left( \mathbf{G}_k \odot \mathbf{W}_k \right)^2
\end{equation}
where \({\mathbf{I}_k}\) is the importance score matrix for \(\mathbf{W}_k\), computed as the squared product of each neuron's weight and its gradient.

In the FL scenario, transmitting full gradient data imposes a burden on BCCs. To mitigate this, gradients are approximated using weight changes. By storing the old global model weight \(\mathbf{W}_{k-1}\), the gradient is approximated as \(\mathbf{G}_k \approx \mathbf{W}_k - \mathbf{W}_{k-1}\), and neuron importance is calculated as:
\begin{equation}
    {\mathbf{I}_k} \approx \left( (\mathbf{W}_k - \mathbf{W}_{k-1}) \odot \mathbf{W}_k \right)^2
    \label{eq:pruning_metric}
\end{equation}

Then, based on Eq.~\ref{eq:pruning_metric}, we compute neuron importance scores,  
sort neurons accordingly, and derive masks consistent with the target density.  
In asynchronous settings, however, frequent aggregation makes per-round sorting  
computationally expensive. To reduce overhead, masks are updated only every few rounds.  

\begin{algorithm}[h]
\caption{Extract the sub-models}
\label{algorithm:Model_Pruning}
\begin{algorithmic}
\STATE \textbf{Given:} global model $\mathbf{W}_k$, old model $\mathbf{W}_{k-1}$, and client densities $\mathbf{P}_k=\{\rho_{i,k}\}_{i=1}^C$
\STATE compute importance scores $\mathbf{I}_k$ by Eq.~\ref{eq:pruning_metric}
\STATE sort neurons by $\mathbf{I}_k$ (descending)
\FORALL{client $i$}
  \STATE assign top-$\rho_{i,k}$ fraction of neurons as $m_{i,k}$
\ENDFOR
\STATE \textbf{return} $\mathbf{M}$
\end{algorithmic}
\end{algorithm}

During these intervals, the server utilizes the masks in \(\mathbf{M}\) to prune the global model to the corresponding client models, thereby reducing the computational overhead.

The client densities list is represented as \( \mathbf{P} \), calculated using the GMR strategy. The algorithm for Federated Model Pruning is detailed in Algorithm~\ref{algorithm:Model_Pruning}, with the output being the corresponding masks \( \mathbf{M} \). The neurons are then sorted based on their importance scores, the masks are filled with sorted neurons to meet the target density and saved.

For simplicity, we assume that \(\mathbf{P}\) is sorted in descending order, meaning the client models satisfy the nested relation:
\((\mathbf{w}^k \odot \mathbf{m}_1) \supseteq (\mathbf{w}^k \odot \mathbf{m}_2) \supseteq \dots \supseteq (\mathbf{w}^k \odot \mathbf{m}_M)\).
The bottleneck of the algorithm lies in the sorting operation, with an overall time complexity of \( O(|\mathbf{w}^k| \log |\mathbf{w}^k|) \).

\section{Incremental Model Splitting: Transmitting Varying-Size Models to Clients.}
\label{sec:IMS}
Moreover, unlike MHFL methods that transmit the full model and require clients to extract their sub-models locally, we avoid sending each sub-model separately. Instead, we employ unstructured pruning to partition the global model based on the nested relation among different densities. The pruned global model is decomposed into a set of non-overlapping \emph{increments}, 
$\Delta W_k=\{\Delta w_{j,k}\}_{j=1}^M$, 
with corresponding index sets 
$\text{Idx}_k=\{\text{idx}_{i,k}\}_{i=1}^C$. 
Each client reconstructs its target sub-model by summing the required increments:
\[
\mathbf{w}_{i,k'_i}=\sum_{j\in \text{idx}_{i,k}} \Delta \mathbf{w}_{j,k}.
\]
This Incremental Model Splitting (IMS) design removes redundancy and substantially reduces server-to-client bandwidth, at the cost of light splitting and merging operations performed once per round. The complete pseudocode is provided in Algorithm~\ref{algorithm:IMS}.

\begin{algorithm}[h]
\caption{Incremental Model Splitting}
\label{algorithm:IMS}
\begin{algorithmic}
\STATE \textbf{Given:} \(\mathbf{W}_k\) and client density set \({\mathbf{P}}_k\)
\STATE \textbf{Server routine IMSs(\(\mathbf{W}_k, {\mathbf{P}}_k\)):}
\STATE \hspace{1em}$\mathbf{M} \gets \textbf{FMP}(\mathbf{W}_k, \mathbf{W}_{k-1}, \mathbf{P}_k)$
\STATE \hspace{1em}\(\mathbf{M}_{s}, I = \text{sort}(\text{density}(\mathbf{M}), \text{descending=False})\)
\FOR{$j = 1$ \textbf{to} \(\mathrm{len}(\mathbf{M}_k)\)}
  \STATE \hspace{1em}\(\Delta \mathbf{W}_k [j] = \mathbf{W}_k \odot (\mathbf{M}_\mathrm{s}[j] - \mathbf{M}_\mathrm{s}[j-1])\)
  \STATE \hspace{1em}\(\mathbf{Idx}_k [I[j]] = [1,j]\)
\ENDFOR
\STATE \hspace{1em}\textbf{return} \(\Delta \mathbf{W}_k, \mathbf{Idx}_k\)
\STATE \textbf{Client routine IMSc(\(\mathbf{idx}_{i,k}\)):}
\STATE \hspace{1em}\(\{\Delta \mathbf{w}_{i,k}\}_{i=1}^{\mathbf{idx}_{i,k}} = \text{Download}(\mathbf{idx}_{i,k})\)
\STATE \hspace{1em}\(\mathbf{w}_{i,k} = \sum_{i \in \mathbf{idx}_{i,k}} \Delta \mathbf{w}_{i,k}\)
\STATE \hspace{1em}\textbf{return} \(\mathbf{w}_i^k\)
\end{algorithmic}
\end{algorithm}

Inside \texttt{IMSs}, the mask set $\mathbf{M}$ is obtained from densities $\mathbf{P}_k$ via \texttt{FMP}. 
In practice, $\mathbf{M}$ is not recomputed every round due to cost; it is refreshed only every $k_{\text{rest}}$ rounds and reused in between:
\[
\mathbf{M} \leftarrow 
\begin{cases}
\texttt{Extract}(\mathbf{W}_k,\mathbf{W}_{k-1},\mathbf{P}_k), & \text{if } k \bmod k_{\text{rest}} = 0, \\
\mathbf{M} \ \text{(reuse previous)}, & \text{otherwise}.
\end{cases}
\]
\texttt{IMSs} then uses the current $\mathbf{M}$ to form increments and indices. 
We embed Algorithm~\ref{algorithm:Model_Pruning} inside \texttt{IMSs} only for brevity in the top-level pseudocode; 
implementation-wise, it can equivalently be invoked outside and passed as input.

\section{Convergence analysis}
\label{sec:convergence_analysis}

\begin{table*}[t]
\centering
\caption{Notation used in the convergence analysis.}
\label{tab:conv_notation}
\small
\renewcommand{\arraystretch}{1.05}
\begin{tabular}{p{0.14\textwidth}p{0.31\textwidth}p{0.14\textwidth}p{0.31\textwidth}}
\toprule
Symbol & Meaning & Symbol & Meaning \\
\midrule
$i,g,n,k,\tau$ & client, structural group, coordinate, round, and local SGD step & $\mathcal{T}$ & number of local SGD steps per round \\
$C,\mathcal{C}$ & number of clients and client set & $W_k$ & global model at round $k$ \\
$\mathbf{w}_{i,k,\tau}$ & client-$i$ local model after $\tau$ local steps in round $k$ & $m_{i,k},m_{g,k}$ & client mask and shared group mask at round $k$ \\
$\rho_{i,k},\rho_{g,k}$ & client/group density & $C_k^{(n)}$ & active-client set on coordinate $n$, i.e., $\{i:m_{i,k}^{(n)}=1\}$ \\
$|C_k^{(n)}|$ & coordinate-wise coverage count & $S_{g,k}$ & retained-coordinate set of group $g$, i.e., $\{n:m_{g,k}^{(n)}=1\}$ \\
$c_{g,k}^*$ & minimum coordinate coverage in group $g$, i.e., $\min_{n\in S_{g,k}} |C_k^{(n)}|$ & $f_1,f_2,f_3$ & density-scaling functions for drift, variance, and Non-IID bias \\
$L,G,\sigma^2,\zeta^2$ & smoothness, gradient bound, noise bound, and Non-IID bias bound & & \\
\bottomrule
\end{tabular}
\end{table*}

We consider $C$ clients and a global model whose coordinates
we index by neurons $n\in\{1,\dots,N\}$ (with $N=d$). MHFL assigns different-sized sub-models to clients via binary masks. To isolate the effect of \emph{model sparsity} (and avoid staleness confounds), we conduct the analysis in the \emph{synchronous} setting,
Client $i$ at round $k$ holds a binary mask $m_{i,k}\in\{0,1\}^{N}$.
Let the active–client set at coordinate $n$ be
\[
C_k^{(n)}:=\{\,i:\ m_{i,k}^{(n)}=1\,\},\qquad |C_k^{(n)}|\ge 1.
\]

To make the proof easier to follow, we organize it around one master
inequality and three auxiliary bounds. The proof of Theorem~\ref{thm:ma_one_round} proceeds in
four steps: (i) derive a one-round smoothness inequality for MA, (ii) use
group-wise regrouping to expose the coverage factors induced by heterogeneous
masks, (iii) bound the resulting drift, variance, and Non-IID bias terms, and
(iv) substitute these bounds back into the master inequality. Theorem~\ref{thm:ma_telescope} then
follows by summing the one-round bound over rounds. The subsections below
follow exactly this order.

\subsection{Aggregation operators and proof roadmap}
\label{sec: two_stage}
Below we give the three aggregation rules used in our analysis.

\medskip\noindent
\paragraph{Mask-aware aggregation (MA).}
Simplify BuffMaskFedAvg to a \textbf{mask-aware aggregation (MA)} operator.
MA averages \emph{only} over clients that retain each coordinate:
\begin{equation}
\label{eq:maa}
W_{k+1}^{(n)}
\;=\;
\frac{1}{|C_k^{(n)}| }\sum_{i\in C_k^{(n)}} w_{i,k}^{(n)}
\quad\Longleftrightarrow\quad
W_{k+1}
=
\Bigl(\sum_{i=1}^{C} w_{i,k}\odot m_{i,k}\Bigr)\odot \pi_k,
\;\;
\pi_k^{(n)}:=\frac{1}{|C_k^{(n)}|}.
\end{equation}

\medskip\noindent
\paragraph{Gradient-average aggregation (GA).}
For comparison, GA applies masked \emph{gradients} but still divides by the
\emph{total} number of clients:
\begin{equation}
\label{eq:ga}
W_{k+1}
\;=\;
W_k
\;-\;
\gamma\cdot \frac{1}{C}
\sum_{i=1}^{C}\sum_{\tau=1}^{\mathcal{T}}
\nabla F_i\!\bigl(w_{i,k,\tau-1},\xi_{i,\tau-1}\bigr)\odot m_{i,k},
\end{equation}
so the per–coordinate step size scales with $|C_k^{(n)}|/C$.

\medskip\noindent
\paragraph{FedAvg (FA) with zero padding.}
Standard FedAvg averages local models; with structural heterogeneity it is
commonly implemented by zero–padding pruned coordinates:
\begin{equation}
\label{eq:fa}
W_{k+1}
\;=\;
\frac{1}{C}\sum_{i=1}^{C}\bigl(w_{i,k}\odot m_{i,k}\bigr),
\qquad\text{equivalently}\qquad
W_{k+1}^{(n)}=\frac{1}{C}\sum_{i=1}^{C} w_{i,k}^{(n)}\,\mathbf{1}\{n\in m_{i,k}\}.
\end{equation}
When $|C_k^{(n)}|\ll C$, many zeros dilute the average, shrinking
$W_{k+1}^{(n)}$ toward $0$; as density changes (e.g., under restoration),
the effective averaging scale $|C_k^{(n)}|/C$ \emph{jumps}, which can
destabilize training.

Unlike MA/GA, FA with zero padding does \emph{not} correspond to a stochastic
gradient step on any fixed smooth objective. Indeed,
\[
W_{k+1}^{(n)}
=\frac{1}{C}\sum_{i=1}^{C} w_{i,k}^{(n)}\,\mathbf{1}\{n\in m_{i,k}\}
=\frac{|C_k^{(n)}|}{C}\;
\underbrace{\Bigl(\tfrac{1}{|C_k^{(n)}|}\sum_{i\in C_k^{(n)}} w_{i,k}^{(n)}\Bigr)}_{\text{MA average at coord.\ }n}.
\]
Thus FA equals the MA average multiplied by a \emph{coverage–dependent} factor
$|C_k^{(n)}|/C\in[0,1]$, i.e., a coordinate/round–dependent contraction that
depends on participation patterns rather than objective geometry. Consequently,
FA cannot be written as $W_{k+1}=W_k-\gamma\,\widehat{g}_k$ with
$\mathbb{E}[\widehat{g}_k]=\nabla F(W_k)$ for a fixed $F$, and is therefore not
amenable to a clean gradient–based convergence analysis in the mask–misaligned
regime. This motivates focusing on MA/GA in our theory.

\paragraph{Group-wise two-stage aggregation.}

Similar to the analytical steps in \citet{zhou2023every}, we examine convergence from the perspective of neuron-wise aggregation. However, we do not agree with the concept of the \textit{minimum coverage index} proposed in \citet{zhou2023every}, as it merely reflects the scaled minimum density among all clients. 
Instead, we argue that the heterogeneity of model densities across clients and communication rounds exerts an accumulative effect on convergence. 

To validate this hypothesis, we leverage the nested structure induced by shared pruning strategies and redefine the model aggregation process as a two-stage procedure. Specifically, clients are first grouped into \( \mathcal{G} \) structural groups \( \{ \mathcal{C}_g \}_{g=1}^{\mathcal{G}} \), where each group shares a common mask \( m_{g,k} \) and density \( \rho_{g,k} \). Let \( \mathcal{G}_k^{(n)} \) denote the set of groups whose masks contain neuron \( n \).

Then the two-stage aggregation is given by:  
(1) Within-group aggregation (using traditional FedAvg since structures are uniform), followed by  
(2) Parameter-wise aggregation across groups to obtain the global model:
\begin{equation}
\mathbf{w}_k^{(n)} = \frac{1}{|C_k^{(n)}|} \sum_{g \in \mathcal{G}_k^{(n)}} |\mathcal{C}_g| \cdot \mathbf{w}_{g,k}^{(n)},
\quad \text{where } 
\mathbf{w}_{g,k} := \frac{1}{|\mathcal{C}_g|} \sum_{i \in \mathcal{C}_g} \mathbf{w}_{i,k} \odot m_{g,k}.
\end{equation}

This formulation is equivalent to Eq.~\ref{eq:maa}, but recast to emphasize the structural grouping.  
For comparison, the corresponding GA becomes:

\begin{equation}
\mathbf{w}_k^{(n)} = \frac{1}{C} \sum_{g \in \mathcal{G}_k^{(n)}} |\mathcal{C}_g| \cdot \mathbf{w}_{g,k}^{(n)},
\quad \text{where } 
\mathbf{w}_{g,k} := \frac{1}{|\mathcal{C}_g|} \sum_{i \in \mathcal{C}_g} \mathbf{w}_{i,k} \odot m_{g,k}.
\end{equation}

\subsubsection{Step 1: Deriving the one-round master inequality}

\textbf{Step 1 (L-smoothness).} Under Assumption 1, for any round \( k \), we have:
\begin{equation}
    F(\mathbf{W}_{k+1}) \le F(\mathbf{W}_k) + \langle \nabla F(\mathbf{W}_k), \mathbf{W}_{k+1} - \mathbf{W}_k \rangle + \frac{L}{2} \|\mathbf{W}_{k+1} - \mathbf{W}_k\|^2. 
\end{equation}

Taking expectation on both sides:
\begin{equation}
\mathbb{E}[F(\mathbf{W}_{k+1})] - \mathbb{E}[F(\mathbf{W}_k)] \le \mathbb{E}\langle \nabla F(\mathbf{W}_k), \mathbf{W}_{k+1} - \mathbf{W}_k \rangle + \frac{L}{2} \mathbb{E} \|\mathbf{W}_{k+1} - \mathbf{W}_k\|^2.
\end{equation}

\textbf{Step 2.} We then analyze the global update expression. For any region \( n \), we have:
\begin{equation}
\mathbf{W}_k^{(n)} - \mathbf{W}_{k+1}^{(n)} = \frac{1}{|C_k^{(n)}|} \sum_{i \in C_k^{(n)}} \sum_{\tau=1}^\mathcal{T} \gamma \nabla F_i^{(n)}(\mathbf{w}_{i,k,\tau-1}, \xi_{i,\tau-1}).
\end{equation}
\textbf{Step 3} We next analyze the inner product.
\(\mathbb{E} \left\langle \nabla F(\mathbf{W}_k), \mathbf{W}_{k+1} - \mathbf{W}_k \right\rangle\)
by considering a sum of inner products over $n$ regions. We have:

\begin{align}
&\mathbb{E} \left\langle \nabla F(\mathbf{W}_k), \mathbf{W}_{k+1} - \mathbf{W}_k \right\rangle \notag \\
&= \sum_{n=1}^{N} \mathbb{E} \left\langle \nabla F^{(n)}(\mathbf{W}_k), \mathbf{W}_{k+1}^{(n)} - \mathbf{W}_k^{(n)} \right\rangle  \notag \\
&\text{\small (According to the properties of the inner product, split the inner product into $K$ structural regions)} \notag \\
&= \sum_{n=1}^{N} \mathbb{E} \left\langle \nabla F^{(n)}(\mathbf{W}_k), 
- \frac{1}{|C_k^{(n)}|} \sum_{i \in C_k^{(n)}} \sum_{\tau=1}^\mathcal{T} \gamma \nabla F_i^{(n)}(\mathbf{w}_{i,k,\tau-1}, \xi_{i,\tau-1}) 
\right\rangle \notag \\
&= \sum_{n=1}^{N} \mathbb{E} \left\langle \nabla F^{(n)}(\mathbf{W}_k), 
- \frac{1}{|C_k^{(n)}|} \sum_{i \in C_k^{(n)}} \sum_{\tau=1}^\mathcal{T} \gamma \, 
\mathbb{E} \left[ \nabla F_i^{(n)}(\mathbf{w}_{i,k,\tau-1}, \xi_{i,\tau-1}) \mid \mathbf{W}_k \right]
\right\rangle \notag \\
&\text{\small (Use the tower rule of expectation over $\mathbf{W}_k$)} \notag \\
&= \sum_{n=1}^{N} \mathbb{E} \left\langle \nabla F^{(n)}(\mathbf{W}_k), 
- \frac{1}{|C_k^{(n)}|} \sum_{i \in C_k^{(n)}} \sum_{\tau=1}^\mathcal{T} \gamma \nabla F_i^{(n)}(\mathbf{w}_{i,k,\tau-1})
\right\rangle  \notag \\
&\text{\small (Now gradients are conditional means)} \notag \\
&= - \sum_{n=1}^{N} \mathbb{E} \left\langle \nabla F^{(n)}(\mathbf{W}_k), 
\frac{1}{|C_k^{(n)}|} \sum_{i \in C_k^{(n)}} \sum_{\tau=1}^\mathcal{T} \gamma 
\left( \nabla F^{(n)}(\mathbf{W}_k) + \nabla F_i^{(n)}(\mathbf{w}_{i,k,\tau-1}) - \nabla F^{(n)}(\mathbf{W}_k) \right)
\right\rangle \notag \\
&\text{\small (Insert $\nabla F^{(n)}(\mathbf{W}_k) - \nabla F^{(n)}(\mathbf{W}_k) = 0$ to isolate descent direction)} \notag \\
&= -\sum_{n=1}^{N} \mathbb{E} \left\langle \nabla F^{(n)}(\mathbf{W}_k), \gamma \mathcal{T} \nabla F^{(n)}(\mathbf{W}_k) \right\rangle 
\text{\small (Because $\frac{1}{|C_k^{(n)}|} \sum_{i \in C_k^{(n)}} 
\nabla F^{(n)}(\mathbf{W}_k) = \nabla F^{(n)}(\mathbf{W}_k)$) } \notag \\ 
&\quad - \sum_{n=1}^{N} \mathbb{E} \left\langle \nabla F^{(n)}(\mathbf{W}_k), 
\frac{1}{|C_k^{(n)}|} \sum_{i \in C_k^{(n)}} \sum_{\tau=1}^\mathcal{T} \gamma 
\left( \nabla F_i^{(n)}(\mathbf{w}_{i,k,\tau-1}) - \nabla F^{(n)}(\mathbf{W}_k) \right) \right\rangle    \label{eq:ananysis_inner_product} \\
&\text{\small (Separate the true descent term and the deviation from it)}  \notag 
\end{align}

For the first term on the right-hand side of Eq.~\ref{eq:ananysis_inner_product}, it is easy to see that:

\begin{align}
       & - \sum_{n=1}^{N} \mathbb{E} \left\langle \nabla F^{(n)}(\mathbf{W}_k), \gamma \mathcal{T} \nabla F^{(n)}(\mathbf{W}_k) \right\rangle  \notag \\
=& -\gamma \mathcal{T} \sum_{n=1}^{N} \mathbb{E} \left\| \nabla F^{(n)}(\mathbf{W}_k) \right\|^2 
= -\gamma \mathcal{T} \mathbb{E} \left\| \nabla F(\mathbf{W}_k) \right\|^2.
\label{eq:inner_product_first}
\end{align}

For the second term on the right-hand side of Eq.~\ref{eq:ananysis_inner_product}, we apply the Cauchy--Schwarz inequality:
\(\langle a, b \rangle \le \frac{1}{2} \|a\|^2 + \frac{1}{2} \|b\|^2\)
to obtain:

\begin{align}
&- \sum_{n=1}^{N} \mathbb{E} \left\langle \nabla F^{(n)}(\mathbf{W}_k), 
\frac{1}{|C_k^{(n)}|} \sum_{i \in C_k^{(n)}} \sum_{\tau=1}^{\mathcal{T}} \gamma 
\left( \nabla F_i^{(n)}(\mathbf{w}_{i,k,\tau-1}) - \nabla F^{(n)}(\mathbf{W}_k) \right)
\right\rangle  \notag \\
=& - \sum_{n=1}^{N} \gamma \mathcal{T} \cdot \mathbb{E} \left\langle \nabla F^{(n)}(\mathbf{W}_k), 
\frac{1}{\mathcal{T} |C_k^{(n)}|} \sum_{i \in C_k^{(n)}} \sum_{\tau=1}^{\mathcal{T}} 
\left( \nabla F_i^{(n)}(\mathbf{w}_{i,k,\tau-1}) - \nabla F^{(n)}(\mathbf{W}_k) \right) 
\right\rangle \notag \\
\le& \frac{\gamma \mathcal{T}}{2} \sum_{n=1}^{N} \mathbb{E} \left\| \nabla F^{(n)}(\mathbf{W}_k) \right\|^2 \notag \\
&+ \frac{\gamma \mathcal{T}}{2} \sum_{n=1}^{N} 
\mathbb{E} \left\| \frac{1}{\mathcal{T} |C_k^{(n)}|} 
\sum_{i \in C_k^{(n)}} \sum_{\tau=1}^{\mathcal{T}} 
\left( \nabla F_i^{(n)}(\mathbf{w}_{i,k,\tau-1}) - \nabla F^{(n)}(\mathbf{W}_k) \right)
\right\|^2. \label{eq:inner_product_second}
\end{align}

Combining Eq.~\ref{eq:ananysis_inner_product}, Eq.~\ref{eq:inner_product_first}, and Eq.~\ref{eq:inner_product_second}, we obtain:
\begin{align}
&\mathbb{E} \left\langle \nabla F(\mathbf{W}_k), \mathbf{w}_{k+1} - \mathbf{w}_k \right\rangle \notag \\ 
\le& - \gamma \mathcal{T} \mathbb{E} \left\| \nabla F(\mathbf{W}_k) \right\|^2  
 + \frac{\gamma \mathcal{T}}{2} \sum_{n=1}^{N} \mathbb{E} \left\| \nabla F^{(n)}(\mathbf{W}_k) \right\|^2  \notag \\ 
 &+ \frac{\gamma \mathcal{T}}{2} \sum_{n=1}^{N} 
\mathbb{E} \left\| \frac{1}{\mathcal{T} |C_k^{(n)}|} 
\sum_{i \in C_k^{(n)}} \sum_{\tau=1}^{\mathcal{T}} 
\left( \nabla F_i^{(n)}(\mathbf{w}_{i,k,\tau-1}) - \nabla F^{(n)}(\mathbf{W}_k) \right)
\right\|^2  \notag  \\ 
\le& - \frac{\gamma \mathcal{T}}{2} \sum_{n=1}^{N} \mathbb{E} \left\| \nabla F^{(n)}(\mathbf{W}_k) \right\|^2 \notag\\
&+ \frac{\gamma \mathcal{T}}{2} \sum_{n=1}^{N} 
\mathbb{E} \left\| \frac{1}{\mathcal{T} |C_k^{(n)}|} 
\sum_{i \in C_k^{(n)}} \sum_{\tau=1}^{\mathcal{T}} 
\left( \nabla F_i^{(n)}(\mathbf{w}_{i,k,\tau-1}) - \nabla F^{(n)}(\mathbf{W}_k) \right)
\right\|^2. \label{eq:first_term}    
\end{align}

\textbf{Step 4: }Analysis of the upperbound for $\frac{L}{2} \mathbb{E} \|\mathbf{W}_{k+1} - \mathbf{W}_k\|^2$.

\begin{align}
&\frac{L}{2} \mathbb{E} \|\mathbf{w}_{k+1} - \mathbf{w}_k\|^2  \notag   \\
=& \frac{L}{2} \sum_{n=1}^{N}\mathbb{E} \left\| 
\frac{1}{|C_k^{(n)}|} \sum_{i \in C_k^{(n)}} \sum_{\tau=1}^\mathcal{T} \gamma \nabla F_i^{(n)}(\mathbf{w}_{i,k,\tau-1}, \xi_{i,\tau-1})
\right\|^2 \notag \\
\le& \frac{3L}{2} \sum_{n=1}^{N}\mathbb{E} \left\|
\frac{1}{|C_k^{(n)}|} \sum_{i \in C_k^{(n)}} \sum_{\tau=1}^\mathcal{T} \gamma 
\left[ \nabla F_i^{(n)}(\mathbf{w}_{i,k,\tau-1}, \xi_{i,\tau-1}) - \nabla F_i^{(n)}(\mathbf{w}_{i,k,\tau-1}) \right]
\right\|^2 \notag \\
&+ \frac{3L}{2} \sum_{n=1}^{N}\mathbb{E} \left\|
\frac{1}{|C_k^{(n)}|} \sum_{i \in C_k^{(n)}} \sum_{\tau=1}^\mathcal{T} \gamma 
\left[ \nabla F_i^{(n)}(\mathbf{w}_{i,k,\tau-1}) - \nabla F_i^{(n)}(\mathbf{W}_k) \right]
\right\|^2 \notag \\
&+ \frac{3L}{2} \sum_{n=1}^{N}\mathbb{E} \left\|
\frac{1}{|C_k^{(n)}|} \sum_{i \in C_k^{(n)}} \sum_{\tau=1}^\mathcal{T} \gamma 
\nabla F_i^{(n)}(\mathbf{W}_k)
\right\|^2. \notag  \\
\le& \frac{3L\mathcal{T}^2\gamma^2}{2} \sum_{n=1}^{N}\mathbb{E} \left\|
\frac{1}{\mathcal{T}|C_k^{(n)}|} \sum_{i \in C_k^{(n)}} \sum_{\tau=1}^\mathcal{T} 
\left[ \nabla F_i^{(n)}(\mathbf{w}_{i,k,\tau-1}, \xi_{i,\tau-1}) - \nabla F_i^{(n)}(\mathbf{w}_{i,k,\tau-1}) \right]
\right\|^2 \notag \\
&+ \frac{3L\mathcal{T}^2\gamma^2}{2} \sum_{n=1}^{N}\mathbb{E} \left\|
\frac{1}{\mathcal{T}|C_k^{(n)}|} \sum_{i \in C_k^{(n)}} \sum_{\tau=1}^\mathcal{T} 
\left[ \nabla F_i^{(n)}(\mathbf{w}_{i,k,\tau-1}) - \nabla F_i^{(n)}(\mathbf{W}_k) \right]
\right\|^2 \notag \\
&+ \frac{3L\mathcal{T}^2\gamma^2}{2} \sum_{n=1}^{N}\mathbb{E} \left\|
\frac{1}{|C_k^{(n)}|} \sum_{i \in C_k^{(n)}}  
\nabla F_i^{(n)}(\mathbf{W}_k)
\right\|^2.  \label{eq:L2_term_1} 
\end{align}

\textbf{Step 5} Combining the two upperbounds, we have:

\begin{align}
&\mathbb{E}[F(\mathbf{w}_{k+1})] - \mathbb{E}[F(\mathbf{W}_k)] \notag    \\
\le&\mathbb{E} \left\langle \nabla F(\mathbf{W}_k), \mathbf{w}_{k+1} - \mathbf{w}_k \right\rangle  
+\frac{L}{2} \mathbb{E} \|\mathbf{W}_{k+1} - \mathbf{W}_k\|^2 \notag \\
\le& - \frac{\gamma \mathcal{T}}{2} \sum_{n=1}^{N} \mathbb{E} \left\| \nabla F^{(n)}(\mathbf{W}_k) \right\|^2 \notag \\ 
&+ \frac{\gamma \mathcal{T}}{2} \sum_{n=1}^{N} 
\mathbb{E} \left\| \frac{1}{\mathcal{T} |C_k^{(n)}|} 
\sum_{i \in C_k^{(n)}} \sum_{\tau=1}^{\mathcal{T}} 
\left( \nabla F_i^{(n)}(\mathbf{w}_{i,k,\tau-1}) - \nabla F^{(n)}(\mathbf{W}_k) \right)
\right\|^2 \notag \\
&+\frac{3L\mathcal{T}^2\gamma^2}{2} \sum_{n=1}^{N}\mathbb{E} \left\|
\frac{1}{\mathcal{T}|C_k^{(n)}|} \sum_{i \in C_k^{(n)}} \sum_{\tau=1}^\mathcal{T} 
\left[ \nabla F_i^{(n)}(\mathbf{w}_{i,k,\tau-1}, \xi_{i,\tau-1}) - \nabla F_i^{(n)}(\mathbf{w}_{i,k,\tau-1}) \right]
\right\|^2 \notag \\
&+ \frac{3L\mathcal{T}^2\gamma^2}{2} \sum_{n=1}^{N}\mathbb{E} \left\|
\frac{1}{\mathcal{T}|C_k^{(n)}|} \sum_{i \in C_k^{(n)}} \sum_{\tau=1}^\mathcal{T} 
\left[ \nabla F_i^{(n)}(\mathbf{w}_{i,k,\tau-1}) - \nabla F_i^{(n)}(\mathbf{W}_k) \right]
\right\|^2 \notag \\
&+ \frac{3L\mathcal{T}^2\gamma^2}{2} \sum_{n=1}^{N}\mathbb{E} \left\|
\frac{1}{|C_k^{(n)}|} \sum_{i \in C_k^{(n)}}  
\nabla F_i^{(n)}(\mathbf{W}_k)
\right\|^2. 
\label{eq:final_initial_term} 
\end{align}

\subsubsection{Step 2: Bounding the drift, variance, and bias terms}

We next upper-bound the three auxiliary terms in
Eq.~\ref{eq:final_initial_term}. The key idea is to regroup clients by shared
mask structure, so that each coordinate-wise average can be controlled through
group-level quantities and the minimum within-group coverage \(c_{g,k}^*\).

Similar to \citet{zhou2023every}, we leverage the fact that the $L_2$ norm of the gradient over the entire model can be decomposed as the sum of norms over different parameter regions (i.e., model blocks $n = 1, \ldots, N$). This allows us to analyze the full weight matrix rather than focusing on sub-vectors.

However, unlike \citet{zhou2023every}, we do not decompose the model drift into the model reduction noise and the local training drift when characterizing the difference between the local and global models at any local epoch $\tau$.

Our rationale is that, under mask-aware aggregation, each sub-model only contributes to its corresponding sub-region of the global model. Hence, the reduction in model size during server splitting does not reflect actual model degradation. We therefore define model drift solely as the local training drift:
\begin{equation}
\left\| \mathbf{w}_{i,k,\tau} \odot \mathbf{m}_{i,k} - \mathbf{W}_k \odot \mathbf{m}_{i,k} \right\|.
\end{equation}
In this formulation, the model reduction noise is excluded because sub-models do not affect the masked-out parameters. The convergence degradation arises from partial updates and limited gradient coverage, not from global model reduction. For brevity, we write $|C_g|:=|\mathcal{C}_g|$ in the proof section below.
\begin{lemma}
\label{lem:2}\quad
\textbf{Group-wise Client Model Deviation Bound}
\end{lemma}
For group \(g\), define the group-average masked model at local step \(\tau\) as
\[
\mathbf{w}_{g,k,\tau}:=\frac{1}{|C_g|}\sum_{i\in C_g}\mathbf{w}_{i,k,\tau}\odot m_{g,k}.
\]
Then
\begin{align}
\frac{1}{\mathcal{T}} \sum_{\tau=1}^{\mathcal{T}} \mathbb{E} \left\|  \left(\mathbf{w}_{g,k,\tau-1} - \mathbf{W}_k \right) \odot m_{g, k} \right\|^2
&\leq \frac{\gamma^2\mathcal{T}^2}{3} f^2_1(\rho_{g,k}) G^2.
\label{eq:lemma2_bound}
\end{align}

\textbf{Proof.}
By Jensen’s inequality over the clients in group \(g\),
\begin{align}
&\frac{1}{\mathcal{T}} \sum_{\tau=1}^{\mathcal{T}} \mathbb{E} \left\| \left(\mathbf{w}_{g,k,\tau-1} - \mathbf{W}_k \right) \odot m_{g, k} \right\|^2 \notag \\
=&\frac{1}{\mathcal{T}} \sum_{\tau=1}^{\mathcal{T}} \mathbb{E} \left\| \frac{1}{|C_g|} \sum_{i \in C_g}
\left[\left(\mathbf{w}_{i,k,\tau-1} - \mathbf{W}_k \right) \odot m_{g, k}\right] \right\|^2 \notag \\
\leq& \frac{1}{|C_g|} \sum_{i \in C_g} \frac{1}{\mathcal{T}} \sum_{\tau=1}^{\mathcal{T}} \mathbb{E} \left\| \left(\mathbf{w}_{i,k,\tau-1} - \mathbf{W}_k \right) \odot m_{g, k} \right\|^2.
\label{eq:lemma2_jensen}
\end{align}
For any client \(i\in C_g\), the masked SGD recursion gives
\[
\left(\mathbf{w}_{i,k,\tau-1} - \mathbf{W}_k\right)\odot m_{g,k}
= -\gamma \sum_{s=0}^{\tau-2} \nabla F_i(\mathbf{w}_{i,k,s},\xi_{i,s})\odot m_{g,k},
\]
where the sum is empty when \(\tau=1\). Therefore,
\begin{align}
&\frac{1}{\mathcal{T}} \sum_{\tau=1}^{\mathcal{T}} \mathbb{E} \left\| \left(\mathbf{w}_{i,k,\tau-1} - \mathbf{W}_k \right) \odot m_{g, k} \right\|^2 \notag \\
\leq& \frac{\gamma^2}{\mathcal{T}} \sum_{\tau=1}^{\mathcal{T}} (\tau-1)
\sum_{s=0}^{\tau-2} \mathbb{E} \left\| \nabla F_i(\mathbf{w}_{i,k,s},\xi_{i,s}) \odot m_{g,k} \right\|^2 \notag \\
\leq& \frac{\gamma^2}{\mathcal{T}} \sum_{\tau=1}^{\mathcal{T}} (\tau-1)^2 f_1^2(\rho_{g,k}) G^2 \notag \\
\leq& \frac{\gamma^2\mathcal{T}^2}{3} f_1^2(\rho_{g,k}) G^2.
\label{eq:lemma2_single_client}
\end{align}
Substituting Eq.~\ref{eq:lemma2_single_client} into Eq.~\ref{eq:lemma2_jensen} completes the proof.

\begin{lemma}
\label{lem:3}\quad
\textbf{Group-aware Gradient Drift Bound under Parameter-wise Aggregation}
\end{lemma}
Let
\[
\nabla F_g(\mathbf{w}) := \frac{1}{|C_g|}\sum_{i\in C_g} \nabla F_i(\mathbf{w}),
\qquad
\nabla F_g^{(n)}(\mathbf{w}) := \bigl[\nabla F_g(\mathbf{w})\bigr]^{(n)}.
\]
Then the aggregated gradient-drift term satisfies
\begin{align}
\sum_{n=1}^N\mathbb{E} \left\| \frac{1}{\mathcal{T}} \sum_{\tau=1}^{\mathcal{T}} \frac{1}{|C_{k}^{(n)}|} \sum_{i \in C_k^{(n)}} \left[\nabla F_i^{(n)}(\mathbf{w}_{i,k,\tau-1}) - \nabla F_i^{(n)}(\mathbf{W}_k)\right]  \right\|^2 
\leq \frac{L^2\gamma^2\mathcal{T}^2}{3} \sum_{g=1}^{\mathcal{G}} \frac{|C_g|}{c_{g,k}^*} f^2_1(\rho_{g,k}) G^2.
\label{eq:lemma5}
\end{align}

\textbf{Proof.}
By Jensen’s inequality over local epochs,
\begin{align}
&\sum_{n=1}^N\mathbb{E} \left\| \frac{1}{\mathcal{T}} \sum_{\tau=1}^{\mathcal{T}} \frac{1}{|C_{k}^{(n)}|} \sum_{i \in C_k^{(n)}} \left[\nabla F_i^{(n)}(\mathbf{w}_{i,k,\tau-1}) - \nabla F_i^{(n)}(\mathbf{W}_k)\right]  \right\|^2 \notag \\
\leq& \frac{1}{\mathcal{T}} \sum_{\tau=1}^{\mathcal{T}} \sum_{n=1}^N\mathbb{E} \left\| \frac{1}{|C_{k}^{(n)}|} \sum_{i \in C_k^{(n)}} \left[\nabla F_i^{(n)}(\mathbf{w}_{i,k,\tau-1}) - \nabla F_i^{(n)}(\mathbf{W}_k)\right]  \right\|^2.
\label{eq:lemma3_jensen_tau}
\end{align}
Because all clients in the same structural group share the same mask,
\begin{align}
\frac{1}{|C_{k}^{(n)}|} \sum_{i \in C_k^{(n)}} \left[\nabla F_i^{(n)}(\mathbf{w}_{i,k,\tau-1}) - \nabla F_i^{(n)}(\mathbf{W}_k)\right]
= \sum_{g \in \mathcal{G}_k^{(n)}} \frac{|C_g|}{|C_k^{(n)}|}
\left[\nabla F_g^{(n)}(\mathbf{w}_{g,k,\tau-1}) - \nabla F_g^{(n)}(\mathbf{W}_k)\right].
\label{eq:lemma3_group_regroup}
\end{align}
Applying weighted Jensen’s inequality to Eq.~\ref{eq:lemma3_group_regroup}, with weights
\(\frac{|C_g|}{|C_k^{(n)}|}\) that sum to one, yields
\begin{align}
&\frac{1}{\mathcal{T}} \sum_{\tau=1}^{\mathcal{T}} \sum_{n=1}^N\mathbb{E} \left\| \sum_{g \in \mathcal{G}_k^{(n)}} \frac{|C_g|}{|C_k^{(n)}|}
\left[\nabla F_g^{(n)}(\mathbf{w}_{g,k,\tau-1}) - \nabla F_g^{(n)}(\mathbf{W}_k)\right] \right\|^2 \notag \\
\leq& \frac{1}{\mathcal{T}} \sum_{\tau=1}^{\mathcal{T}} \sum_{n=1}^N \sum_{g \in \mathcal{G}_k^{(n)}} \frac{|C_g|}{|C_k^{(n)}|}
\mathbb{E} \left\| \nabla F_g^{(n)}(\mathbf{w}_{g,k,\tau-1}) - \nabla F_g^{(n)}(\mathbf{W}_k) \right\|^2 \notag \\
=& \frac{1}{\mathcal{T}} \sum_{\tau=1}^{\mathcal{T}} \sum_{g=1}^{\mathcal{G}} \sum_{n \in S_{g,k}} \frac{|C_g|}{|C_k^{(n)}|}
\mathbb{E} \left\| \nabla F_g^{(n)}(\mathbf{w}_{g,k,\tau-1}) - \nabla F_g^{(n)}(\mathbf{W}_k) \right\|^2.
\label{eq:lemma3_swap}
\end{align}
By Assumption~5.1 and \(c_{g,k}^* := \min_{n\in S_{g,k}} |C_k^{(n)}|\),
\begin{align}
&\sum_{n \in S_{g,k}} \frac{|C_g|}{|C_k^{(n)}|}
\mathbb{E} \left\| \nabla F_g^{(n)}(\mathbf{w}_{g,k,\tau-1}) - \nabla F_g^{(n)}(\mathbf{W}_k) \right\|^2 \notag \\
\leq& \frac{|C_g|}{c_{g,k}^*}
\mathbb{E} \left\| \left( \nabla F_g(\mathbf{w}_{g,k,\tau-1}) - \nabla F_g(\mathbf{W}_k) \right) \odot m_{g,k} \right\|^2 \notag \\
\leq& \frac{L^2 |C_g|}{c_{g,k}^*}
\mathbb{E} \left\| \left( \mathbf{w}_{g,k,\tau-1} - \mathbf{W}_k \right) \odot m_{g,k} \right\|^2.
\label{eq:lemma3_lipschitz}
\end{align}
Combining Eqs.~\ref{eq:lemma3_jensen_tau}--\ref{eq:lemma3_lipschitz} and invoking Lemma~\ref{lem:2}, we obtain
\begin{align}
\sum_{n=1}^N\mathbb{E} \left\| \frac{1}{\mathcal{T}} \sum_{\tau=1}^{\mathcal{T}} \frac{1}{|C_{k}^{(n)}|} \sum_{i \in C_k^{(n)}} \left[\nabla F_i^{(n)}(\mathbf{w}_{i,k,\tau-1}) - \nabla F_i^{(n)}(\mathbf{W}_k)\right]  \right\|^2 
\leq \frac{L^2\gamma^2\mathcal{T}^2}{3} \sum_{g=1}^{\mathcal{G}} \frac{|C_g|}{c_{g,k}^*} f_1^2(\rho_{g,k}) G^2.
\end{align}

\begin{lemma}
\label{lem:4}\quad
\textbf{Group-aware Gradient Variance Bound under Parameter-wise Aggregation}
\end{lemma}
Under Assumption~5.3, for any round \(k\), the deviation between the aggregated stochastic gradient and its full-batch counterpart satisfies
\begin{align}
&\sum_{n=1}^N\mathbb{E} \left\| \frac{1}{\mathcal{T}} \sum_{\tau=1}^{\mathcal{T}} \frac{1}{|C_{k}^{(n)}|} \sum_{i \in C_k^{(n)}}
\Bigl[\nabla F_i^{(n)}(\mathbf{w}_{i,k,\tau-1}, \xi_{i,\tau-1}) - \nabla F_i^{(n)}(\mathbf{w}_{i,k,\tau-1})\Bigr] \right\|^2 \notag\\
&\leq
\sum_{g=1}^{\mathcal{G}} \frac{|C_g|}{(c_{g,k}^*)^2} \cdot \frac{f^2_2(\rho_{g,k}) \sigma^2}{\mathcal{T}}.
\label{eq:variance_result}
\end{align}

\textbf{Proof.}
Define the client-level and group-level gradient noises as
\[
\delta_{i,k,\tau}:=\nabla F_i(\mathbf{w}_{i,k,\tau-1},\xi_{i,\tau-1})-\nabla F_i(\mathbf{w}_{i,k,\tau-1}),
\qquad
\delta_{g,k,\tau}:=\frac{1}{|C_g|}\sum_{i\in C_g}\delta_{i,k,\tau}.
\]
By Assumption~5.3, \(\mathbb{E}[\delta_{i,k,\tau}\mid \mathbf{w}_{i,k,\tau-1}]=0\), and the client noises are independent across clients. Hence,
\begin{align}
\mathbb{E}\|\delta_{g,k,\tau}\odot m_{g,k}\|^2
&= \frac{1}{|C_g|^2}\sum_{i\in C_g} \mathbb{E}\|\delta_{i,k,\tau}\odot m_{g,k}\|^2 \notag\\
&\leq \frac{f_2^2(\rho_{g,k})\sigma^2}{|C_g|}.
\label{eq:group_noise_bound}
\end{align}
Moreover, for each active coordinate \(n\),
\[
\frac{1}{|C_k^{(n)}|}\sum_{i\in C_k^{(n)}}\delta_{i,k,\tau}^{(n)}
= \sum_{g\in\mathcal{G}_k^{(n)}} \frac{|C_g|}{|C_k^{(n)}|}\, \delta_{g,k,\tau}^{(n)}.
\]
Applying Jensen’s inequality over \(\tau\), then using the independence and zero mean of the group noises, gives
\begin{align}
&\sum_{n=1}^N\mathbb{E} \left\| \frac{1}{\mathcal{T}} \sum_{\tau=1}^{\mathcal{T}} \sum_{g\in\mathcal{G}_k^{(n)}} \frac{|C_g|}{|C_k^{(n)}|}\, \delta_{g,k,\tau}^{(n)} \right\|^2 \notag\\
\leq& \frac{1}{\mathcal{T}} \sum_{\tau=1}^{\mathcal{T}} \sum_{n=1}^N \sum_{g\in\mathcal{G}_k^{(n)}} \frac{|C_g|^2}{|C_k^{(n)}|^2} \, \mathbb{E}\|\delta_{g,k,\tau}^{(n)}\|^2 \notag\\
=& \frac{1}{\mathcal{T}} \sum_{\tau=1}^{\mathcal{T}} \sum_{g=1}^{\mathcal{G}} \sum_{n\in S_{g,k}} \frac{|C_g|^2}{|C_k^{(n)}|^2} \, \mathbb{E}\|\delta_{g,k,\tau}^{(n)}\|^2 \notag\\
\leq& \frac{1}{\mathcal{T}} \sum_{\tau=1}^{\mathcal{T}} \sum_{g=1}^{\mathcal{G}} \frac{|C_g|^2}{(c_{g,k}^*)^2} \, \mathbb{E}\|\delta_{g,k,\tau}\odot m_{g,k}\|^2 \notag\\
\leq& \sum_{g=1}^{\mathcal{G}} \frac{|C_g|}{(c_{g,k}^*)^2} \cdot \frac{f_2^2(\rho_{g,k})\sigma^2}{\mathcal{T}},
\end{align}
where the last step uses Eq.~\ref{eq:group_noise_bound}.

\begin{lemma}
\label{lem:5}\quad
\textbf{Group-aware Gradient Bias Bound under Parameter-wise Aggregation}
\end{lemma}
Let \(\nabla F_g(\mathbf{W}) := \frac{1}{|C_g|}\sum_{i\in C_g} \nabla F_i(\mathbf{W})\). Under Assumption~5.4,
\begin{align}
\sum_{n=1}^N\mathbb{E} \left\| \frac{1}{|C_{k}^{(n)}|} \sum_{i \in C_k^{(n)}}
\left[\nabla F_i^{(n)}(\mathbf{W}_k) - \nabla F^{(n)}(\mathbf{W}_k)\right] \right\|^2
\leq \sum_{g=1}^{\mathcal{G}} \frac{|C_g|}{c_{g,k}^{*}} f_3^2(\rho_{g,k}) \zeta^2.
\label{eq:final_bias}
\end{align}

\textbf{Proof.}
Using the client bias notation \(b_i(\mathbf{W}_k):=\nabla F_i(\mathbf{W}_k)-\nabla F(\mathbf{W}_k)\), we first bound the group-average bias:
\begin{align}
&\mathbb{E} \left\| \left[\nabla F_g(\mathbf{W}_k) - \nabla F(\mathbf{W}_k)\right] \odot m_{g,k} \right\|^2 \notag\\
=& \mathbb{E} \left\| \frac{1}{|C_g|}\sum_{i\in C_g} b_i(\mathbf{W}_k) \odot m_{g,k} \right\|^2 \notag\\
\leq& \frac{1}{|C_g|}\sum_{i\in C_g} \mathbb{E}\|b_i(\mathbf{W}_k)\odot m_{g,k}\|^2 \notag\\
\leq& f_3^2(\rho_{g,k})\zeta^2.
\label{eq:group_bias}
\end{align}
Now regroup the coordinate-wise MA average by structural groups:
\begin{align}
&\sum_{n=1}^N\mathbb{E} \left\| \frac{1}{|C_{k}^{(n)}|} \sum_{i \in C_k^{(n)}}
\left[\nabla F_i^{(n)}(\mathbf{W}_k) - \nabla F^{(n)}(\mathbf{W}_k)\right] \right\|^2 \notag\\
=& \sum_{n=1}^N\mathbb{E} \left\| \sum_{g \in \mathcal{G}_k^{(n)}} \frac{|C_g|}{|C_k^{(n)}|}
\left[\nabla F_g^{(n)}(\mathbf{W}_k) - \nabla F^{(n)}(\mathbf{W}_k)\right] \right\|^2 \notag\\
\leq& \sum_{n=1}^N \sum_{g \in \mathcal{G}_k^{(n)}} \frac{|C_g|}{|C_k^{(n)}|}
\mathbb{E} \left\| \nabla F_g^{(n)}(\mathbf{W}_k) - \nabla F^{(n)}(\mathbf{W}_k) \right\|^2 \notag\\
=& \sum_{g=1}^{\mathcal{G}} \sum_{n \in S_{g,k}} \frac{|C_g|}{|C_k^{(n)}|}
\mathbb{E} \left\| \nabla F_g^{(n)}(\mathbf{W}_k) - \nabla F^{(n)}(\mathbf{W}_k) \right\|^2 \notag\\
\leq& \sum_{g=1}^{\mathcal{G}} \frac{|C_g|}{c_{g,k}^{*}}
\mathbb{E} \left\| \left[\nabla F_g(\mathbf{W}_k) - \nabla F(\mathbf{W}_k)\right]\odot m_{g,k} \right\|^2 \notag\\
\leq& \sum_{g=1}^{\mathcal{G}} \frac{|C_g|}{c_{g,k}^{*}} f_3^2(\rho_{g,k}) \zeta^2,
\end{align}
where the last step uses Eq.~\ref{eq:group_bias}.

\subsubsection{Step 3: From auxiliary lemmas to the one-round bound}
\label{sec:convergence_threome_1}

We now substitute the previous lemmas into the master inequality
Eq.~\ref{eq:final_initial_term} and recover the one-round descent statement in
Theorem~\ref{thm:ma_one_round}.

\textbf{An intermediate form before shorthand notation.}
Substituting the drift and variance bounds into
Eq.~\ref{eq:final_initial_term}, while keeping the coordinate-wise aggregation
term unexpanded, first gives
\begin{align}
&\mathbb{E}\big[F(W_{k+1})\big]-\mathbb{E}\big[F(W_k)\big]  \notag\\
\le{}\;&
\underbrace{- \frac{\gamma \mathcal{T}}{2} \, \mathbb{E}\|\nabla F(W_k)\|^{2}}_{\text{from Eq.~\ref{eq:final_initial_term}}}
\notag \\
&+
\underbrace{\frac{3L\mathcal{T}^2\gamma^2}{2} \, \mathbb{E}\sum_{n=1}^{N}\Bigl\|\tfrac{1}{|C_k^{(n)}|}\sum_{j\in C_k^{(n)}}\nabla F_j^{(n)}(W_k)\Bigr\|^{2}}_{\text{kept unexpanded; IID/Non-IID split later}}
\notag \\
&+
\underbrace{\frac{L^2\gamma^3\mathcal{T}^3(3L\mathcal{T}\gamma+1)}{6}
\sum_{g=1}^{\mathcal{G}} \frac{|C_g|}{c_{g,k}^{*}} f_1^2(\rho_{g,k}) G^2}_{\text{from Lemma~\ref{lem:3}}}
\notag \\
&+
\underbrace{\frac{3L\mathcal{T}\gamma^2}{2}
\sum_{g=1}^{\mathcal{G}} \frac{|C_g|}{(c_{g,k}^{*})^{2}} f_2^2(\rho_{g,k}) \sigma^2}_{\text{from Lemma~\ref{lem:4}}}.
\label{eq:appendix_one_round_precompact}
\end{align}
Eq.~\ref{eq:appendix_one_round_precompact} is the direct intermediate result
obtained from Eq.~\ref{eq:final_initial_term} after substituting the auxiliary
lemmas.

\textbf{Introducing shorthand notation.}
Using \(G_0=\mathcal{T}\gamma\), \(J_0=L\mathcal{T}\gamma\),
\(I_0=L\gamma\), and \(H_0=L^2\mathcal{T}^2\gamma^2(3L\mathcal{T}\gamma+1)\),
Eq.~\ref{eq:appendix_one_round_precompact} can be rewritten in the compact form
used in the main text. This compact form is exactly
Theorem~\ref{thm:ma_one_round}:
\begin{align}
&\mathbb{E}\big[F(W_{k+1})\big]-\mathbb{E}\big[F(W_k)\big]  \notag\\
\le &
- \frac{G_0}{2} \, \mathbb{E}\|\nabla F(W_k)\|^{2} \notag \\
&+ \frac{3G_0J_0}{2} \, \mathbb{E}\sum_{n=1}^{N}\Bigl\|\tfrac{1}{|C_k^{(n)}|}\sum_{j\in C_k^{(n)}}\nabla F_j^{(n)}(W_k)\Bigr\|^{2} \notag \\
&+ \frac{G_0 H_0}{6} \sum_{g=1}^{\mathcal{G}} \frac{|C_g|}{c_{g,k}^{*}} f_1^2(\rho_{g,k}) G^2
+ \frac{3 G_0 I_0}{2} \sum_{g=1}^{\mathcal{G}} \frac{|C_g|}{(c_{g,k}^{*})^{2}} f_2^2(\rho_{g,k}) \sigma^2.
	\label{eq:appendix_compact_one_round}
\end{align}
Eq.~\ref{eq:appendix_compact_one_round} is the appendix restatement of
Theorem~\ref{thm:ma_one_round}. The next step only expands its second term
under IID and Non-IID data.

We next rewrite the third term on the right-hand side of Eq.~\ref{eq:L2_term_1}
in terms of \(\nabla F(\mathbf{W}_k)\). If the data are IID, then for all
$n \in \mathcal{N}_k^{(n)}$, we have
\(\nabla F_i(\mathbf{W}_k) = \nabla F(\mathbf{W}_k)\). Hence, the third term
in Eq.~\ref{eq:L2_term_1} simplifies as:
\begin{align}
\frac{3L\mathcal{T}^2\gamma^2}{2} \sum_{n=1}^{N}\mathbb{E} \left\|
\frac{1}{|C_k^{(n)}|} \sum_{i \in C_k^{(n)}}  
\nabla F_i^{(n)}(\mathbf{W}_k)
\right\|^2
&= \frac{3L \mathcal{T}^2 \gamma^2}{2} \mathbb{E} \left\| \nabla F(\mathbf{W}_k) \right\|^2 
\label{eq:IID_client_gradient}
\end{align}
Eq.~\ref{eq:IID_client_gradient} is exactly the IID reduction of the second term
in Eq.~\ref{eq:appendix_compact_one_round}. Substituting it back into
Eq.~\ref{eq:appendix_compact_one_round} gives an IID specialization of
Theorem~\ref{thm:ma_one_round}, written explicitly in Eq.~\ref{eq:iid_singlebound} below.

If the data are non-IID, by inserting and subtracting the reference point $\nabla F^{(n)}(\mathbf{W}_k)$ and using bias decomposition:
\begin{align}
&\frac{3L\mathcal{T}^2\gamma^2}{2} \sum_{n=1}^{N}\mathbb{E} \left\|
\frac{1}{|C_k^{(n)}|} \sum_{i \in C_k^{(n)}}  
\nabla F_i^{(n)}(\mathbf{W}_k)
\right\|^2. \notag \\
=& \frac{3L\mathcal{T}^2\gamma^2}{2} \sum_{n=1}^{N} \mathbb{E} \left\| \nabla F^{(n)}(\mathbf{W}_k) 
+\sum_{i \in C_k^{(n)}} \frac{1}{|C_k^{(n)}|} 
\left( \nabla F_i^{(n)}(\mathbf{W}_k) - \nabla F^{(n)}(\mathbf{W}_k) \right)
\right\|^2 \notag \\
\le& 3L \mathcal{T}^2 \gamma^2 \sum_{n=1}^{N} \mathbb{E} \left\| \nabla F^{(n)}(\mathbf{W}_k) \right\|^2
+ 3L \mathcal{T}^2 \gamma^2 \sum_{n=1}^{N} \mathbb{E} \left\| 
\sum_{i \in C_k^{(n)}} \frac{1}{|C_k^{(n)}|} 
\left( \nabla F_i^{(n)}(\mathbf{W}_k) - \nabla F^{(n)}(\mathbf{W}_k) \right)
\right\|^2 \notag \\
\le& 3L \mathcal{T}^2 \gamma^2 \cdot \mathbb{E} \left\| \nabla F(\mathbf{W}_k) \right\|^2 
+ 3L \mathcal{T}^2 \gamma^2 \cdot \sum_{g=1}^{\mathcal{G}} \frac{|C_g|}{c_{g,k}^*} f_3^2(\rho_{g,k}) \zeta^2. 
\label{eq:nonIID_client_gradient}
\end{align}
Eq.~\ref{eq:nonIID_client_gradient} is the corresponding non-IID expansion of
the same term. Substituting it into Eq.~\ref{eq:appendix_compact_one_round}, and
then combining the result with Lemma~\ref{lem:5}, gives a non-IID
specialization of Theorem~\ref{thm:ma_one_round}, written explicitly in
Eq.~\ref{eq:noniid_singlebound} below.

Therefore, under IID data, combining Eq.~\ref{eq:IID_client_gradient} and Eq.~\ref{eq:final_initial_term} with Lemmas~\ref{lem:2}–\ref{lem:4}, we obtain the explicit IID specialization:
\begin{align}
&\mathbb{E}[F(\mathbf{w}_{k+1})] - \mathbb{E}[F(\mathbf{W}_k)] \notag\\
\le& \underbrace{-\frac{\gamma \mathcal{T}}{2} \cdot \mathbb{E} \| \nabla F(\mathbf{W}_k) \|^2}_{\text{descent term}} 
+ \underbrace{ \frac{L^2 \gamma^3 \mathcal{T}^3 }{6}     \sum_{g=1}^\mathcal{G}\frac{|C_g|}{c_{g,k}^*} f^2_1(\rho_{g,k}) G^2 }_{\text{inner product drift}} + \underbrace{ \frac{3L \mathcal{T} \gamma^2}{2} \sum_{g=1}^\mathcal{G} \frac{|C_g|}{(c_{g,k}^*)^2} f_2^2(\rho_{g,k}) \sigma^2 }_{\text{variance term}} 
\notag \\
&+ \underbrace{ \frac{L^3 \mathcal{T}^4 \gamma^4}{2} \sum_{g=1}^\mathcal{G} \frac{|C_g|}{c_{g,k}^*} f_1^2(\rho_{g,k}) G^2 }_{\text{model drift}} + \underbrace{ \frac{3L \mathcal{T}^2 \gamma^2}{2} \cdot \mathbb{E} \| \nabla F(\mathbf{W}_k) \|^2 }_{\text{gradient noise term}}.\notag \\
 =& \left( \frac{\gamma \mathcal{T} (3L \mathcal{T} \gamma - 1)}{2} \right) \cdot \mathbb{E} \| \nabla F(\mathbf{W}_k) \|^2 + \frac{L^2 \gamma^3 \mathcal{T}^3 (3L\mathcal{T}\gamma +1)}{6}
\cdot \sum_{g=1}^\mathcal{G} \frac{|C_g|}{c_{g,k}^*} f_1^2(\rho_{g,k}) G^2 \notag \\
&+ \frac{3L \mathcal{T} \gamma^2}{2} \sum_{g=1}^\mathcal{G} \frac{|C_g|}{(c_{g,k}^*)^2} f_2^2(\rho_{g,k}) \sigma^2.
\label{eq:iid_singlebound}
\end{align}

If the data are non-IID, 
combining Eq.~\ref{eq:nonIID_client_gradient} and Eq.~\ref{eq:final_initial_term} with Lemmas~\ref{lem:2}–\ref{lem:5}, we obtain the explicit non-IID specialization:
\begin{align}
&\mathbb{E}[F(\mathbf{w}_{k+1})] - \mathbb{E}[F(\mathbf{W}_k)] \notag  \\
\le&\underbrace{-\frac{\gamma \mathcal{T}}{2} \cdot \mathbb{E} \| \nabla F(\mathbf{W}_k) \|^2}_{\text{descent term}} + \underbrace{ \frac{L^2 \gamma^3 \mathcal{T}^3 }{6}     \sum_{g=1}^\mathcal{G}\frac{|C_g|}{c_{g,k}^*} f^2_1(\rho_{g,k}) G^2 }_{\text{inner product drift}} 
+ \underbrace{ \frac{3L\mathcal{T}\gamma^2}{2} \sum_{g=1}^\mathcal{G} \frac{|C_g|}{(c_{g,k}^*)^2} f_2^2(\rho_{g,k}) \sigma^2 }_{\text{variance}} \notag \\ 
&+ \underbrace{ \frac{L^3\mathcal{T}^4\gamma^4}{2} \sum_{g=1}^\mathcal{G} \frac{|C_g|}{c_{g,k}^*} f_1^2(\rho_{g,k}) G^2 }_{\text{model drift}} + \underbrace{ 3L \mathcal{T}^2 \gamma^2 \mathbb{E} \| \nabla F(\mathbf{W}_k) \|^2 }_{\text{gradient noise}} + \underbrace{ 3L \mathcal{T}^2 \gamma^2 \sum_{g=1}^\mathcal{G} \frac{|C_g|}{c_{g,k}^*} f_3^2(\rho_{g,k}) \zeta^2 }_{\text{bias term}}. \notag  \\
 =& \left( \frac{\gamma \mathcal{T} (6L \mathcal{T} \gamma - 1)}{2} \right) \cdot \mathbb{E} \| \nabla F(\mathbf{W}_k) \|^2 + \frac{L^2 \gamma^3 \mathcal{T}^3 (3L\mathcal{T}\gamma +1)}{6}
\cdot \sum_{g=1}^\mathcal{G} \frac{|C_g|}{c_{g,k}^*} f_1^2(\rho_{g,k}) G^2 \notag \\
&+ \frac{3L \mathcal{T} \gamma^2}{2} \sum_{g=1}^\mathcal{G} \frac{|C_g|}{(c_{g,k}^*)^2} f_2^2(\rho_{g,k}) \sigma^2 + 3L \mathcal{T}^2 \gamma^2 \sum_{g=1}^\mathcal{G} \frac{|C_g|}{c_{g,k}^*} f_3^2(\rho_{g,k}) \zeta^2.
\label{eq:noniid_singlebound}
\end{align}

\subsubsection{Step 4: Telescoping over rounds to prove Theorem~\ref{thm:ma_telescope}}
\label{sec:convergence_threome_2}
\textbf{Telescoping sum.} 
We now apply the upper bound for 
$\mathbb{E}\left\langle \nabla F(\mathbf{W}_k), \mathbf{W}_{k+1} - \mathbf{W}_k \right\rangle$ 
and the bound for $\frac{L}{2} \mathbb{E} \| \mathbf{W}_{k+1} - \mathbf{W}_k \|^2$, 
and plug them into the smoothness inequality. 
Summing both sides of the inequality over $k = 1, \ldots, K$ yields:
\begin{align}
\mathbb{E}[F(\mathbf{W}_{K+1})] - \mathbb{E}[F(\mathbf{W}_1)] 
&= \sum_{k=1}^K \left( \mathbb{E}[F(\mathbf{W}_{k+1})] - \mathbb{E}[F(\mathbf{W}_k)] \right)
\end{align}

Based on Eq.~\ref{eq:iid_singlebound}, substituting the detailed expressions of each term under the IID setting, we obtain:
\begin{align}
&\mathbb{E}[F(\mathbf{W}_{k+1})] - \mathbb{E}[F(\mathbf{W}_1)] \notag \\
\le 
&\left( \frac{\gamma \mathcal{T} (3L \mathcal{T} \gamma - 1)}{2} \right)
\sum_{k=1}^K \mathbb{E} \| \nabla F(\mathbf{W}_k) \|^2 + \sum_{k=1}^K
\left[
\frac{L^2 \gamma^3 \mathcal{T}^3 (3L\mathcal{T}\gamma +1)}{6}
\sum_{g=1}^\mathcal{G} \frac{|C_g|}{c_{g,k}^*} f_1^2(\rho_{g,k}) G^2
\right] \notag \\
&+ \sum_{k=1}^K 
\left[
\frac{3L \mathcal{T} \gamma^2}{2} 
\sum_{g=1}^\mathcal{G} \frac{|C_g|}{(c_{g,k}^*)^2} f_2^2(\rho_{g,k}) \sigma^2
\right].
\end{align}
We choose learning rate $\gamma \leq \frac{1}{6L\mathcal{T}}$ (which satisfies $\gamma \leq \frac{1}{3L\mathcal{T}}$) to ensure descent.
With this choice, the descent coefficient becomes negative, and we can move the descent term to the left-hand side:

\begin{align}
\frac{\mathcal{T} \gamma}{4} \sum_{k=1}^K \mathbb{E} \| \nabla F(\mathbf{W}_k) \|^2 
\le 
& \mathbb{E}[F(\mathbf{W}_1)] - \mathbb{E}[F(\mathbf{W}_{k+1})]  \notag \\
&+ \sum_{k=1}^K 
\left[
\frac{L^2 \gamma^3 \mathcal{T}^3 (3L\mathcal{T}\gamma +1)}{6}
\sum_{g=1}^\mathcal{G} \frac{|C_g|}{c_{g,k}^*} f_1^2(\rho_{g,k}) G^2
\right] \notag \\
&+ \sum_{k=1}^K 
\left[
\frac{3L \mathcal{T} \gamma^2}{2} 
\sum_{g=1}^\mathcal{G} \frac{|C_g|}{(c_{g,k}^*)^2} f_2^2(\rho_{g,k}) \sigma^2
\right].
\end{align}

Using the fact that $\mathbb{E}[F(\mathbf{W}_{k+1})]$ is non-negative and dividing both sides above by $\frac{\mathcal{T} \gamma K}{4}$, we have:

\begin{align}
\frac{1}{K} \sum_{k=1}^K \mathbb{E} \| \nabla F(\mathbf{W}_k) \|^2 
\le 
& \frac{4}{\mathcal{T} \gamma K}\mathbb{E}[F(\mathbf{W}_1)]  \notag \\
&+ \frac{3L^2 \gamma^2 \mathcal{T}^2 (3L\mathcal{T}\gamma +1)}{2K}\sum_{k=1}^K 
\sum_{g=1}^\mathcal{G} \frac{|C_g|}{c_{g,k}^*} f_1^2(\rho_{g,k}) G^2
 \notag \\
&+ \frac{6L \gamma}{K}\sum_{k=1}^K 
\sum_{g=1}^\mathcal{G} \frac{|C_g|}{(c_{g,k}^*)^2} f_2^2(\rho_{g,k}) \sigma^2
\label{eq:final_convergence_IID}
\end{align}

For non-IID data distribution, based on Eq.~\ref{eq:noniid_singlebound}, similar to the process of the IID setting, by choosing learning rate
 $\gamma \leq \frac{1}{12L\mathcal{T}}$ (which satisfies $\gamma \leq \frac{1}{6L\mathcal{T}}$) to ensure descent and using the fact that $\mathbb{E}[F(\mathbf{W}_{K+1})]$ is non-negative, we have:

 \begin{align}
\frac{1}{K} \sum_{k=1}^K \mathbb{E} \| \nabla F(\mathbf{W}_k) \|^2 
\le 
& \frac{4}{\mathcal{T} \gamma K}\mathbb{E}[F(\mathbf{W}_1)]  \notag \\
&+ \frac{3L^2 \gamma^2 \mathcal{T}^2 (3L\mathcal{T}\gamma +1)}{2K}\sum_{k=1}^K 
\sum_{g=1}^\mathcal{G} \frac{|C_g|}{c_{g,k}^*} f_1^2(\rho_{g,k}) G^2
 \notag \\
&+ \frac{6L \gamma}{K}\sum_{k=1}^K 
\sum_{g=1}^\mathcal{G} \frac{|C_g|}{(c_{g,k}^*)^2} f_2^2(\rho_{g,k}) \sigma^2
\notag \\
&+ \frac{12L \mathcal{T} \gamma}{K} \sum_{k=1}^K 
\sum_{g=1}^\mathcal{G} \frac{|C_g|}{c_{g,k}^*} f_3^2(\rho_{g,k}) \zeta^2.
\label{eq:final_convergence_non_IID}
\end{align}

\subsection{Convergence analysis based on gradient average aggregation}
\label{sec:convergnce_gradient_average}
The previous subsection completes the MA proof. We next analyze GA under the
same proof template to clarify how the aggregation rule changes the effective
descent direction and error terms.

Similar to the process of MA:
\[
\mathbb{E}[F(\mathbf{W}_{k+1})] - \mathbb{E}[F(\mathbf{W}_k)] \le \mathbb{E}\langle \nabla F(\mathbf{W}_k), \mathbf{w}_{k+1} - \mathbf{w}_k \rangle + \frac{L}{2} \mathbb{E} \|\mathbf{W}_{k+1} - \mathbf{W}_k\|^2.
\]
But for gradient average(GA), we have:
\[
\mathbf{W}_k - \mathbf{W}_{k+1} = \frac{1}{C} \sum_{i=1}^C \sum_{\tau=1}^\mathcal{T} \gamma \nabla F_i(\mathbf{w}_{i,k,\tau-1}, \xi_{i,\tau-1})\odot m_{i,k}.
\]
Then, first analyze the inner product, we have:

\begin{align}
&\mathbb{E} \left\langle \nabla F(\mathbf{W}_k), \mathbf{W}_{k+1} - \mathbf{W}_k \right\rangle \notag \\
=& \mathbb{E} \left\langle \nabla F(\mathbf{W}_k), 
- \frac{1}{C} \sum_{i=1}^C \sum_{\tau=1}^\mathcal{T} \gamma \nabla F_i(\mathbf{w}_{i,k,\tau-1}, \xi_{i,\tau-1}) \odot m_{i,k}
\right\rangle \notag \\
=&  \mathbb{E} \left\langle \nabla F(\mathbf{W}_k), 
-\frac{1}{C} \sum_{i=1}^C \sum_{\tau=1}^\mathcal{T} \gamma \, 
\mathbb{E} \left[ \nabla F_i(\mathbf{w}_{i,k,\tau-1}, \xi_{i,\tau-1}) \mid \mathbf{w}_k \right]\odot m_{i,k}
\right\rangle \notag \\
=& \mathbb{E} \left\langle \nabla F(\mathbf{W}_k), 
- \frac{1}{C} \sum_{i=1}^C \sum_{\tau=1}^\mathcal{T} \gamma \nabla F_i(\mathbf{w}_{i,k,\tau-1})\odot m_{i,k}
\right\rangle \notag \\
=& - \mathbb{E} \left\langle \nabla (\mathbf{W}_k), 
\frac{1}{C}  \sum_{i=1}^C \sum_{\tau=1}^\mathcal{T} \gamma 
\left( \left(\nabla F_i(\mathbf{w}_{i,k,\tau-1}) - \nabla F(\mathbf{W}_k) + \nabla F(\mathbf{W}_k) \right)  \odot m_{i,k}  \right)
\right\rangle \notag \\
=& - \mathbb{E} \left\langle \nabla F(\mathbf{W}_k), \gamma \mathcal{T} \frac{1}{C}  \sum_{i=1}^C \nabla F(\mathbf{W}_k)\odot m_{i,k} \right\rangle \notag \\
&- \mathbb{E} \left\langle \nabla F(\mathbf{W}_k), 
\frac{1}{C}  \sum_{i=1}^C \sum_{\tau=1}^\mathcal{T} \gamma 
\left( \nabla F_i(\mathbf{w}_{i,k,\tau-1}) - \nabla F(\mathbf{W}_k) \right)\odot m_{i,k} \right\rangle  \notag \\
\leq&  -\gamma \mathcal{T} \mathbb{E} \left\| \frac{1}{C}  \sum_{i=1}^C \nabla F(\mathbf{W}_k) \odot m_{i,k} \right\|^2 + \frac{\gamma \mathcal{T}}{2} \mathbb{E} \left\| \frac{1}{C}  \sum_{i=1}^C \nabla F(\mathbf{W}_k) \odot m_{i,k}\right\|^2 \notag \\ 
&+  \frac{\gamma \mathcal{T}}{2} 
\mathbb{E} \left\| \frac{1}{C} 
\sum_{i=1}^C \sum_{\tau=1}^{\mathcal{T}} 
\left( \nabla F_i(\mathbf{w}_{i,k,\tau-1})- \nabla F(\mathbf{W}_k) \right)\odot m_{i,k} 
\right\|^2 \notag \\
\leq&  -\frac{\gamma \mathcal{T}}{2} \mathbb{E} \left\| \frac{1}{C}  \sum_{i=1}^C \nabla F(\mathbf{W}_k) \odot m_{i,k}\right\|^2 \notag \\
&+  \frac{\gamma \mathcal{T}}{2} 
\mathbb{E} \left\| \frac{1}{C} 
\sum_{i=1}^C \sum_{\tau=1}^{\mathcal{T}} 
\left( \nabla F_i(\mathbf{w}_{i,k,\tau-1})- \nabla F(\mathbf{W}_k) \right)\odot m_{i,k} 
\right\|^2. \label{eq:avginner_product} 
\end{align}

Upperbound for $\frac{L}{2} \mathbb{E} \|\mathbf{w}_{k+1} - \mathbf{w}_k\|^2$
\begin{align}
\frac{L}{2} \mathbb{E} \|\mathbf{w}_{k+1} - \mathbf{w}_k\|^2 
&= \frac{L}{2} \mathbb{E} \left\| 
\frac{1}{C}\frac{1}{\mathcal{T}} \sum_{i=1}^C \sum_{\tau=1}^\mathcal{T} \gamma \nabla F_i(\mathbf{w}_{i,k,\tau-1}, \xi_{i,\tau-1})
\right\|^2 \notag \\
&\le \frac{3L\mathcal{T}^2\gamma^2}{2} \mathbb{E} \left\|
\frac{1}{C}\frac{1}{\mathcal{T}} \sum_{i=1}^C \sum_{\tau=1}^\mathcal{T} 
\left[ \nabla F_i(\mathbf{w}_{i,k,\tau-1}, \xi_{i,\tau-1}) - \nabla F_i(\mathbf{w}_{i,k,\tau-1}) \right]\odot m_{i,k}
\right\|^2 \notag \\
&+ \frac{3L\mathcal{T}^2\gamma^2}{2} \mathbb{E} \left\|
\frac{1}{C}\frac{1}{\mathcal{T}} \sum_{i=1}^C \sum_{\tau=1}^\mathcal{T} 
\left[ \nabla F_i(\mathbf{w}_{i,k,\tau-1}) - \nabla F_i(\mathbf{W}_k) \right]\odot m_{i,k}
\right\|^2 \notag \\
&+ \frac{3L\mathcal{T}^2\gamma^2}{2} \mathbb{E} \left\|
\frac{1}{C}\sum_{}^C
\nabla F_i(\mathbf{W}_k)\odot m_{i,k}
\right\|^2. 
\label{eq:avgL2_term_1}
\end{align}

Then, combining the two upperbounds, we have:
\begin{align}
\mathbb{E}[F(\mathbf{w}_{k+1})] - \mathbb{E}[F(\mathbf{W}_k)] &\le\mathbb{E} \left\langle \nabla F(\mathbf{W}_k), \mathbf{w}_{k+1} - \mathbf{w}_k \right\rangle  
+\frac{L}{2} \mathbb{E} \|\mathbf{W}_{k+1} - \mathbf{W}_k\|^2 \notag \\
\le& - \frac{\gamma \mathcal{T}}{2} \mathbb{E} \| \frac{1}{C}\sum_{i=1}^\mathcal{C} \nabla F(W_k) \odot m_{i,k} \|^2 \notag \\ 
& +  \frac{\gamma \mathcal{T}}{2} 
\mathbb{E} \left\| \frac{1}{\mathcal{T} C} 
\sum_{}^C \sum_{\tau=1}^{\mathcal{T}} 
\left( \nabla F_i(\mathbf{w}_{i,k,\tau-1}) - \nabla F(\mathbf{W}_k) \right)\odot m_{i,k}
\right\|^2 \notag \\
&+ \frac{3L\mathcal{T}^2\gamma^2}{2} \mathbb{E} \left\|
\frac{1}{C}\frac{1}{\mathcal{T}} \sum_{n=1}^C \sum_{\tau=1}^\mathcal{T} 
\left[ \nabla F_i(\mathbf{w}_{i,k,\tau-1}, \xi_{i,\tau-1}) - \nabla F_i(\mathbf{w}_{i,k,\tau-1}) \right]\odot m_{i,k}
\right\|^2 \notag \\
&+ \frac{3L\mathcal{T}^2\gamma^2}{2} \mathbb{E} \left\|
\frac{1}{C}\frac{1}{\mathcal{T}} \sum_{n=1}^C \sum_{\tau=1}^\mathcal{T} 
\left[ \nabla F_i(\mathbf{w}_{i,k,\tau-1}) - \nabla F_i(\mathbf{W}_k) \right]\odot m_{i,k}
\right\|^2 \notag \\
&+ \frac{3L\mathcal{T}^2\gamma^2}{2} \mathbb{E} \left\|
\frac{1}{C}\sum_{n=1}^C
\nabla F_i(\mathbf{W}_k)\odot m_{i,k}
\right\|^2.   
\label{eq:avg_final_2}
\end{align}

Then, similar to the MA process, by choosing local learning rate
\(\gamma<\frac{1}{6L\mathcal{T}}\) for IID data and
\(\gamma<\frac{1}{12L\mathcal{T}}\) for Non-IID data, we have:

\begin{align}
\frac{1}{K} \sum_{k=1}^K \mathbb{E} \| \frac{1}{C}\sum_{i=1}^\mathcal{C} \nabla F(W_k) \odot m_{i,k} \|^2 
\le 
& \frac{4\mathbb{E}[F(W_1)]}{\mathcal{T} \gamma K}
+ \underbrace{
\frac{3L^2 \gamma^2 \mathcal{T}^2 (3L\mathcal{T}\gamma + 1)}{2KC} 
\sum_{k=1}^K \sum_{i=1}^C 
f_1^2(\rho_{i,k}) G^2
}_{\text{Model Drift Term}} \notag\\
&+ \underbrace{
\frac{6L \mathcal{T} \gamma^2}{KC^2} 
\sum_{k=1}^K \sum_{i=1}^C 
f_2^2(\rho_{i,k}) \sigma^2
}_{\text{Variance Term}}+ \underbrace{
\frac{12L \mathcal{T}^2 \gamma^2}{KC} 
\sum_{k=1}^K \sum_{i=1}^C 
f_3^2(\rho_{i,k}) \zeta^2
}_{\text{Bias Term (Non-IID only)}}.
\label{eq:fedAvg_convergence}
\end{align}

\subsection{Comparison and Final Analysis of the Convergence of Two Aggregation Rules}
\label{sec:convergence_final_analysis}
With the MA bound established in
Eqs.~\ref{eq:final_convergence_IID}--\ref{eq:final_convergence_non_IID} and
the GA bound in Eq.~\ref{eq:fedAvg_convergence}, we can now compare the two
aggregation rules and explain why MA is better aligned with GMR.

\textbf{Interpretation.}
In GA, the server aggregates sparse gradients as 
\(\tfrac{1}{C} \sum_{i=1}^C \nabla F_i(W_k) \odot m_{i,k}\), 
where both the gradients and the associated error terms (model drift and variance) are reduced in magnitude because they are sparsified by the masks, with the scaling governed by \(f(\rho_{i,k})\). 
By contrast, MA simulates global gradients by aggregating partial client gradients, 
where each retained coordinate is effectively \emph{amplified} by a factor 
\(\tfrac{|\mathcal{C}_g|}{c_{g,k}^*}\). 
This factor arises because MA normalizes only over covering clients, 
rather than over all $C$ clients as GA does. 
Since $c_{g,k}^* \leq C$, we obtain
\(\sum_{g=1}^G \tfrac{|\mathcal{C}_g|}{c_{g,k}^*} f^2(\rho_{g,k}) 
\;\geq\; \sum_{g=1}^G \tfrac{|\mathcal{C}_g|}{C} f^2(\rho_{g,k}) 
= \tfrac{1}{C} \sum_{i=1}^C f^2(\rho_{i,k}).\)
Thus, MA trades a more faithful coordinate reconstruction for larger
coverage-weighted error terms. This comparison is exactly the quantity-level
explanation behind Theorem~\ref{thm:ma_one_round}.

Combining Eq.~\ref{eq:fedAvg_convergence} with the MA analysis above also
clarifies the interpretation of Theorem~\ref{thm:ma_telescope}: both GA and MA converge to a
stationary neighborhood determined by the average density and coverage pattern,
but MA preserves coordinate-wise normalization and therefore interacts with
restoration more predictably.

\textbf{Stationary neighborhood view.}
Both GA and MA converge to a small neighborhood of a stationary point of standard FL. The distance is better captured by the average sub-model density across clients and time:
\(\frac{1}{K}\sum_{k=1}^K \sum_{g=1}^{\mathcal{G}} [\frac{|C_g|}{c_{g,k}^*} f_1^2(\rho_{g,k}) + \frac{|C_g|}{(c_{g,k}^*)^2} f_2^2(\rho_{g,k})],\)
and higher densities (e.g., under GMR) increase coverage, decrease these functionals, and tighten the neighborhood.
Although GA and MA operate over the same model search space in theory, their aggregation strategies differ fundamentally. GA aggregates gradients on all regions, effectively computing:
\begin{equation}
    \nabla F(W_k) \approx \frac{1}{C} \sum_{i=1}^C \nabla F_i(W_k) \odot m_{i,k},
\label{eq:GA_client_gradient}
\end{equation}

where each mask \(m_{i,k}\) restricts the gradient to the pruned sub-model. This leads to a \textit{shrunk and overly averaged} update direction, which does not reflect the true full-model gradient.

By contrast, MA reconstructs the full gradient more faithfully through partial participation and per-parameter aggregation:
\begin{equation}
\nabla F(W_k^{(n)}) \approx \frac{1}{|C_k^{(n)}|} \sum_{i \in C_k^{(n)}} \nabla F_i^{(n)}(W_k),
\label{eq:MA_client_gradient}
\end{equation}
which preserves more diverse update signals across the parameter space.

As a result, gradient FedAvg may converge more slowly or to inferior optima compared to MA, especially in heterogeneous settings where model overlaps are limited and sub-models vary significantly.

\subsection{Connection with the GMR}
\paragraph{Optimization subspace view.}
Building on Eq.~\ref{eq:GA_client_gradient} and Eq.~\ref{eq:MA_client_gradient}, we derive the client-specific gradient expressions. This allows us to formally relate the attainable solutions of full-model training and model-heterogeneous training to their respective effective search subspaces.
Let
\[
\mathcal{S}_{\text{mask}}
=\mathcal{S}_{\text{gradient}}
:= \mathrm{span}\Bigl\{\;\nabla F_i(W_k)\odot m_{i,k}\;:\; i=1,\ldots,C\Bigr\},
\qquad
\mathcal{S}_{\text{full}}:=\mathbb{R}^d .
\]
Thus both submodel-based strategies (GA and MA) update within the same restricted space,
\(\mathcal{S}_{\text{gradient}}=\mathcal{S}_{\text{mask}}\subseteq \mathbb{R}^d\),
whereas full-model FL searches in \(\mathbb{R}^d\).
By standard convex-optimization arguments, optimizing over a strict subspace yields a weakly worse optimum than optimizing over the full space; hence
\[
F(W^*_{\text{full}})\ \le\ F(W^*_{\text{gradient}}),
\qquad
F(W^*_{\text{full}})\ \le\ F(W^*_{\text{mask}}),
\]
with strict inequality whenever the full-model minimizer does not lie in the affine set reachable by masked updates.
Under GMR’s nested masks, the span \(\mathcal{S}_{\text{mask}}\) expands over rounds, enlarging the feasible search region.
Therefore, when submodels plateau, restoring capacity increases \(\mathcal{S}_{\text{mask}}\), enabling further descent and tighter proximity to the full-model optimum.

\paragraph{Effect of Different Aggregation Methods on GMR.}
\label{different_aggregation_method_effect}
When \emph{model restoration} (density increase) occurs, GA/FA lack \emph{per-coordinate scale normalization}, so update magnitudes jump up or down and can dilute or offset GMR’s gains.

\textbf{FedAvg (FA).} (zeros for missing neurons) The $n$-th coordinate at round $k$ is
\[
W_{k+1}^{(n)}=\frac{1}{C}\sum_{i=1}^C w_{i,k}^{(n)},\qquad
w_{i,k}^{(n)}=0\ \text{if pruned.}
\]
Before restoration, only $|C_k^{(n)}|$ clients are active on coordinate $n$, so averaging over zeros effectively \emph{shrinks} the corresponding weight. Once restoration occurs, $|C_k^{(n)}|$ increases and the effective averaging scale jumps from $|C_k^{(n)}|/C$ to $|C_{k+1}^{(n)}|/C$, leading to step-size discontinuities and instability. More critically, some neurons may collapse to zero during early aggregation; once removed, they cannot be effectively retrained even after restoration.

Our ablation results corroborate this effect. On FEMNIST with Conv2D, FA remains trainable but shows degraded accuracy, since only a fraction of neurons consistently contribute to similar tasks. On CIFAR-10 with VGG11, FA collapses entirely: the loss falls rapidly to a trivial value within a few rounds and fails to recover, even when neurons are later restored. In contrast, on ImageNet100 with ResNet, the residual connections allow restored neurons to re-integrate, preventing collapse and enabling recovery. These results explain the divergent behavior of FA across datasets and architectures.

\textbf{Gradient-Average (GA).} (Sum only active gradients but still divide by $C$)
\[
\Delta W_k^{(n)}\approx \frac{\gamma}{C}\sum_{i=1}^C g_{i,k}^{(n)}\,\mathbf{1}\{n\in m_{i,k}\}
\quad\Rightarrow\quad
\mathbb{E}\bigl|\Delta W_k^{(n)}\bigr| \propto \gamma\,\frac{|C_k^{(n)}|}{C}.
\]
Compared to FA, GA only rescales the magnitude of gradients without altering neuron weight, which explains why GA achieves comparable or even superior convergence once models are restored, in contrast to MA. However, under GMR, restoration not only changes model capacity but also increases $|C_k^{(n)}|$, thereby \emph{linearly amplifying} the per-coordinate update step. This amplification introduces higher variance and jitter, ultimately diminishing the benefit of larger models.

\textbf{Mask-aware (MA).} MA normalizes per coordinate by $|C_k^{(n)}|$ (average over active clients only), keeping the update scale \emph{invariant} across restorations. It avoids FA’s weight shrinkage and GA’s step amplification, hence combines stably and effectively with GMR.

Importantly, once all clients are fully restored ($\rho_i=1$ for all $i$), every coordinate is active on all clients ($|C_k^{(n)}|=C$). In this regime, FA, GA, and MA all reduce to the standard FedAvg update
\[
W_{k+1}^{(n)} \;=\; \frac{1}{C}\sum_{i=1}^C w_{i,k}^{(n)},
\]
and are therefore equivalent in both formulation and convergence behavior.

\textbf{GA/FA benefit less from restoration}. Though model restoration enlarges the representational capacity and enables better convergence of the global model, 
FA and GA benefit less from restoration because they lack proper normalization on weights or gradients. 
In contrast, MA explicitly normalizes per-coordinate updates, making it more compatible with GMR.

FA aggregates with zero-padding,
\[
W_{k+1}^{(n)}=\tfrac{1}{C}\sum_{i=1}^C w_{i,k}^{(n)}.
\]
When $|C_k^{(n)}|\!\ll C$, zeros dominate and weights shrink; as GMR raises density ($|C_k^{(n)}|\!\uparrow$), shrinkage abates but the effective averaging scale jumps from $|C_k^{(n)}|/C$ to $|C_{k+1}^{(n)}|/C$, inducing abrupt step-size changes. 

GA updates
\[
\Delta W_k^{(n)}=\tfrac{\gamma}{C}\sum_{i=1}^C g_{i,k}^{(n)}\,\mathbf{1}\{n\in m_{i,k}\},
\]
so restoration linearly enlarges the per-coordinate step ($\propto |C_k^{(n)}|/C$), increasing variance. In contrast, MA normalizes by the active coverage and effectively averages over $|C_k^{(n)}|$ rather than $C$, keeping the update scale stable across restorations and preserving GMR’s gains.
\subsection{Asynchronous extension: MHFL-aware perturbation of the synchronous bound}
\label{sec:async_extension}

The main analysis above is intentionally carried out in the synchronous regime.
Here we only provide a preliminary asynchronous extension built on top of this
synchronous reference. The purpose is not to establish a separate full
asynchronous convergence theorem, but to formalize the following effect: within
the same wall-clock interval as one synchronous round, bandwidth-abundant groups
can contribute additional updates, while bandwidth-constrained clients (BCCs)
may contribute relatively older and weaker updates. This creates a tradeoff
between extra update opportunities and the errors induced by participation skew,
relative dilution, and staleness.

We use the synchronous round duration as the reference time window. Let
\(\Delta_k\) denote the duration of the \(k\)-th synchronous round, i.e., the
time determined by the slowest participating group in that round. Because MHFL
uses heterogeneous sub-models, the comparison must be carried out
\emph{coordinate-wise}. For every active coordinate \(n\) and structural group
\(g\in\mathcal{G}_{k}^{(n)}:=\{g:n\in S_{g,k}\}\), define the masked
group-average gradient
\begin{equation}
\bar g_{g,k}^{(n)}(W)
:=
\frac{1}{|C_g|}
\sum_{i\in C_g}\nabla F_i^{(n)}(W),
\qquad
s_{g,k}^{(n)}:=\frac{|C_g|}{|C_k^{(n)}|}.
\label{eq:async_group_grad}
\end{equation}
Thus the synchronous MA reference on coordinate \(n\) is
\begin{equation}
G_{k,\mathrm{sync}}^{(n)}
:=
\sum_{g\in\mathcal{G}_{k}^{(n)}} s_{g,k}^{(n)}\bar g_{g,k}^{(n)}(W_k)
=
\frac{1}{|C_k^{(n)}|}\sum_{j\in C_k^{(n)}}\nabla F_j^{(n)}(W_k).
\label{eq:async_sync_ref}
\end{equation}

Now consider the asynchronous arrivals received during the same interval
\(\Delta_k\). Since this interval is as long as the synchronous round, every
structural group contributes at least one usable update, while faster
bandwidth-abundant groups may contribute multiple updates. Suppose group \(g\)
contributes \(R_{g,k}\ge 1\) usable arrivals, with weights
\(\omega_{g,k}^{(r)}\ge 0\), delays \(\tau_{g,k}^{(r)}\), and total
multiplicity
\begin{equation}
\bar m_{g,k}:=\sum_{r=1}^{R_{g,k}}\omega_{g,k}^{(r)}\ge 1.
\end{equation}
Rather than forcing the multiple stale arrivals to be the gradient at a single
historical model, we define their exact window-averaged direction as
\begin{equation}
\widetilde g_{g,k}^{(n)}
:=
\frac{1}{\bar m_{g,k}}
\sum_{r=1}^{R_{g,k}}\omega_{g,k}^{(r)}
\bar g_{g,k}^{(n,r)}(W_{k-\tau_{g,k}^{(r)}}).
\label{eq:async_effective_group}
\end{equation}
This definition is exact for the aggregated direction over the window and is
sufficient for comparing asynchronous and synchronous directions. The
asynchronous coordinate-wise aggregate is then
\begin{equation}
G_{k,\mathrm{async}}^{(n)}
:=
\sum_{g\in\mathcal{G}_{k}^{(n)}}
s_{g,k}^{(n)}\bar m_{g,k}\,
\widetilde g_{g,k}^{(n)}.
\label{eq:async_ma_update}
\end{equation}
Define the coordinate-wise average participation multiplicity
\begin{equation}
\bar m_k^{(n)}
:=
\sum_{g\in\mathcal{G}_{k}^{(n)}}
s_{g,k}^{(n)}\bar m_{g,k}.
\label{eq:async_mbar}
\end{equation}

\paragraph{Proposition E.1.}
\label{prop:async_ma}
For every active coordinate \(n\), the asynchronous aggregate in
Eq.~\ref{eq:async_ma_update} admits the exact decomposition
\begin{align}
G_{k,\mathrm{async}}^{(n)}
=\;&
G_{k,\mathrm{sync}}^{(n)}
\notag\\
&+
\underbrace{
\bigl(\bar m_k^{(n)}-1\bigr)G_{k,\mathrm{sync}}^{(n)}
}_{\text{extra participation}}
\notag\\
&+
\underbrace{
\sum_{g\in\mathcal{G}_{k}^{(n)}}
s_{g,k}^{(n)}\bigl(\bar m_{g,k}-\bar m_k^{(n)}\bigr)
\Bigl(\bar g_{g,k}^{(n)}(W_k)-G_{k,\mathrm{sync}}^{(n)}\Bigr)
}_{\text{participation skew}}
\notag\\
&+
\underbrace{
\sum_{g\in\mathcal{G}_{k}^{(n)}}
s_{g,k}^{(n)}\bar m_{g,k}
\Bigl(\widetilde g_{g,k}^{(n)}
-\bar g_{g,k}^{(n)}(W_k)\Bigr)
}_{\text{staleness}}.
\label{eq:async_three_way}
\end{align}

\textbf{Proof.}
Starting from Eq.~\ref{eq:async_ma_update}, insert and subtract
\(\bar g_{g,k}^{(n)}(W_k)\) inside each group contribution:
\begin{align}
G_{k,\mathrm{async}}^{(n)}
=\;&
\sum_{g\in\mathcal{G}_{k}^{(n)}}
s_{g,k}^{(n)}\bar m_{g,k}\bar g_{g,k}^{(n)}(W_k)
\notag\\
&+
\sum_{g\in\mathcal{G}_{k}^{(n)}}
s_{g,k}^{(n)}\bar m_{g,k}
\Bigl(\widetilde g_{g,k}^{(n)}
-\bar g_{g,k}^{(n)}(W_k)\Bigr).
\label{eq:async_step_1}
\end{align}
Next, decompose the multiplicity as
\[
\bar m_{g,k}=1+\bigl(\bar m_k^{(n)}-1\bigr)+\bigl(\bar m_{g,k}-\bar m_k^{(n)}\bigr),
\]
and apply it to the first term in Eq.~\ref{eq:async_step_1}:
\begin{align}
&\sum_{g\in\mathcal{G}_{k}^{(n)}}
s_{g,k}^{(n)}\bar m_{g,k}\bar g_{g,k}^{(n)}(W_k)
\notag\\
=\;&
\sum_{g\in\mathcal{G}_{k}^{(n)}}
s_{g,k}^{(n)}\bar g_{g,k}^{(n)}(W_k)
\notag\\
&+
\bigl(\bar m_k^{(n)}-1\bigr)
\sum_{g\in\mathcal{G}_{k}^{(n)}}
s_{g,k}^{(n)}\bar g_{g,k}^{(n)}(W_k)
\notag\\
&+
\sum_{g\in\mathcal{G}_{k}^{(n)}}
s_{g,k}^{(n)}\bigl(\bar m_{g,k}-\bar m_k^{(n)}\bigr)
\bar g_{g,k}^{(n)}(W_k).
\label{eq:async_step_2}
\end{align}
Because
\[
\sum_{g\in\mathcal{G}_{k}^{(n)}}
s_{g,k}^{(n)}\bigl(\bar m_{g,k}-\bar m_k^{(n)}\bigr)=0,
\]
the last term in Eq.~\ref{eq:async_step_2} can be centered around the
synchronous reference \(G_{k,\mathrm{sync}}^{(n)}\), which yields exactly
Eq.~\ref{eq:async_three_way}. \qed

\paragraph{Connection to the current MHFL scaling.}
Eq.~\ref{eq:async_three_way} shows how the preliminary asynchronous structure
perturbs the synchronous MA direction within the same time window.

\textbf{(i) Extra participation.}
The term \((\bar m_k^{(n)}-1)G_{k,\mathrm{sync}}^{(n)}\) captures the additional
updates contributed by bandwidth-abundant groups within the same wall-clock
interval. When these additional directions remain aligned with the current
synchronous descent direction, they can accelerate optimization. This is the
main benefit of using an asynchronous workflow on top of the synchronous
reference.

\textbf{(ii) Participation skew.}
The centered term
\[
\sum_{g\in\mathcal{G}_{k}^{(n)}}
s_{g,k}^{(n)}\bigl(\bar m_{g,k}-\bar m_k^{(n)}\bigr)
\Bigl(\bar g_{g,k}^{(n)}(W_k)-G_{k,\mathrm{sync}}^{(n)}\Bigr)
\]
has the same structure as the masked Non-IID bias term in
Lemma~\ref{lem:5}, but it is additionally weighted by the imbalance in the
number of arrivals across groups. Defining
\[
V_{m,k}^{(n)}
:=
\sum_{g\in\mathcal{G}_{k}^{(n)}}
s_{g,k}^{(n)}
\bigl(\bar m_{g,k}-\bar m_k^{(n)}\bigr)^2,
\]
weighted Cauchy--Schwarz gives
\[
\left\|
\sum_{g\in\mathcal{G}_{k}^{(n)}}
s_{g,k}^{(n)}
\bigl(\bar m_{g,k}-\bar m_k^{(n)}\bigr)
\Bigl(\bar g_{g,k}^{(n)}(W_k)-G_{k,\mathrm{sync}}^{(n)}\Bigr)
\right\|^2
\le
V_{m,k}^{(n)}
\sum_{g\in\mathcal{G}_{k}^{(n)}}
s_{g,k}^{(n)}
\Bigl\|
\bar g_{g,k}^{(n)}(W_k)-G_{k,\mathrm{sync}}^{(n)}
\Bigr\|^2.
\]
Thus, after regrouping over coordinates, the skew term preserves the same
MHFL coverage structure as the synchronous bias term,
\[
\sum_{g=1}^{\mathcal{G}}
\frac{|C_g|}{c_{g,k}^{*}}\,f_3^2(\rho_{g,k})\,\zeta^2.
\]
However, its magnitude is amplified by the multiplicity imbalance
\(V_{m,k}^{(n)}\). Therefore, more frequent uploads from bandwidth-abundant
groups are not always beneficial; they can bias the effective direction toward
those groups.

\textbf{(iii) Staleness.}
For the staleness term, Lipschitz smoothness gives
\[
\bigl\|
\widetilde g_{g,k}^{(n)}
-\bar g_{g,k}^{(n)}(W_k)
\bigr\|
\le
\frac{1}{\bar m_{g,k}}
\sum_{r=1}^{R_{g,k}}
\omega_{g,k}^{(r)}
L\|W_{k-\tau_{g,k}^{(r)}}-W_k\|.
\]
Using the same bounded local-drift argument as in Lemmas~\ref{lem:2}--\ref{lem:3},
this gives a drift-type penalty with the same MHFL coverage structure as the
synchronous drift term:
\[
\sum_{g=1}^{\mathcal{G}}
\frac{|C_g|}{c_{g,k}^{*}}\,f_1^2(\rho_{g,k})\,G^2
\cdot
\Theta_{g,k}^{\mathrm{stale}},
\qquad
\Theta_{g,k}^{\mathrm{stale}}
:=
\frac{1}{\bar m_{g,k}}
\sum_{r=1}^{R_{g,k}}
\omega_{g,k}^{(r)}\bigl(\tau_{g,k}^{(r)}\bigr)^2 .
\]
This term is more harmful for BCCs because their uploads are more likely to
arrive late within the same window.

\textbf{(iv) Relative dilution of BCC updates.}
If the server normalizes buffered arrivals over the window, the effective
normalized weight of group \(g\) on coordinate \(n\) is
\[
\alpha_{g,k}^{(n)}
:=
\frac{s_{g,k}^{(n)}\bar m_{g,k}}
{\bar m_k^{(n)}}.
\]
For a bandwidth-constrained group \(b\) with only one arrival,
\(\bar m_{b,k}=1\), while bandwidth-abundant groups may satisfy
\(\bar m_{g,k}>1\). If at least one such faster group is active on coordinate
\(n\), then \(\bar m_k^{(n)}>1\) and
\[
\alpha_{b,k}^{(n)}
=
\frac{s_{b,k}^{(n)}}{\bar m_k^{(n)}}
<
s_{b,k}^{(n)}.
\]
Thus, within the same time interval, BCC updates can be relatively diluted by
additional arrivals from faster groups. This is the MHFL-specific reason why
asynchrony does not always improve performance even though it increases the
number of updates.

Consequently, the asynchronous one-window relation can be viewed as the
synchronous bound plus explicit perturbations:
\begin{align}
&\mathbb{E}\big[F(W_{k+1})\big]-\mathbb{E}\big[F(W_k)\big]
\notag\\
\le{}&
\textnormal{RHS of Eq.~\ref{eq:main_descent_bound}}
-\mathcal{G}_{k}^{\mathrm{gain}}
+\mathcal{P}_{k}^{\mathrm{skew}}
+\mathcal{P}_{k}^{\mathrm{stale}},
\label{eq:async_one_round}
\end{align}
where \(\mathcal{G}_{k}^{\mathrm{gain}}\) is induced by the extra aligned
participation term in Eq.~\ref{eq:async_three_way},
\(\mathcal{P}_{k}^{\mathrm{skew}}\) follows the synchronous Non-IID bias
coverage structure with an additional multiplicity-skew factor, and
\(\mathcal{P}_{k}^{\mathrm{stale}}\) follows the synchronous drift coverage
structure with an additional delay-dependent factor.

\paragraph{Remark E.2.}
\label{cor:async_telescope}
This preliminary asynchronous extension should be interpreted as a perturbation
view rather than a complete asynchronous convergence theorem. Summing
Eq.~\ref{eq:async_one_round} over \(k=1,\ldots,K\) yields the corresponding
telescoped perturbation relative to Theorem~\ref{thm:ma_telescope}:
\[
\textnormal{RHS of Theorem~\ref{thm:ma_telescope}}
-\frac{1}{K}\sum_{k=1}^{K}\mathcal{G}_{k}^{\mathrm{gain}}
+\frac{1}{K}\sum_{k=1}^{K}\mathcal{P}_{k}^{\mathrm{skew}}
+\frac{1}{K}\sum_{k=1}^{K}\mathcal{P}_{k}^{\mathrm{stale}}.
\]
Hence, under MHFL, asynchrony brings a three-way tradeoff. In the same
wall-clock time, it can use more updates from bandwidth-abundant groups, which
may accelerate optimization. At the same time, it can make BCC updates stale,
relatively dilute their contribution, and skew the aggregate direction toward
faster groups. The synchronous setting is recovered when
\(\bar m_{g,k}=1\) and \(\tau_{g,k}^{(1)}=0\) for all groups.

\section{Experiments settings}
\label{experiemnt_settings}
The FL simulation consists of one server and ten clients, where heterogeneity is introduced by varying client bandwidth. 
The server's average upload speed is fixed at 20 MB/s, while the download bandwidth is assumed to be sufficiently large. 
The client bandwidth allocation is summarized in Table~\ref{tab:bandwidth}, where each entry denotes the pair of (download/upload) speeds in MB/s.  

\begin{table}[h!] 
\centering 
\caption{Bandwidth allocation for clients. Each cell shows the number of clients assigned to the corresponding bandwidth setting. 
The preset client model density is determined according to the bandwidth for the comparison methods.} 
\label{tab:bandwidth}
\begin{tabular}{c c c c c c}
    \hline
    Bandwidth (Download / Upload, MB/s) & 20/5 & 10/2.5 & 4/1 & 2/0.5 & 1/0.25 \\ 
    Preset model density & 1.0 & 0.5 & 0.20 & 0.10 & 0.05 \\ \hline
    Low    & 2 & 2 & 2 & 2 & 2 \\ 
    Medium & 1 & 1 & 2 & 3 & 3 \\ 
    High   & 1 & 1 & 1 & 1 & 6 \\ \hline
\end{tabular}
\end{table}

To simulate bandwidth variability, we introduce additional randomness using a log-exponential distribution, which better reflects real-world conditions. Consequently, BCCs require more time to complete each round of FL, consistent with their slower communication capabilities.

\textbf{Training setup.} The training hyperparameters are summarized in Table~\ref{tab:eval-config}. For FedGMR, we further fine-tune the patience parameter under different settings.

\textbf{Model architectures.} The architectural details of the employed models are provided in Table~\ref{tab:arch}.


\begin{table}[t]
\centering
\caption{Evaluation configurations. $k$ denotes the index of the fastest-client rounds in the semi-asynchronous setting.}
\label{tab:eval-config}
\resizebox{\linewidth}{!}{%
\setlength{\tabcolsep}{6pt}
\renewcommand{\arraystretch}{1.15}
\begin{tabular}{lcccc}
\toprule
 & \textbf{FEMNIST} & \textbf{CIFAR-10} & \textbf{ImageNet-100} & \textbf{StackOverflow} \\
\midrule
Learning rate at round $k$ 
  & $0.25$ 
  & $0.25$ 
  & $0.05 \cdot 0.5^{\tfrac{k}{10000}}$ 
  & $2\times 10^{-3}$ (MultiStepLR: $\{2500,5000,7500\}$, $\gamma=0.5$) \\
Mini-batch size / local iterations
  & $20\,/\,5$ & $20\,/\,5$ & $20\,/\,5$ 
  & $16\,/\,10$ \\
Restoration interval $k_{\mathrm{rest}}$ (rounds)
  & $25$ & $50$ & $50$ 
  & $25$ \\

Total training time (sec)
  & $80{,}000$ & $250{,}000$ & $280{,}000$ 
  &  $80{,}000$ \\
\bottomrule
\end{tabular}
}
\end{table}

\begin{table}[t]
\centering
\caption{Model architectures and parameter sizes.}
\label{tab:arch}
\setlength{\tabcolsep}{5pt}
\renewcommand{\arraystretch}{1.2}
\resizebox{\linewidth}{!}{%
\begin{tabular}{l p{2.8cm} p{4.4cm} p{4.6cm} p{4.8cm}}
\toprule
\textbf{Architecture} & \textbf{Conv-2} & \textbf{VGG-11} & \textbf{ResNet-18} & \textbf{Transformer-2L (Ours)} \\
\midrule
Convolutional 
  & 32, pool; 64, pool 
  & 64, pool; 128, pool; $2 \times 256$, pool; $2 \times 512$, pool; $2 \times 512$, pool
  & 64, pool; $2 \times [64,64]$; $2 \times [128,128]$; $2 \times [256,256]$; $2 \times [512,512]$
  & token emb.\ 96; self-attn (4 heads); FFN 384; $\times 2$ layers; dropout 0.1 (vocab=10k, max\_len=128) \\
Fully-connected 
  & 2048, 62 (input: 3136) 
  & 512, 512, 10 (input: 512) 
  & avgpool, 100 (input: 512)
  & LM head: linear (96 $\rightarrow$ 10k) + softmax \\
Conv/FC/all params 
  & 52.1K / 6.6M / 6.6M 
  & 9.2M / 530.4K / 9.8M 
  & 11.2M / 102.6K / 11.3M
  & -- / -- / 17.26M \\
\bottomrule
\end{tabular}
}
\end{table}

\paragraph{StackOverflow language modeling.}
We evaluate a 2-layer Transformer on the StackOverflow simulation dataset \cite{bonawitz2019towards}. We use a 10k-word vocabulary with sequence length 128 and up to 50 samples per user. We subsample 6,000 users for training and 1,500 users for evaluation, yielding 299,968 training samples and 31,571 evaluation samples. The model uses embedding size 96, hidden size 384, and 4 attention heads, where the width is scaled by a client-specific model rate to enable heterogeneous training.
\textbf{Metric.} We report \emph{token-level next-token top-1 accuracy}: given input tokens $(x_1,\ldots,x_{T-1})$, the model predicts $x_t$ for each position $t\in\{2,\ldots,T\}$ under teacher forcing, and accuracy is computed as the fraction of positions where $\arg\max p(\cdot \mid x_{<t}) = x_t$. Padding positions are masked out when computing the metric. We aggregate accuracy over all tokens in the evaluation set.

\section{Ablation experiments with different parts of FedGMR}
\label{sec:ablation}
\subsection{Ablation setup.}  
We decompose FedGMR into its key components to isolate their contributions:
(1) the core GMR mechanism,  
(2) asynchronous aggregation, and  
(3) the optional buffer and IMS modules, which improve practical efficiency but
incur additional overhead. For fair comparison, all variants maintain the same restoration sensitivity so
that differences arise solely from removing each module rather than modifying
the restoration schedule or density trajectory.

\subsection{Result analysis.} 
As shown in Tab.~\ref{tab:ablation}, the ablation results are consistent with the previous analysis in Fig.~\ref{fig:Ab}: \textbf{GMR} is the primary source of improvement, and \textbf{asynchrony} is beneficial when restoration introduces latency imbalance. Since these effects are already discussed in the main text, we focus here on the two auxiliary modules, \textbf{buffering} and \textbf{incremental model splitting (IMS)}.

\textbf{IMS (communication efficiency).}
IMS is designed to reduce \emph{server-to-client} communication by exploiting the nested structure of sub-models: when a client’s density increases, the server sends only the required increments rather than the entire sub-model. In most cases, this reduces per-round transmission time and is therefore beneficial at scale. However, its impact on final accuracy can exhibit mild fluctuations across tasks and regimes (e.g., due to randomness in training dynamics or when communication is not the dominant bottleneck). We further provide a dedicated study showing that IMS consistently reduces communication cost (Appendix~\ref{sec:eff_ims}).

\textbf{Buffering (stability under asynchrony).}
The buffer aims to stabilize asynchronous aggregation by smoothing the set of participating client updates. Its effect is more nuanced: in some settings, removing the buffer can accelerate early progress because the global trajectory is more quickly driven by high-capacity (full-model) clients. Nevertheless, without buffering the aggregation can become less stable after restoration, and we often observe a small but consistent drop when both buffering and IMS are removed (see \textbf{w/o (Buff, IMS)}). For example, on ImageNet-100 (High IID), removing the buffer reduces accuracy from $60.84\%$ to $56.50\%$, suggesting that buffering can improve robustness under strong system heterogeneity.

\textbf{Robustness without auxiliary modules.}
Importantly, even without buffering and IMS, FedGMR remains competitive and in most cases still outperforms prior baselines, indicating that the core gains stem from GMR (with MA) rather than from auxiliary engineering components.

\begin{table*}[t]
\centering
\caption{Ablation of FedGMR components across datasets and heterogeneity levels. }
\renewcommand{\arraystretch}{0.5}
\resizebox{\textwidth}{!}{
\begin{tabular}{llccccccccll}
\toprule
\multirow{2}{*}{Hetero.} & \multirow{2}{*}{Method}
& \multicolumn{2}{c}{FEMNIST (70k)} 
& \multicolumn{2}{c}{CIFAR-10 (220k)}
& \multicolumn{2}{c}{ImageNet-100 (250k)} & \multicolumn{2}{c}{StackOverflow(70k)}\\
\cmidrule(lr){3-4}\cmidrule(lr){5-6}\cmidrule(lr){7-8}\cmidrule(lr){9-10}
& & IID & Non-IID & IID & Non-IID & IID & Non-IID  & IID&Non-IID\\
\midrule
\multirow{6}{*}{High}
& FedGMR 
& \textbf{82.67} {\tiny$\pm$0.21} & 81.86 {\tiny$\pm$0.19}
& 84.52 {\tiny$\pm$0.14} & \textbf{81.68} {\tiny$\pm$0.28}
& \textbf{60.84} {\tiny$\pm$0.41} & 58.01 {\tiny$\pm$0.50}
& \textbf{30.00} {\tiny$\pm$0.01} & 30.07 {\tiny$\pm$0.02} \\

& w/o Asyn
& 80.59 {\tiny$\pm$0.36} & 78.80 {\tiny$\pm$0.54}
& 84.36 {\tiny$\pm$0.28} & 80.53 {\tiny$\pm$0.22}
& 58.66 {\tiny$\pm$0.74} & 57.85 {\tiny$\pm$1.50}
& 29.84 {\tiny$\pm$0.02} & \textbf{30.11} {\tiny$\pm$0.03} \\

& w/o GMR
& 82.18 {\tiny$\pm$0.10} & 79.91 {\tiny$\pm$0.18}
& 81.65 {\tiny$\pm$0.26} & 71.93 {\tiny$\pm$0.45}
& 53.49 {\tiny$\pm$0.53} & 48.28 {\tiny$\pm$0.73}
& 29.68 {\tiny$\pm$0.02} & 29.76 {\tiny$\pm$0.02} \\

& w/o Buff
& 82.19 {\tiny$\pm$0.13} & 82.08 {\tiny$\pm$0.25}
& 83.98 {\tiny$\pm$0.18} & 79.97 {\tiny$\pm$0.26}
& 60.62 {\tiny$\pm$0.42} & 58.38 {\tiny$\pm$0.52}
& 29.89 {\tiny$\pm$0.01} & 30.02 {\tiny$\pm$0.01} \\

& w/o IMS
& 82.27 {\tiny$\pm$0.18} & \textbf{82.08} {\tiny$\pm$0.16}
& \textbf{84.54} {\tiny$\pm$0.22} & 80.11 {\tiny$\pm$0.32}
& 57.79 {\tiny$\pm$0.95} & 58.05 {\tiny$\pm$0.52}
& 29.87 {\tiny$\pm$0.01} & 29.96 {\tiny$\pm$0.01} \\

& w/o (Buff, IMS)
& 82.07 {\tiny$\pm$0.23} & 81.33 {\tiny$\pm$0.24}
& 84.35 {\tiny$\pm$0.22} & 80.24 {\tiny$\pm$0.29}
& 56.50 {\tiny$\pm$1.30} & \textbf{58.40} {\tiny$\pm$0.38}
& 29.88 {\tiny$\pm$0.01} & 29.96 {\tiny$\pm$0.01} \\

\midrule
\multirow{6}{*}{Medium}
& FedGMR
& \textbf{82.94} {\tiny$\pm$0.13} & 82.35 {\tiny$\pm$0.24}
& 85.31 {\tiny$\pm$0.22} & \textbf{82.92} {\tiny$\pm$0.32}
& \textbf{62.17} {\tiny$\pm$0.45} & 60.27 {\tiny$\pm$0.63}
& \textbf{30.21} {\tiny$\pm$0.01} & \textbf{30.22} {\tiny$\pm$0.01} \\

& w/o Asyn
& 81.44 {\tiny$\pm$0.50} & 80.59 {\tiny$\pm$1.00}
& \textbf{85.39} {\tiny$\pm$0.15} & 82.20 {\tiny$\pm$0.39}
& 61.55 {\tiny$\pm$0.54} & 56.69 {\tiny$\pm$0.75}
& 30.15 {\tiny$\pm$0.03} & 30.10 {\tiny$\pm$0.04} \\

& w/o GMR
& 81.85 {\tiny$\pm$0.15} & 80.04 {\tiny$\pm$0.29}
& 84.20 {\tiny$\pm$0.21} & 80.20 {\tiny$\pm$0.30}
& 56.94 {\tiny$\pm$0.50} & 52.86 {\tiny$\pm$0.49}
& 29.92 {\tiny$\pm$0.02} & 29.91 {\tiny$\pm$0.01} \\

& w/o Buff
& 82.44 {\tiny$\pm$0.15} & 82.43 {\tiny$\pm$0.11}
& 85.36 {\tiny$\pm$0.12} & 82.29 {\tiny$\pm$0.33}
& 61.44 {\tiny$\pm$0.32} & 59.62 {\tiny$\pm$0.40}
& 30.19 {\tiny$\pm$0.01} & 30.08 {\tiny$\pm$0.01} \\

& w/o IMS
& 82.38 {\tiny$\pm$0.24} & \textbf{82.44} {\tiny$\pm$0.21}
& 84.91 {\tiny$\pm$0.15} & 81.10 {\tiny$\pm$0.38}
& 61.21 {\tiny$\pm$0.45} & \textbf{60.41} {\tiny$\pm$0.41}
& 30.13 {\tiny$\pm$0.01} & 30.13 {\tiny$\pm$0.01} \\

& w/o (Buff, IMS)
& 82.39 {\tiny$\pm$0.17} & 82.17 {\tiny$\pm$0.18}
& 85.09 {\tiny$\pm$0.20} & 81.74 {\tiny$\pm$0.21}
& 61.06 {\tiny$\pm$0.38} & 59.04 {\tiny$\pm$0.35}
& 30.13 {\tiny$\pm$0.01} & 30.12 {\tiny$\pm$0.01} \\

\midrule
\multirow{6}{*}{Low}
& FedGMR
& \textbf{83.85} {\tiny$\pm$0.19} & \textbf{82.89} {\tiny$\pm$0.30}
& \textbf{85.99} {\tiny$\pm$0.18} & 83.76 {\tiny$\pm$0.23}
& \textbf{63.89} {\tiny$\pm$0.50} & 62.37 {\tiny$\pm$0.42}
& \textbf{30.55} {\tiny$\pm$0.01} & \textbf{30.43} {\tiny$\pm$0.01} \\

& w/o Asyn
& 81.21 {\tiny$\pm$0.42} & 77.62 {\tiny$\pm$1.20}
& 65.40 {\tiny$\pm$7.10} & 66.64 {\tiny$\pm$6.10}
& 60.60 {\tiny$\pm$0.74} & 60.30 {\tiny$\pm$0.52}
& 30.31 {\tiny$\pm$0.03} & 30.28 {\tiny$\pm$0.01} \\

& w/o GMR
& 82.13 {\tiny$\pm$0.15} & 79.91 {\tiny$\pm$0.18}
& 84.81 {\tiny$\pm$0.14} & 81.09 {\tiny$\pm$0.31}
& 57.40 {\tiny$\pm$0.68} & 55.38 {\tiny$\pm$0.80}
& 30.14 {\tiny$\pm$0.02} & 30.14 {\tiny$\pm$0.01} \\

& w/o Buff
& 83.52 {\tiny$\pm$0.19} & 82.61 {\tiny$\pm$0.28}
& 85.61 {\tiny$\pm$0.18} & 83.63 {\tiny$\pm$0.23}
& 63.19 {\tiny$\pm$0.41} & \textbf{62.46} {\tiny$\pm$0.53}
& 30.27 {\tiny$\pm$0.01} & 30.33 {\tiny$\pm$0.02} \\

& w/o IMS
& 83.23 {\tiny$\pm$0.22} & 82.75 {\tiny$\pm$0.24}
& 85.65 {\tiny$\pm$0.17} & \textbf{83.82} {\tiny$\pm$0.16}
& 62.51 {\tiny$\pm$0.60} & 61.80 {\tiny$\pm$0.39}
& 30.37 {\tiny$\pm$0.01} & 30.23 {\tiny$\pm$0.01} \\

& w/o (Buff, IMS)
& 83.25 {\tiny$\pm$0.16} & 82.35 {\tiny$\pm$0.27}
& 85.52 {\tiny$\pm$0.17} & 83.23 {\tiny$\pm$0.22}
& 63.45 {\tiny$\pm$0.35} & 61.90 {\tiny$\pm$0.37}
& 30.34 {\tiny$\pm$0.01} & 30.23 {\tiny$\pm$0.01} \\

\bottomrule
\end{tabular}}
\label{tab:ablation}
\end{table*}

\subsection{Effect of IMS.} 
\label{sec:eff_ims}

 We evaluate the efficiency of IMS on the FEMNIST dataset under a synchronous setting for 1000 communication rounds, and the total training time (in seconds) was recorded. The advantage of IMS becomes more pronounced as the number of clients increases. Without IMS, training time grows rapidly with more clients—for example, from 8934s (5 clients) to 24104s (30 clients) under high heterogeneity—due to repeated transmission of similar sub-models.

In contrast, with IMS, the increase in training time remains nearly linear and much slower—for example, from 8375s to 8883s under the same setting—highlighting IMS’s effectiveness in reusing shared parameters and mitigating redundancy during server-to-client communication.
\begin{table}[htbp]
\centering
\begin{tabular}{lcccccc}
\toprule
\multirow{2}{*}{Method} 
& \multicolumn{3}{c}{High Heterogeneity} 
& \multicolumn{3}{c}{Low Heterogeneity} \\
\cmidrule(lr){2-4}\cmidrule(lr){5-7}
& 5 Clients & 10 Clients & 30 Clients & 5 Clients & 10 Clients & 30 Clients \\
\midrule
IMS     & 8375  & 8651  & 8883  & 8551  & 8738  & 8978  \\
w/o IMS & 8934  & 11999 & 24104 & 10966 & 15999 & 36136 \\
\bottomrule
\end{tabular}
\caption{Communication cost comparison (seconds) with and without IMS under different 
heterogeneity levels and client counts.}
\label{tab:ims_cost}
\end{table}

\section{Additional Experimental Analysis}
\label{sec:additional_experimental_analysis}

This section collects several additional analyses motivated by questions about the generality of GMR, the sensitivity of the restoration trigger, whether early stopping is essential, and how restoration changes communication behavior under heterogeneous bandwidths. Together, these results show that the gains of FedGMR mainly come from restoration itself, remain stable under reasonable trigger choices, and are consistent with the time-based evaluation used in the main paper.

\subsection{Applying GMR to Other MHFL Methods}
\label{sec:gmr_other_methods_appendix}

\textbf{Motivation.} One natural question is whether the improvement comes from the restoration mechanism itself or from some implementation detail specific to FedGMR.

\textbf{What it shows.} We apply GMR to HeteroFL, FjORD, FedRolex, and FIARSE on FEMNIST, covering MHFL methods based on both structured and unstructured sub-model construction. To separate the gain from restoration from that of the supporting implementation, Table~\ref{tab:gmr_other_methods_plugin} compares three configurations: the original method, the method with the auxiliary asynchronous workflow (\textbf{+Aux}), and the complete configuration with both the workflow and GMR (\textbf{+Aux+GMR}). Adding GMR consistently improves performance over \textbf{+Aux}, including from $80.42$ to $82.29$ for HeteroFL, from $80.36$ to $81.58$ for FedRolex, and from $79.56$ to $81.97$ for FIARSE. These results show that the gain from model capacity restoration transfers across different MHFL methods rather than depending on a particular pruning or aggregation design.

\begin{figure}[t]
\centering
\includegraphics[width=0.82\textwidth]{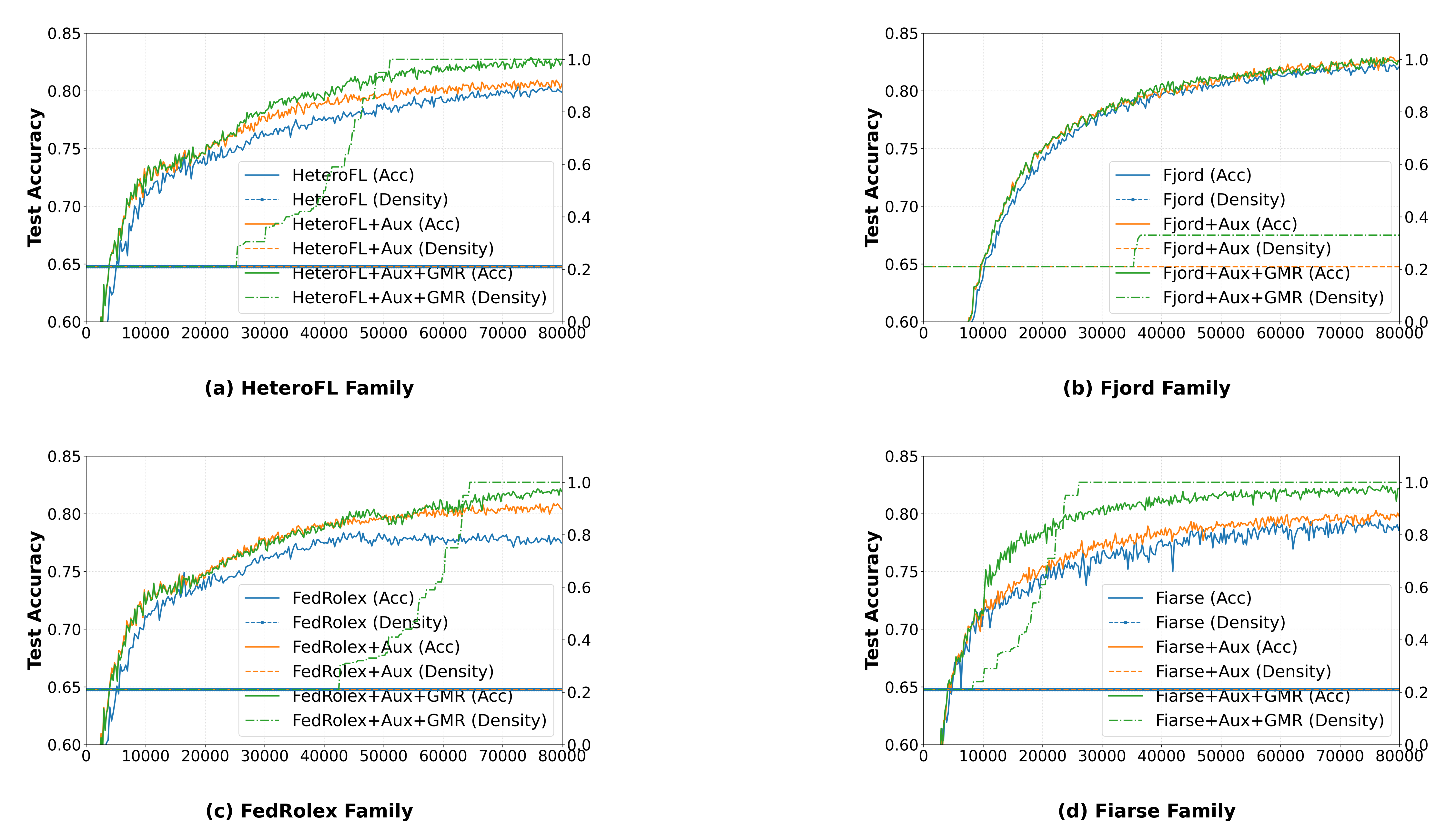}
\caption{Applying GMR to other MHFL methods on FEMNIST. The improvement remains visible across different baselines.}
\label{fig:gmr_other_methods}
\end{figure}

\begin{table}[t]
\centering
\caption{Applying GMR as a plug-in to different MHFL methods on FEMNIST. Results are reported as mean $\pm$ standard deviation (\%).}
\label{tab:gmr_other_methods_plugin}
\small
\begin{tabular}{lccc}
\toprule
Method & Original & +Aux & +Aux+GMR \\
\midrule
HeteroFL & $79.80\pm0.26$ & $80.42\pm0.20$ & $\mathbf{82.29\pm0.22}$ \\
FjORD & $81.76\pm0.17$ & $82.16\pm0.26$ & $\mathbf{82.20\pm0.29}$ \\
FedRolex & $77.83\pm0.27$ & $80.36\pm0.19$ & $\mathbf{81.58\pm0.23}$ \\
FIARSE & $78.77\pm0.43$ & $79.56\pm0.25$ & $\mathbf{81.97\pm0.20}$ \\
\bottomrule
\end{tabular}
\end{table}

\subsection{Patience Analysis}
\label{sec:patience_analysis_appendix}

\textbf{Motivation.} Since the main method uses a stagnation-based trigger, an important question is whether the performance gain is overly sensitive to the patience hyperparameter.

\textbf{What it shows.} Table~\ref{tab:patience_summary_appendix} summarizes the sensitivity of FedGMR to the patience setting across datasets. Overall, FedGMR consistently improves over the no-restoration baseline across normal patience ranges, while degradation appears only under extreme settings. This indicates that the restoration benefit is robust and does not rely on one narrowly tuned patience choice.

\begin{table}[t]
\centering
\caption{Cross-dataset sensitivity summary for the patience setting (Acc@time-line, \%).}
\label{tab:patience_summary_appendix}
\small
\resizebox{\textwidth}{!}{
\begin{tabular}{lcccc}
\toprule
Dataset & Baseline (w/o GMR) & FedGMR sweep range & Best FedGMR & Best gain \\
\midrule
FEMNIST (70k s) & 79.69 & 79.63--81.86 & 81.86 & +2.17 \\
CIFAR-10 (220k s) & 71.93 & 76.17--81.68 & 81.68 & +9.75 \\
ImageNet-100 (250k s, non-IID) & 48.28 & 50.22--58.01 & 58.01 & +9.73 \\
StackOverflow (70k s) & 29.76 & 29.31--30.07 & 30.07 & +0.31 \\
\bottomrule
\end{tabular}}
\end{table}

\begin{figure}[t]
\centering
\includegraphics[width=0.82\textwidth]{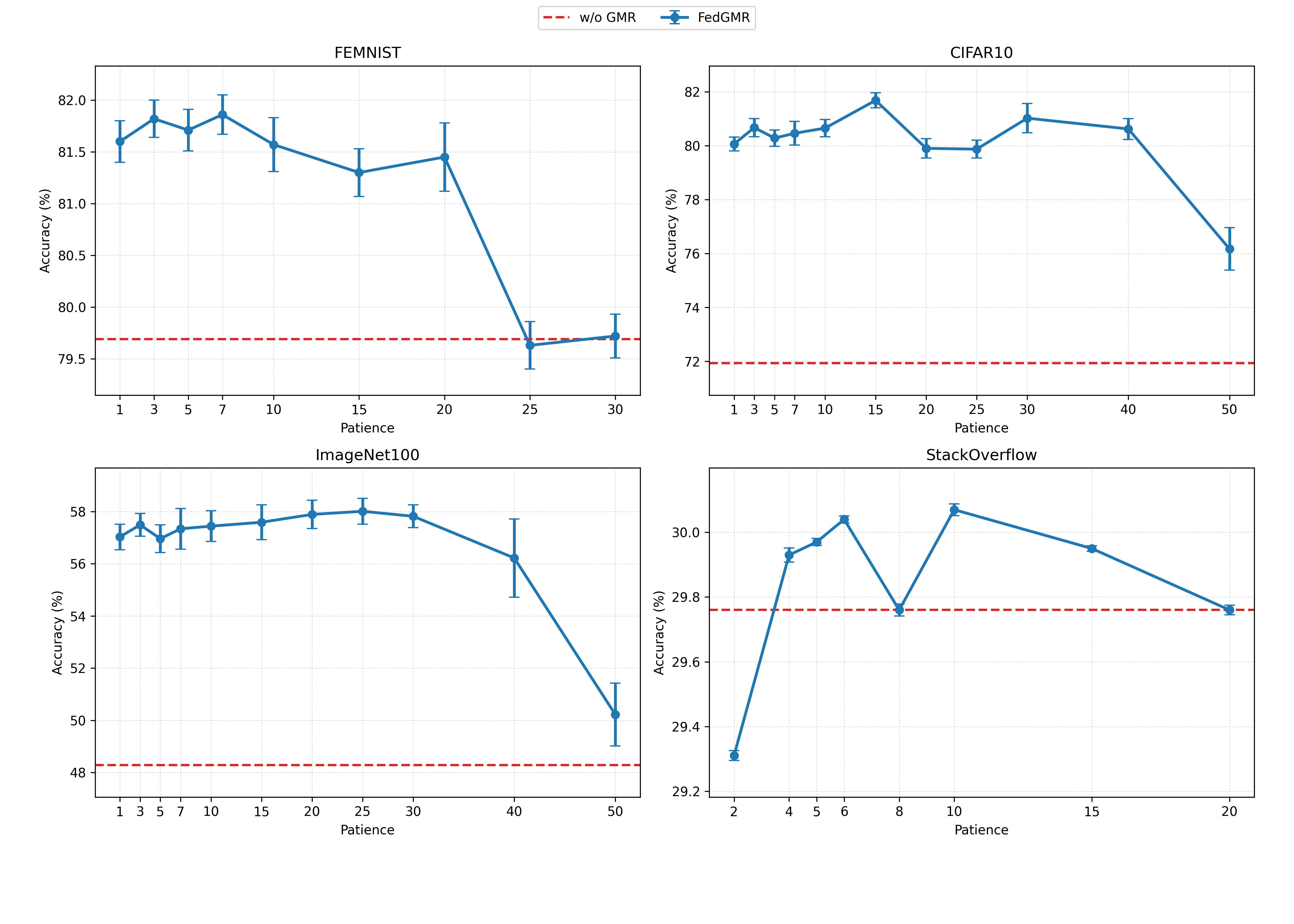}
\caption{Patience analysis under high heterogeneity. FedGMR remains beneficial across a broad range of patience values.}
\label{fig:patience_analysis_appendix}
\end{figure}

\subsection{Fixed-Time Restoration}
\label{sec:fixed_time_restoration_appendix}

\textbf{Motivation.} Another question is whether server-side early stopping is essential, or whether the gain mainly comes from gradually restoring model capacity itself.

\begin{figure}[t]
\centering
\includegraphics[width=\textwidth]{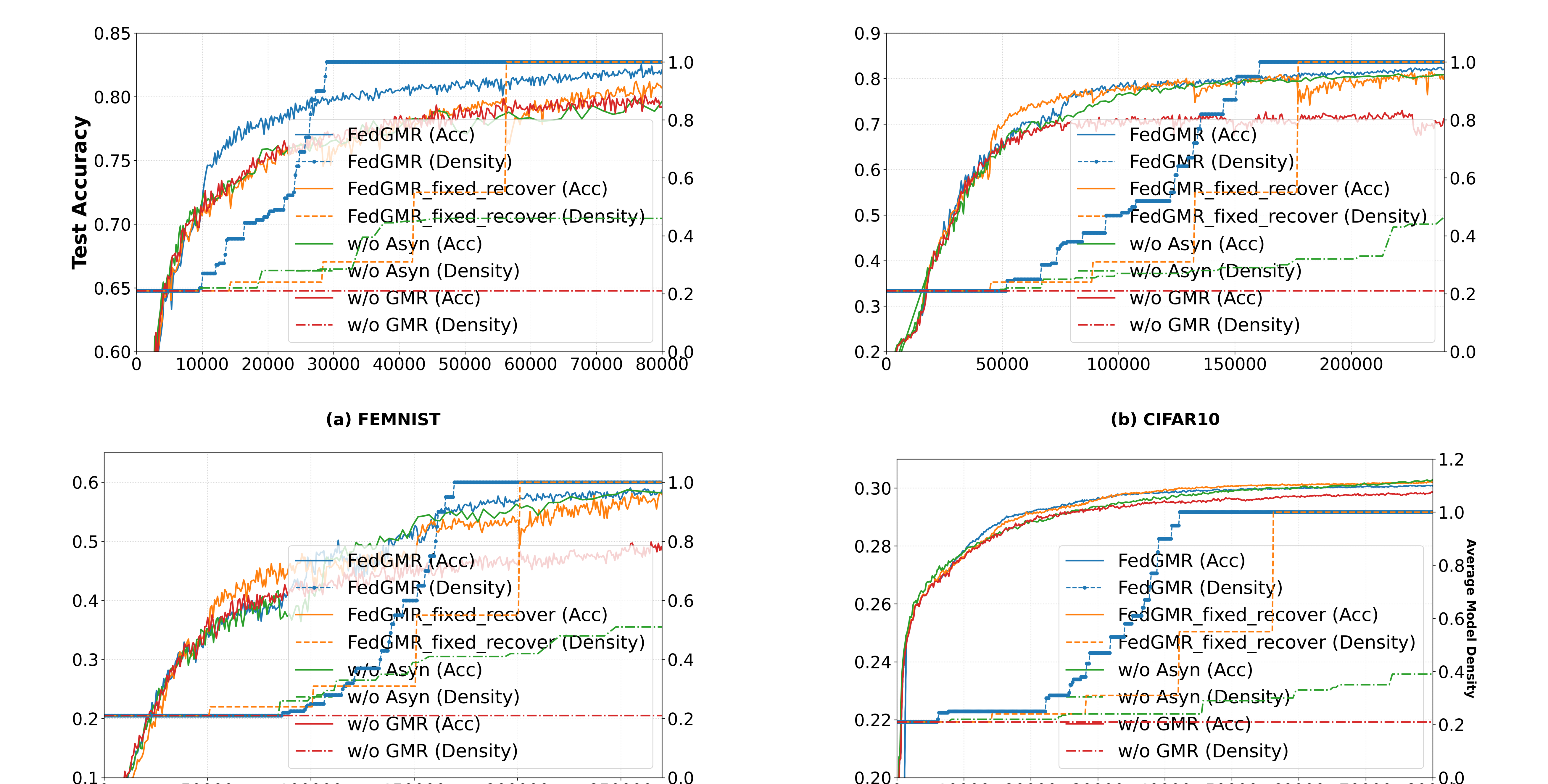}
\caption{Fixed-time restoration results. Even without server-side early stopping, gradual restoration still improves performance.}
\label{fig:fixed_time_restoration_appendix}
\end{figure}

\textbf{What it shows.} Besides the stagnation trigger used in the main method, we also tested a fixed-time variant that restores by training progress at $0.2/0.4/0.6/0.8/1.0$ with the ladder $0.05/0.1/0.2/0.5/1.0$. This removes early stopping entirely, yet still preserves the overall benefit of restoration. Therefore, the advantage of GMR does not depend solely on one trigger heuristic. For clarity, we report the updated curves in Fig.~\ref{fig:fixed_time_restoration_appendix} and list the corresponding Acc@time summary separately in Table~\ref{tab:fixed_time_restoration_summary_appendix}.

\begin{table}[t]
\centering
\caption{Acc@time summary (\%) for the fixed-time restoration analysis.}
\label{tab:fixed_time_restoration_summary_appendix}
\small
\resizebox{\textwidth}{!}{
\begin{tabular}{lcccc}
\toprule
Dataset & GMR (ES) & GMR (Fixed) & w/o Asyn & w/o GMR \\
\midrule
FEMNIST (70k s) & $81.71 \pm 0.20$ & $80.07 \pm 0.32$ & $78.80 \pm 0.54$ & $79.51 \pm 0.35$ \\
CIFAR-10 (220k s) & $81.68 \pm 0.28$ & $80.42 \pm 0.48$ & $80.53 \pm 0.22$ & $71.93 \pm 0.45$ \\
ImageNet100 (250k s) & $58.01 \pm 0.50$ & $56.38 \pm 0.93$ & $57.85 \pm 1.50$ & $48.28 \pm 0.73$ \\
StackOverflow (70k s) & $30.04 \pm 0.0088$ & $30.15 \pm 0.018$ & $30.11 \pm 0.034$ & $29.76 \pm 0.019$ \\
\bottomrule
\end{tabular}}
\end{table}

\subsection{Communication Example}
\label{sec:communication_example_appendix}

\textbf{Motivation.} A further question is why we report wall-clock time rather than a single per-round communication number. In FedGMR, communication per participation changes with both client bandwidth and the restored model density.

\textbf{What it shows.} Table~\ref{tab:communication_example_appendix} gives a concrete FEMNIST example under fixed-time restoration. The serialized full model is $25.98$ MB, and one participation includes one download and one upload. For slower clients, restoration increases the communicated model size and can reduce their update frequency. At the same time, those later, larger-capacity updates are more informative than indefinitely repeating very small-model updates. This illustrates why time-based evaluation is more appropriate for MHFL with restoration.

\begin{table}[t]
\centering
\caption{Communication example on FEMNIST under fixed-time restoration.}
\label{tab:communication_example_appendix}
\small
\resizebox{\textwidth}{!}{
\begin{tabular}{lccc}
\toprule
Item & 1.0x client & 0.2x client & 0.05x client \\
\midrule
Bandwidth & $20\downarrow / 5\uparrow$ MB/s & $4\downarrow / 1\uparrow$ MB/s & $1\downarrow / 0.25\uparrow$ MB/s \\
Before first restoration & $\rho=1.0$; $25.98$ MB model; $\sim 1$ train/round & $\rho=0.2$; $5.20$ MB model; $\sim 1$ train/round & $\rho=0.05$; $1.30$ MB model; $\sim 1$ train/round \\
After first restoration & $\rho=1.0$; $25.98$ MB model; $\sim 1$ train/round & $\rho=0.2$; $5.20$ MB model; $\sim 1$ train/round & $\rho=0.1$; $2.60$ MB model; $\sim 1$ train / $2$ rounds \\
\bottomrule
\end{tabular}}
\end{table}


\end{document}